\documentclass[superscriptaddress,aps,prx,preprintnumbers,twocolumn,amsmath,amssymb,showpacs]{revtex4-2}

\usepackage{graphicx}

\usepackage{dcolumn}
\usepackage{bm}
\usepackage{amssymb}
\usepackage{amsmath}
\usepackage{amsthm}
\usepackage{chemarrow}

\usepackage[dvipsnames]{xcolor}
\usepackage{mathrsfs}
\usepackage{float}
\usepackage{physics}
\usepackage{slashed}
\usepackage{bbold}
\usepackage{bbm}

\usepackage{latexsym}

\usepackage{multirow}

\usepackage{enumitem}
\setlist{nolistsep}
\usepackage{etoolbox}
\makeatletter
\preto{\@verbatim}{\topsep=0pt \partopsep=0pt }
\makeatother

\usepackage[colorlinks=true, linkcolor=teal, citecolor=teal]{hyperref}

\usepackage{orcidlink}

\newcommand{\beq}{\begin{equation}}
\newcommand{\eeq}{\end{equation}}
\newcommand{\beqa}{\begin{eqnarray}}
\newcommand{\eeqa}{\end{eqnarray}}
\newcommand{\bit}{\begin{itemize}}
\newcommand{\eit}{\end{itemize}}

\newcommand{\GeV}{\textrm{GeV}}
\newcommand{\TeV}{\textrm{TeV}}
\newcommand{\calO}{\mathcal{O}}

\newcommand{\github}[1]{\href{https://github.com/#1}{\includegraphics[width=10pt]{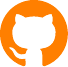}}}

\allowdisplaybreaks

\begin{document}

\preprint{FERMILAB-PUB-25-0731-T, MSUHEP-25-022}

\title{Large Neutrino ``Colliders"}

\author{Yang Bai\,{\orcidlink{0000-0002-2957-7319}}}

\affiliation{Department of Physics, University of Wisconsin-Madison, Madison, WI 53706, USA}
\affiliation{HEP Division, Argonne National Laboratory, Argonne, IL 60439, USA}

\author{Keping Xie\,{\orcidlink{0000-0003-4261-3393}}}

\affiliation{State Key Laboratory of Dark Matter Physics \& Key Laboratory for Particle Astrophysics and Cosmology (MOE) \& Shanghai Key Laboratory for Particle Physics and Cosmology, Tsung-Dao Lee Institute \& School of Physics and Astronomy, Shanghai Jiao Tong University, Shanghai 201210, China\looseness=-1}
\affiliation{Department of Physics and Astronomy, Michigan State University, East Lansing, MI 48824, USA}

\author{Bei Zhou\,{\orcidlink{0000-0003-1600-8835}}}

\affiliation{Theory Division, Fermi National Accelerator Laboratory, Batavia, IL 60510, USA}
\affiliation{Kavli Institute for Cosmological Physics, University of Chicago, Chicago, IL 60637, USA\looseness=-1}

\begin{abstract}
We propose using current and future large-volume neutrino telescopes as ``Large Neutrino Colliders" (L$\nu$Cs) to explore TeV-scale physics beyond the Standard Model. Cosmic neutrinos with energies above 100 PeV colliding with nucleons in the detector reach center-of-mass energies beyond the 14 TeV limit of the Large Hadron Collider (LHC). Using recently predicted and measured high-energy and ultra-high-energy neutrino fluxes from IceCube and KM3NeT, we estimate mass-scale sensitivities for representative new physics scenarios at 1--30 km$^3$ L$\nu$Cs. Our results demonstrate that L$\nu$Cs provide a novel avenue to probe multi-TeV particles with sensitivities comparable to, or even surpassing, those of the LHC.
\end{abstract}

\maketitle

\section{Introduction}
For nearly a century, the search for new particles has driven advances in fundamental physics, beginning with the discovery of the positron by Carl D. Anderson via cosmic-ray detection in a cloud chamber~\cite{Anderson:1933mb}. The development of particle accelerators and head-on colliders has since enabled the construction and experimental validation of the Standard Model (SM)~\cite{Weinberg:2018apv}. Despite its success, the SM fails to account for several key phenomena, including dark matter, neutrino masses, the matter--antimatter asymmetry, dark energy, and the hierarchy between the electroweak and Planck scales. These shortcomings strongly motivate the search for physics beyond the SM (BSM), which may involve new heavy particles at the multi-TeV scale~\cite{ATLAS:2024lda}. Such searches are traditionally pursued at high-energy colliders, including the Large Hadron Collider (LHC) operating at a center-of-mass (COM) energy of 14 TeV, and proposed future machines with even higher energies~\cite{FCC:2018evy,FCC:2018vvp,CEPCStudyGroup:2023quu,CEPCStudyGroup:2025kmw}.

Given the high cost and long lead time associated with designing and constructing future colliders (see Section III-A of the Supplemental Material for comparison), it is worthwhile to explore alternative approaches for probing heavy new particles. In this Letter, we propose adopting naturally occurring high-energy cosmic rays as an effective ``beam" that collides with particles in a detector---forming a ``Cosmic Collider"---to search for BSM particles. While traditional colliders typically involve two accelerated particles, our approach utilizes one cosmic particle and one stationary target. Among the various cosmic-ray species, we focus on high-energy (HE) and ultra-high-energy (UHE) cosmic neutrinos, and investigate the physics potential of a ``Large Neutrino Collider" (L$\nu$C), which offers optimal sensitivity for discovering multi-TeV-scale new particles~\footnote{The L$\nu$C concept is distinct from ``collider neutrinos'', i.e., neutrinos produced in forward directions at the LHC and studied at dedicated facilities such as FASER$\nu$, SND@LHC, and the proposed forward physics facility. See Refs.~\cite{Feng:2022inv, Ariga:2025qup, Machado:2025ktc} for recent reviews and references therein.}. Early explorations of this idea can be found in Refs.~\cite{Alvarez-Muniz:2001efi,Albuquerque:2003mi,Alvarez-Muniz:2002snq,Anchordoqui:2005ey,Anchordoqui:2006wc,Romero:2009vu,Dev:2016uxj,Becirevic:2018uab,Babu:2022fje,Kirk:2023fin}. Only with the recently measured HE and UHE cosmic-neutrino fluxes by IceCube~\cite{IceCube:2024fxo} and KM3NeT~\cite{KM3NeT:2025npi} has the L$\nu$C become a realistic and timely framework for exploring TeV-scale new physics. This Letter provides the first explicit and systematic proposal of this concept, thereby broadening the discovery potential of both current and future neutrino telescopes.

\begin{figure}[t!!]
\includegraphics[width=0.95\columnwidth]{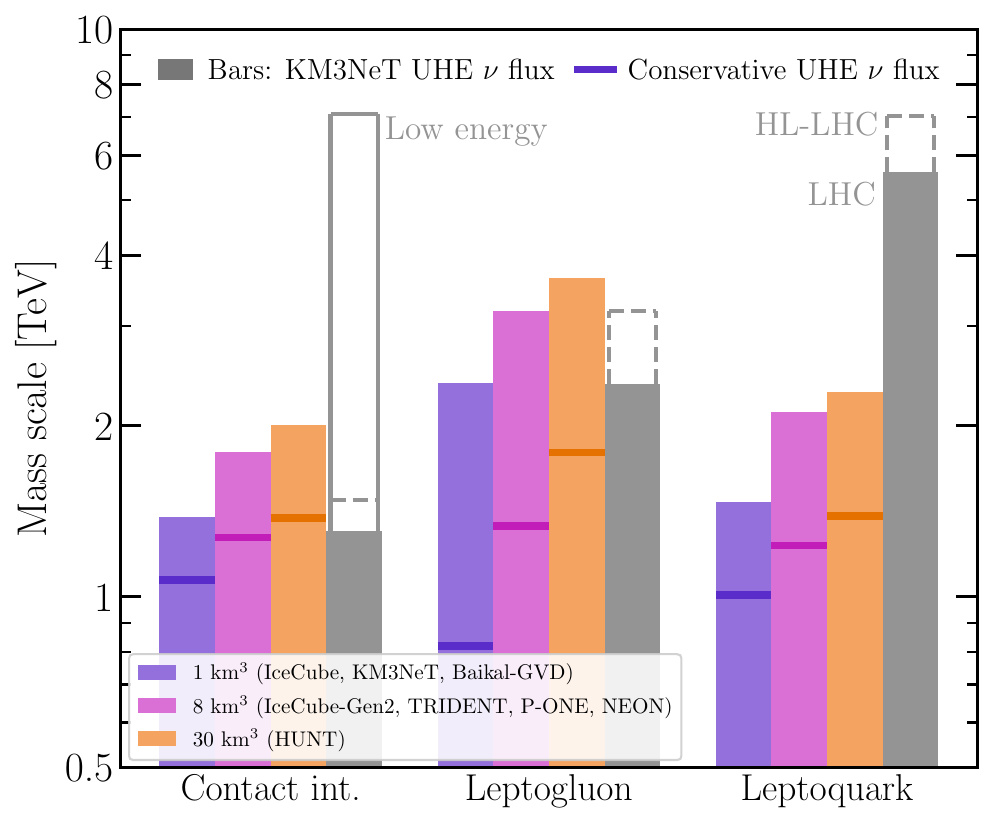}

\caption{Projected mass-scale reaches at different lepton-hadron L$\nu$C configurations, compared with those at the LHC and HL-LHC. The analysis in this study includes both starting and throughgoing muon tracks. 
The projected sensitivities are shown separately for the KM3NeT-normalized UHE flux and a conservative, theoretically motivated UHE flux.
}
\label{fig:histogram}
\end{figure}

Fig.~\ref{fig:histogram} summarizes our results, presenting the attainable mass scales for three benchmark models at neutrino telescopes with different instrumented volumes.
With the KM3NeT-measured UHE flux, the projected sensitivities for a representative contact interaction and the leptogluon model can exceed those of both the LHC and HL-LHC, whereas the leptoquark achieves sensitivities comparable to those of the two terrestrial colliders.

The conservative flux reduces all three reaches, while retaining complementary sensitivity.

This remarkable performance arises from the exceptionally high COM energy of the L$\nu$C, which enables the exploration of very heavy particles, and from its sensitivity to BSM scenarios that induce enhanced neutrino--nucleon or neutrino--electron scattering cross sections, even with relatively low luminosity. The remainder of the Letter now defines the fluxes, detector response, SM backgrounds, statistical test, and benchmark interactions before interpreting these headline reaches.

For orientation, we use three representative detector volumes throughout: (1) 1~km$^3$, corresponding to IceCube~\cite{IceCube_web} and the target instrument volumes of KM3NeT or Baikal-GVD~\cite{KM3Net:2016zxf,Allakhverdyan:2021vkk,Aynutdinov:2023ifk}; (2) 8~km$^3$, representative of IceCube-Gen2~\cite{IceCube-Gen2:2020qha} and other proposed multi-km$^3$ arrays such as TRIDENT~\cite{Ye:2023dch}, P-ONE~\cite{P-ONE:2020ljt}, and NEON~\cite{Zhang:2024slv}; and (3) 30~km$^3$, representative of HUNT-like concepts~\cite{Huang:2023mzt}.
In addition, we study three neutrino-nucleon benchmarks in the headline comparison: (1) a \emph{contact interaction} describes charged-current-like neutrino-quark scattering mediated by particles heavier than the resolved collision energy, and its quoted scale is the effective-field-theory (EFT) cutoff $\Lambda$; (2) a \emph{leptogluon} is a color-octet excited lepton resonantly produced from a neutrino and a gluon, and its quoted scale is $M_{\mu_8}$, the mass of the lighter charged leptogluon $\mu_8$;
(3) a \emph{leptoquark} is a boson coupling directly to a lepton and a quark, and its quoted scale is $M_{S_1}$, the mass of the scalar singlet leptoquark $S_1$. Thus, the bars above compare representative sensitivity reaches under the stated benchmark couplings, rather than a model-independent common mass parameter.

\section{The Large Neutrino Collider Framework}
\subsection{Detector configurations and collider kinematics}

Several large-volume neutrino telescopes are operational or under construction. IceCube has instrumented about 1~km$^3$ at the South Pole~\cite{IceCube_web}; KM3NeT and Baikal-GVD are progressing toward a similar scale~\cite{KM3Net:2016zxf,Allakhverdyan:2021vkk,Aynutdinov:2023ifk}. Proposed next-generation arrays include IceCube-Gen2 ($\simeq8$~km$^3$)~\cite{IceCube-Gen2:2020qha}, TRIDENT ($\simeq7.5$~km$^3$)~\cite{Ye:2023dch}, the multi-km$^3$ P-ONE~\cite{P-ONE:2020ljt}, NEON ($\simeq10$~km$^3$)~\cite{Zhang:2024slv}, and HUNT ($\simeq30$~km$^3$)~\cite{Huang:2023mzt}. We therefore use representative instrumented volumes of 1, 8, and 30~km$^3$. Requiring a 0.2-km boundary layer, as in IceCube starting-event selections, gives fiducial volumes of 0.512, $0.728\times8$, and $0.819\times30$~km$^3$, respectively~\cite{IceCube:2013low,Taboada:2019lfc}. Every benchmark uses a 20-year exposure. HUNT's anticipated 10--100~TeV threshold does not affect the benchmarks considered here, whose selected events lie above $\sim 100$~TeV.

It is instructive to compare certain characteristics of the L$\nu$C with those of traditional colliders such as the LHC. First, the L$\nu$C operates as a multi-beam collider, functioning both as a lepton-hadron collider, $\nu(\bar{\nu}) + N$ with $N = p$ (proton) or $n$ (neutron), and as a lepton-lepton collider, $\nu(\bar{\nu}) + e^-$. This contrasts with the LHC or the Electron-Ion Collider (EIC)~\cite{AbdulKhalek:2021gbh}, which use fixed beam configurations of $p+p$ or $e+\mbox{Ion}$, and the Large Electron-Positron Collider (LEP), which used $e^-+e^+$. Second, the L$\nu$C offers a wide range of COM energies, determined by the energy of HE or UHE neutrinos.
For the $\nu(\bar{\nu}) + N$ configuration, the COM energy is
$\sqrt{s} \approx \sqrt{2m_p E_\nu} \approx 14\,\text{TeV}\,(E_\nu / 100\,\text{PeV})^{1/2}$, which can exceed the LHC's $\sqrt{s}=14$ TeV for sufficiently energetic cosmic neutrinos and is far above that of the EIC. Thus, for very heavy new particles, the L$\nu$C may offer an advantage over the LHC by enabling non-negligible production cross sections. Similarly, for the $\nu(\bar{\nu}) + e^-$ configuration, the COM energy is $\sqrt{s} \approx \sqrt{2m_e E_\nu} \approx 500\,\text{GeV} \times (E_\nu / 250\,\text{PeV})^{1/2}$, which can surpass the LEP's 209 GeV and even the projected COM energy of the proposed FCC-ee and CEPC colliders~\cite{FCC:2018evy,CEPCStudyGroup:2023quu,CEPCStudyGroup:2025kmw}.
Third, the integrated luminosity of the L$\nu$C depends on the sky-averaged neutrino flux $\phi_\nu$, the neutrino energy $E_\nu$, and the number of target nucleons (or electrons), $N_N = V_{\rm det} \rho_{\rm H_2O} / m_p$, where $V_{\rm det}$ is the detector volume, $\rho_{\rm H_2O}$ is the density of water and $m_p$ is the proton mass.

\begin{figure}[t!]
\includegraphics[width=1.0\columnwidth]{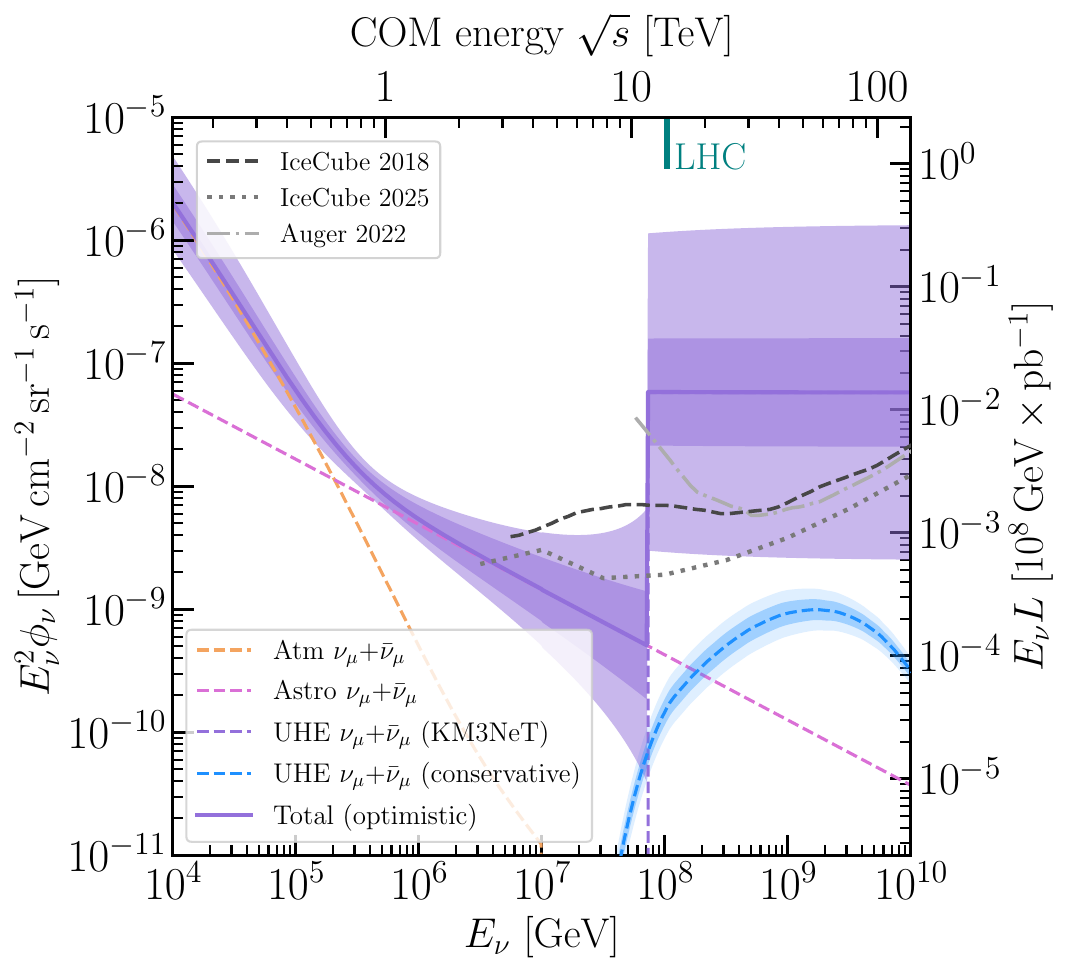}
\caption{
The high-energy (HE) and ultra-high-energy (UHE) neutrino fluxes ($\nu_\mu + \bar{\nu}_\mu$) used in this work, including three main components:
the atmospheric flux, the astrophysical flux measured by IceCube with a power-law spectrum~\cite{IceCube:2024fxo}, and a UHE component.
We consider two possibilities for the UHE component: an optimistic one motivated by the recent KM3NeT measurement~\cite{KM3NeT:2025npi} and a theory-motivated conservative one  (see Ref.~\cite{Machado:2025uhe} and references therein).
{
For visual clarity, the total flux and its uncertainty band are shown only for the optimistic case.
}
See text for details on the models and uncertainties.
{
Also shown are the upper limits from IceCube~\cite{IceCube:2018fhm,IceCube:2025ezc} and Auger~\cite{PierreAuger:2023pjg}.
}
The upper horizontal axis indicates the COM energy, $\sqrt{s} \approx \sqrt{2m_p E_\nu}$, while the right vertical axis shows the product of $E_\nu$ and the L$\nu$C luminosity $L$, assuming a 10\,km$^3$ detector volume and a 10-year observation period.
}
\label{fig:fig_flux}
\end{figure}

For reference, the luminosity is $L = 4\pi E_\nu \phi_\nu N_N T \approx 1.4 \times 10^{-2}\,\text{pb}^{-1} \times (V_{\rm det}/10\,\text{km}^3)(T/10\,\text{yr})$ at $E_\nu = 100$\,PeV for the KM3NeT flux normalization $E_\nu^2 \phi_\nu = 5.8 \times 10^{-8}\,\text{GeV}\,\text{cm}^{-2}\,\text{s}^{-1}\,\text{sr}^{-1}$~\cite{KM3NeT:2025npi}. The conservative flux is lower and energy dependent, as shown in Fig.~\ref{fig:fig_flux}.

Although this luminosity is significantly lower than that of the LHC, the L$\nu$C can still probe BSM physics scenarios with sufficiently large neutrino-scattering cross sections.

\subsection{HE and UHE neutrino fluxes}
Fig.~\ref{fig:fig_flux} shows the HE and UHE neutrino fluxes comprising three components: atmospheric, astrophysical, and an UHE component.

The atmospheric flux, dominant at $E_\nu \lesssim 10^5~\mathrm{GeV}$, is calculated using \texttt{NuFlux}~\cite{nuflux:2024} with the H3a SIBYLL23C model~\cite{MCEq,Gaisser:2011klf,Riehn:2017mfm} and has a typical uncertainty of $\sim$30\%.
At higher energies, the flux is dominated by the diffuse astrophysical component measured by IceCube~\cite{IceCube:2024fxo}, with a power-law spectrum and well-characterized normalization and spectral index. 
At the highest energies, we consider two different UHE neutrino fluxes. The first is the KM3NeT-normalized UHE flux inferred from the $\sim$220~PeV event~\cite{KM3NeT:2025npi}. The second is a lower, theory-motivated conservative flux from cosmogenic neutrinos produced when UHE cosmic rays interact with cosmological photon backgrounds.

For both astrophysical and UHE neutrinos, $1:1:1$ flavor ratios are assumed.  These fluxes, shown in Fig.~\ref{fig:fig_flux}, determine the ``luminosity'' of L$\nu$Cs, with large uncertainties at the highest energies. Specifically, the per-flavor IceCube fit used here is
\begin{equation}
\begin{aligned}
\phi_\nu^{\rm Astro}={}&1.68^{+0.19}_{-0.22}\!\times10^{-18}\\[-2pt]
&\times\left(\frac{E_\nu}{100~{\rm TeV}}\right)^{-2.58^{+0.10}_{-0.09}}
\frac{1}{{\rm GeV\,cm^2\,s\,sr}},
\end{aligned}
\label{eq:flux_astro_main}
\end{equation}
and the KM3NeT UHE normalization is
\begin{equation}
E_\nu^2\phi_\nu^{\rm UHE}
=5.8^{+10.1}_{-3.7}\!\times10^{-8}
~{\rm GeV\,cm^{-2}\,s^{-1}\,sr^{-1}} .
\label{eq:flux_uhe_main}
\end{equation}
The latter comes from one event with median energy 220~PeV and a 90\% interval of 72~PeV--2.6~EeV, so its normalization is treated as a profiled uncertainty rather than as a precisely known beam intensity.

For the conservative flux,
we adopt a cosmic-ray injection index $\Gamma=2.5$, cutoff energy $E_p^{\rm max}=250$~EeV, and moderate source redshift-evolution parameter $m=1.5$.
The $m=0$ and $m=3$ source-evolution cases define the lower and upper edges of a $2\sigma$ envelope around the physical $m=1.5$ central case.
The conservative flux remains below current IceCube and Auger UHE limits~\cite{IceCube:2018fhm,IceCube:2025ezc,PierreAuger:2023pjg}.
For more details, see Ref.~\cite{Machado:2025uhe} and references therein.

\subsection{Neutrino interactions in the SM}

At energies above 1 TeV, the dominant neutrino interaction in the SM is neutrino-nucleus deep-inelastic scattering (DIS)~\cite{Gandhi:1995tf,Xie:2023suk,Reno:2023sdm}, including charged-current (CC) and neutral-current (NC) DIS.
The cross sections per nucleon at 10~TeV are $(3\sim4)\times 10^{-35}~\textrm{cm}^{2}$  for CCDIS and $(1\sim1.5)\times 10^{-35}~\textrm{cm}^2$ for NCDIS, as demonstrated in Fig.~\ref{fig:xsLG}. At lower energies, they grow linearly with $E_\nu$, and then the growth slows down and eventually scales approximately as $E_\nu^{0.3}$ above $\sim 10^7$~GeV, due to the masses of the weak gauge bosons ($\simeq 80$~GeV).
For this work, we adopt the DIS cross sections and uncertainties calculated up to next-to-next-to-next-leading-order (N$^3$LO) in Ref.~\cite{Xie:2023suk}, which is based on CT18 parton-distribution functions (PDFs)~\cite{Hou:2019efy,Xie:2023qbn}. The theoretical uncertainties, including heavy-quark flavor and mass effects, small-$x$ resummation, and nuclear corrections, are found to be 10\%--20\%, which are relatively small with respect to the neutrino flux uncertainty. The Glashow resonance at 6.3~PeV affects $\bar\nu_e$ but not the $\nu_\mu+\bar\nu_\mu$ track sample used for the reach. We nevertheless include subleading neutrino-nucleus $W$-boson production (WBP)~\cite{Seckel:1997kk,Alikhanov:2015kla,Zhou:2019vxt,Zhou:2019frk}, which contributes about 10\% of the total neutrino-nucleus cross section, and final-state photon radiation (FSR)~\cite{Plestid:2024bva}. FSR can lower the reconstructed charged-lepton energy by up to 5\% and increase the accompanying cascade energy by up to 25\%; we apply these corrections to the differential cross section entering the event rates.

{
\subsection{Event samples and sensitivity estimate}

We focus on $\nu_\mu$ charged-current-like events containing a reconstructed muon track. A \emph{starting} track has its interaction vertex inside the fiducial volume; a \emph{throughgoing} track is produced outside the array and enters it after the muon propagates through ice or water. We retain only upgoing throughgoing tracks to suppress the atmospheric-muon background. For each detector volume, we assume 20 years of observation and include Earth absorption.

For starting tracks, the event spectrum is
\begin{align}
 \frac{dN^{\rm st}}{dE_\mu}={}&T V\rho N_A\!\int_{E_\mu}^{\infty}\!dE_\nu
 \frac{d\sigma}{dE_\mu}\nonumber\\[-2pt]
 &\times2\pi\!\int_{-1}^{1}\!d\cos\theta_z\,
 \frac{d\phi_\nu}{dE_\nu}e^{-\tau(E_\nu,\theta_z)} .
 \label{eq:starting_rate}
\end{align}
Here $T$, $V$, $\rho$, and $N_A$ are the exposure time, fiducial volume, target density, and Avogadro constant, respectively. The attenuation exponent $\tau(E_\nu,\theta_z)$ is the neutrino cross section integrated over the Earth column density along zenith angle $\theta_z$. In the signal hypothesis, the cross section in both the production rate and $\tau$ contains SM and BSM terms, whereas the background-only prediction contains the SM terms.

For upgoing throughgoing tracks, the projected area $V^{2/3}$ and the muon propagation range give
\begin{align}
\frac{dN^{\rm thr}}{dE_\mu}={}&
\frac{T N_A V^{2/3}}{\alpha+\beta E_\mu}
\int_{E_\mu}^{\infty}\!dE_\nu
\int_{E_\mu}^{E_\nu}\!dE_\mu'\,
\frac{d\sigma}{dE_\mu'} \nonumber\\[-2pt]
&\times2\pi\!\int_{-1}^{0}\!d\cos\theta_z\,
\frac{d\phi_\nu}{dE_\nu}e^{-\tau(E_\nu,\theta_z)} .
\label{eq:throughgoing_rate}
\end{align}
We use $\alpha=2.54\times10^{-3}$~GeV\,cm$^2$\,g$^{-1}$ and
$\beta=3.93\times10^{-6}$~cm$^2$\,g$^{-1}$ for continuous muon energy loss~\cite{Koehne:2013gpa}, and approximate the outgoing-muon spectrum by
$d\sigma/dE_\mu'\simeq\sigma(E_\nu)\delta[E_\mu'-\langle E_\mu'\rangle(E_\nu)]$.

We bin starting and throughgoing events separately using five reconstructed-muon-energy bins per decade. Flux and interaction uncertainties are represented by Gaussian-constrained nuisance parameters and profiled in the Poisson likelihood,
\begin{equation}
 \mathcal L=\prod_{i,j}{\rm Pois}\!\left[n_{ij}\mid s^{\rm eff}_{ij}+b^{\rm eff}_{ij}\right]
 \mathcal N(\theta^s_{ij})\mathcal N(\theta^b_{ij}) .
 \label{eq:likelihood}
\end{equation}
Here $i$ labels the energy bin and $j\in\{\mathrm{starting},\mathrm{throughgoing}\}$. The profiled predictions are $s^{\rm eff}_{ij}=s_{ij}10^{\theta^s_{ij}}$ and $b^{\rm eff}_{ij}=b_{ij}10^{\theta^b_{ij}}$, with zero-centered Gaussian priors whose widths encode the propagated flux and cross-section uncertainties in that bin. In particular, the IceCube normalization and spectral-index errors, the large KM3NeT UHE normalization error, atmospheric-flux modeling, and DIS/PDF uncertainties set the binwise nuisance widths. We use the Asimov data set $n_{ij}=s_{ij}+b_{ij}$. The median 95\% CL reach is defined by $\sqrt{q_0}=1.96$, where $q_0=-2\ln(\mathcal{L}_0/\mathcal{L}_1)$, and $\mathcal{L}_0$ and $\mathcal{L}_1$ are the profiled likelihoods under the background-only and signal-plus-background hypotheses, respectively. Thus every curve shown below is obtained from a joint profile of both event classes rather than from a fixed-flux event-count threshold.
}

{
\section{Benchmark new-physics scenarios}
}

We use four benchmarks to expose the complementary collision modes. Contact interactions test nonresonant neutrino-quark scattering; leptogluons and leptoquarks are resonances initiated by gluons and quarks, respectively; and $W'/H^-$ production illustrates neutrino-electron collisions. The first three enter the headline comparison in Fig.~\ref{fig:histogram}; the fourth is retained as a useful counterexample for which present hadron colliders remain superior. 

Our benchmarks are chosen to span the main target constituents and interaction structures accessible at L$\nu$Cs, rather than simply to highlight the models for which the projected sensitivity is strongest. In particular, the leptogluon and leptoquark scenarios represent interactions of high-energy neutrinos with gluons and quarks, respectively, while the charged-vector and charged-scalar resonances probe neutrino--electron interactions. The four-fermion SMEFT operators provide a model-independent description of nonresonant effects from heavy new physics. Taken together, these benchmarks cover contact interactions and resonant production, as well as different mediator spins and couplings to electrons, quarks, and gluons. They should therefore be regarded as representative rather than exhaustive.

\subsection{Contact interaction}

We consider dimension-6 four-fermion operators that can induce CC-like interactions. These interactions can effectively capture the effects of a broad class of new physics models, obtained by integrating out heavy states such as those arising in extra-dimensional theories~\cite{Alvarez-Muniz:2001efi,Anchordoqui:2005ey,Alvarez-Muniz:2002snq}, $W^\prime$ gauge bosons, or heavy charged scalars.
The headline benchmark is the SMEFT operator
\begin{equation}
\Delta\mathcal L=\frac{1}{\Lambda^2}\,
\mathcal O_{\ell q,2211}^{(3)},\quad
\mathcal O_{\ell q}^{(3)}
=(\bar\ell_L\bar\sigma_\mu\sigma^i\ell_L)
(\bar q_L\bar\sigma^\mu\sigma^i q_L),
\label{eq:contact_main}
\end{equation}
where 2211 selects second-generation leptons and first-generation quarks. It therefore modifies $\nu_\mu d\to\mu^-u$ and the crossed antineutrino and antiquark channels. Because this left-handed operator has the SM charged-current chirality, its cross section has the form
$\sigma=\sigma_{\rm SM}+\sigma_{\rm int}/\Lambda^2+\sigma_{\rm BSM}/\Lambda^4$.
For the positive coefficient in Eq.~(\ref{eq:contact_main}), $\hat t<0$ makes the interference with $W$ exchange destructive. Other scalar and tensor SMEFT operators do not interfere in the massless limit. We retain only events satisfying the operator-dependent validity criterion $\Lambda\gtrsim\sqrt{\hat s}/\kappa$, with $\kappa=3$ or 4 depending on the operator normalization; a resolved resonance must instead be treated as one of the explicit models below.

\subsection{Leptogluon}

The leptogluons are predicted in many BSM unification scenarios, where quarks and leptons can be composite particles, \emph{e.g.}, the ``Rishon model"~\cite{Harari:1981uh} or some schematic constructions~\cite{Harari:1979gi,Shupe:1979fv}. More recent studies can be found in Refs.~\cite{Dobrescu:2021fny,Assi:2025rjx}.
As the constituents of leptons can also be charged under QCD, one may anticipate the excited states of leptons to carry QCD charges, including the adjoint representation~\cite{Fritzsch:1981zh,Harari:1982xy}. A pioneering phenomenological study can be found in Ref.~\cite{King:1984qk}, followed by Refs.~\cite{Goncalves-Netto:2013nla,Mandal:2016csb,Almeida:2022udp,Han:2025wdy}. Some existing searches at HERA can be found in Ref.~\cite{H1:1993vsn}.

Due to the resonance enhancement and the large gluon PDF within the nucleon, the leptogluon model emerges as an ideal target for L$\nu$Cs. Although leptogluons are generally predicted in quark--lepton composite models, we use a vector-like realization containing a neutral state $\nu_8$ and charged states $\mu_8$ and $\ell_8$. For the hierarchy $M_{\ell_8}>M_{\nu_8}>M_{\mu_8}$, the observable chain is
\begin{equation}
 \nu g\to\nu_8\to\mu_8^-W^+,\qquad \mu_8^-\to\mu^-g,
 \label{eq:lg_chain}
\end{equation}
where the last decay is induced by a chromodipole interaction of strength $a/\Lambda$.
A minimal realization contains a vector-like doublet $L_8=(\nu_8,\ell_8)^T\sim(8,2)_{-1/2}$ and a charged singlet $E_8\sim(8,1)_{-1}$, with
$-\mathcal L\supset M_{L_8}\bar L_{8L}L_{8R}+M_{E_8}\bar E_{8L}E_{8R}
+y_8\bar L_{8L}\Phi E_{8R}+{\rm h.c.}$
The Higgs-induced charged-state mixing permits $M_{\ell_8}>M_{\nu_8}>M_{\mu_8}$ and the gauge decay $\nu_8\to\mu_8W$. Composite dynamics is encoded by
\begin{equation}
-\mathcal L_5=
\frac{a\,g_s}{2\Lambda}\,
\bar\ell_L\sigma^{\mu\nu}G_{\mu\nu}^{A}L_{8R}^{A}
+{\rm h.c.},
\label{eq:chromodipole_main}
\end{equation}
which both produces $\nu_8$ from $\nu g$ and gives $\mu_8\to\mu g$. The quoted reach uses $y_8=3$ and is restricted to the nominal EFT-validity region $\Lambda/a>M_{\mu_8}/3$.

\begin{figure}[t!]
\includegraphics[width=1.0\columnwidth]{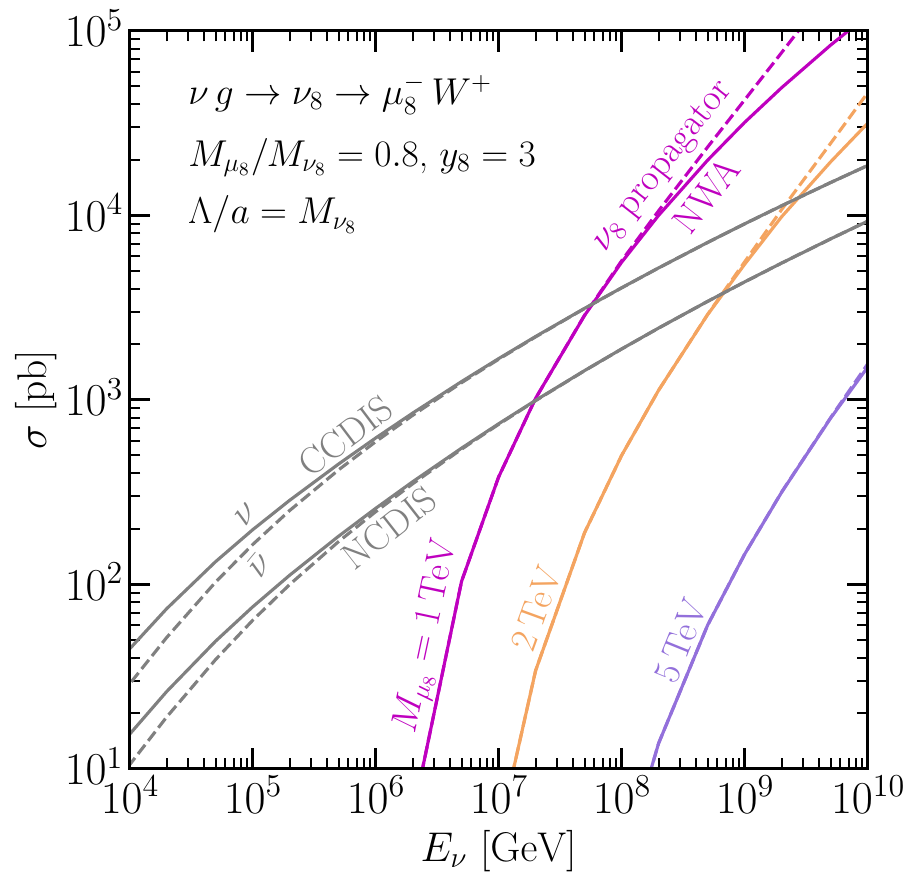}
\caption{
Tree-level production cross sections for neutrinogluon ($\nu_8$) from neutrino--nucleon scattering, which are identical for neutrinos and antineutrinos, followed by the decay $\nu_8 \to \mu_8^- W^+$ (with $\mu_8^-\to \mu^- g$).
We present the results from two calculations: using the full $\nu_8$ propagator (dashed lines) and using the narrow-width approximation (NWA; solid lines).
For comparison, we also show the SM background cross sections for CCDIS and NCDIS of neutrinos and antineutrinos, as labeled.
}
\label{fig:xsLG}
\end{figure}

Fig.~\ref{fig:xsLG} shows the neutrino--nucleon scattering cross section via the parton-level process $\nu \, g \rightarrow \nu_8 \rightarrow \mu^-_8W^+$ as a function of the neutrino energy $E_\nu$, for several representative values of $M_{\mu_8}$ while fixing $M_{\mu_8}/M_{\nu_8} = 0.8$, $y_8 = 3$, and $\Lambda/a = M_{\nu_8}$. For comparison, we present results using both the full $\nu_8$ propagator and the narrow-width approximation (NWA), shown as dashed and solid lines, respectively. The two approaches agree well at lower neutrino energies. We therefore use the NWA for the sensitivity calculation: it isolates on-shell $\nu_8$ production and avoids interpreting the unphysical high-energy growth of the dimension-5 amplitude outside its validity range.

Additionally, the SM CC and NCDIS cross sections intersect with the leptogluon-mediated cross sections at high $E_\nu$. For a given $M_{\mu_8}$, the new physics cross section is significantly suppressed at lower $E_\nu$ due to the reduced gluon PDF within the nucleon at large Bjorken-$x$. The cross section is identical for the neutrinogluon $\nu_8$ and its antiparticle $\bar{\nu}_8$. We introduce one leptogluon multiplet per lepton generation and take the chromodipole couplings flavor diagonal, avoiding lepton-flavor violation. The production cross section shown in Fig.~\ref{fig:xsLG} is then common to the three flavors, while the second-generation multiplet supplies the $\nu_\mu$ CC-like track signal analyzed here.

The neutrinogluon-mediated scattering process, followed by the decay of heavy particles $\mu^-_8 \rightarrow \mu^-g$ or $e^-_8 \rightarrow e^-g$ and $W^+ \rightarrow e^+ \nu_e, \mu^+ \nu_\mu, \tau^+ \nu_\tau$, can lead to a variety of final states, including $e^- g \tau^+\nu_\tau$, $\mu^- g \mu^+ \nu_\mu$, and others. In a neutrino detector, these events can manifest as ``double shower", ``double track", or other signatures, often with unequal energies in the two components. In this work, we do not focus on these novel event topologies. Instead, we select single-track-like events---those resembling standard $\nu_\mu$ CC interactions in the SM---to estimate projected sensitivities. This choice is motivated by the larger effective volume for track-like events and their relatively low background rates compared with those of shower-like events.

\begin{figure}[t!]
\includegraphics[width=0.985\columnwidth]{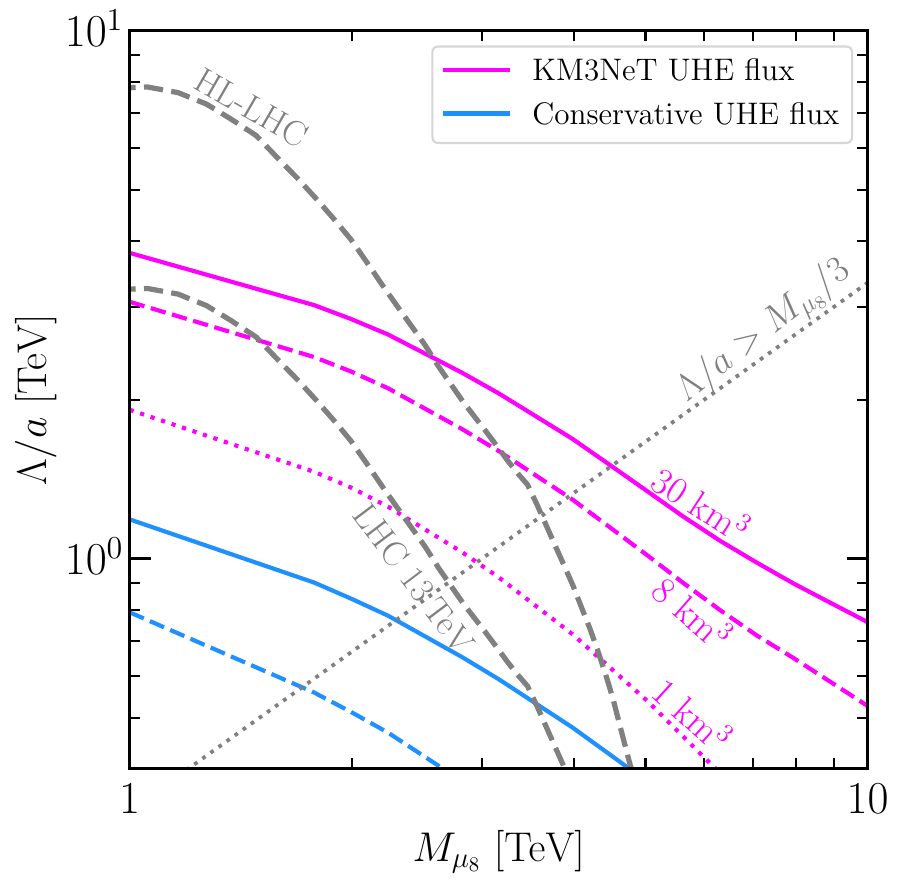}

\caption{Our projected sensitivities for the leptogluon model using starting and throughgoing track events in neutrino telescopes with volumes of 1, 8, and 30~km$^3$, and 20 years of observation.
We show results for two UHE flux scenarios: one using the KM3NeT-normalized UHE flux (pink) and the other using a theory-motivated conservative flux (blue).
For comparison, we also show the sensitivity in the dashed lines for the LHC with an integrated luminosity of 139~fb$^{-1}$~\cite{Almeida:2022udp} and the projected one based on the HL-LHC with 3~ab$^{-1}$~\cite{Almeida:2022udp,ZurbanoFernandez:2020cco}.
The gray dotted line is the guided effective-field-theory lower bound for $\Lambda/a > M_{\mu_8}/3$.
}

\label{Fig_LG_sensitivity}
\end{figure}

\subsection{Leptoquark}

Leptoquarks are bosons that couple directly to a lepton and a quark and occur in quark-lepton composite models and GUTs~\cite{Harari:1979gi,Shupe:1979fv,Georgi:1974sy,Pati:1974yy}. The headline benchmark is the scalar singlet interaction
\begin{equation}
-\mathcal L\supset
y_1^{LL}\,\bar q_L^{\,c} i\sigma_2\ell_L S_1
+y_1^{RR}\,\bar u_R^{\,c}e_R S_1+{\rm h.c.},
\label{eq:lq_main}
\end{equation}
with only the first-quark/second-lepton flavor entry nonzero and
$y_1^{LL}=y_1^{RR}\equiv y$. Its charge-conjugate state is resonantly produced through
$\nu_\mu d\to S_1^{-1/3}\to\mu^-u$. In the massless limit the $d\,\nu_\mu$ and $u\,\mu$ branching fractions are $1/3$ and $2/3$, respectively. The event rate uses the leading-order resonant cross section folded with the quark PDF; complete NLO corrections to inclusive production are below 20\% and are not included. Figure~\ref{fig:histogram} takes $y=3$.

\subsection{Neutrino-electron resonances}

To demonstrate the lepton-lepton mode, we also consider
$\bar\nu_e e^-\to X^-\to\bar\nu_\mu\mu^-$, where $X=W'$ has spin $J=1$ and $X=H^-$ has $J=0$. The Breit-Wigner cross section is
\begin{equation}
\sigma(s)=8\pi(2J+1)\frac{\Gamma_{\rm in}\Gamma_{\rm out}}{M_X^2}
\frac{s}{(s-M_X^2)^2+M_X^2\Gamma_X^2}.
\label{eq:electron_resonance_main}
\end{equation}
For the sequential-SM benchmark $g_{W'}=g_W=0.65$, the 1 and 8~km$^3$ configurations reach only $M_{W'}\simeq70$ and 100~GeV, while the LHC already excludes an SSM $W'$ below about 6~TeV~\cite{ParticleDataGroup:2024cfk}. The limitation is the lower $\bar\nu e$ COM energy and the small electron-target luminosity. Consequently, this benchmark illustrates the collision mode but is not included in Fig.~\ref{fig:histogram}; nonstandard leptophilic resonances could be more favorable.

{
\section{Results and comparison with
terrestrial colliders}
}

For each of the 1, 8, and 30~km$^3$ configurations, we estimate the projected reach for the four representative new-physics scenarios. Figure~\ref{fig:histogram} displays the contact-interaction, leptogluon, and leptoquark results. The $W^\prime/H^-$ conclusion is stated above because it does not compete with existing terrestrial limits in the conventional benchmark.

Fig.~\ref{Fig_LG_sensitivity} shows that with the KM3NeT-normalized UHE flux, next-generation detectors can probe broad leptogluon parameter space and surpass LHC and HL-LHC limits at large masses. The conservative flux gives a weaker but still complementary reach and makes explicit the dependence on the UHE intensity. The LHC sensitivity drops sharply near 4~TeV because of the steeply falling parton luminosity at large Bjorken $x$, whereas the broad UHE neutrino spectrum accesses higher partonic energies. The benchmark selection uses single-track-like events; distinctive double-track or double-shower signatures are not included and could improve the reach in a dedicated analysis.

For the contact-interaction model shown in Fig.~\ref{fig:histogram}, we use Eq.~(\ref{eq:contact_main}) to present the projected L$\nu$C sensitivity to the cutoff scale $\Lambda$. Assuming the KM3NeT-normalized UHE flux, L$\nu$Cs can probe this interaction more sensitively than global SMEFT fits based on current LHC data and projected HL-LHC sensitivities~\cite{deBlas:2022ofj}, although their sensitivity remains weaker than the constraints from low-energy observables~\cite{Falkowski:2017pss}. For larger detector volumes, the theoretically motivated UHE flux yields limits slightly weaker than those obtained using the KM3NeT-normalized UHE flux but remains competitive.

The mass scale shown for the leptogluon model in Fig.~\ref{fig:histogram} corresponds to $M_{\mu_8}$, with the parameters fixed at $\Lambda/a = M_{\mu_8}/2$, $M_{\mu_8}/M_{\nu_8} = 0.8$, and $y_8 = 3$ (see Fig.~\ref{Fig_LG_sensitivity} for coverage of a broader parameter space). Next-generation L$\nu$Cs assuming the KM3NeT-normalized UHE flux can achieve a mass reach exceeding that of both current LHC searches based on an integrated luminosity of 139~fb$^{-1}$~\cite{Almeida:2022udp} and the future HL-LHC with 3~ab$^{-1}$~\cite{Almeida:2022udp,ZurbanoFernandez:2020cco}. By contrast, L$\nu$Cs assuming the conservative UHE neutrino flux have weaker sensitivity than the corresponding LHC searches but still achieve a substantial mass reach.

For the leptoquark scenario, we consider the scalar leptoquark $S_1^{-1/3}$, with the interaction and branching fractions given in Eq.~(\ref{eq:lq_main}), and set $y=3$. Other scalar or vector representations modify the spin factor, the relevant initial-state parton distribution function, and the attenuation through the Earth. The displayed result therefore represents a specific benchmark rather than a generic leptoquark limit. For both flux assumptions, L$\nu$Cs achieve mass reaches that are somewhat lower than, but still comparable to, the corresponding reaches of current LHC searches~\cite{Campana:2024} and future HL-LHC projections.

The hierarchy among the three reaches has a direct physical origin. The contact-interaction cross section grows with partonic collision energy until its EFT-validity cut is reached. Leptogluon production benefits simultaneously from an $s$-channel resonance and the large gluon PDF, giving the strongest particle-mass reach. Leptoquark production is also resonant but samples quark rather than gluon PDFs and, for the selected coupling and track channel, yields fewer observable events. At terrestrial colliders, all heavy-particle rates fall rapidly with parton luminosity at large Bjorken $x$; the broad UHE neutrino spectrum softens this loss for L$\nu$Cs. The comparison is therefore conditional on the displayed couplings, branching fractions, and EFT cuts, rather than a universal ranking of the models.

We have not included longer-term future colliders, such as CEPC, FCC-ee, and a muon collider, in Fig.~\ref{fig:histogram}. These facilities could achieve impressive sensitivity to new physics. For instance, a muon collider could probe leptogluon masses up to 4.9~TeV and leptoquark masses up to 4.0~TeV~\cite{Han:2025wdy}. Given their longer construction timelines and the greater uncertainty surrounding their realization, we do not include their projected limits here.

We acknowledge that the quantitative projections depend on uncertainties in the future astrophysical neutrino flux, detector performance, and event statistics. We therefore view the projected L$\nu$C reach as complementary to, rather than uniformly stronger than, conventional collider and low-energy searches, while illustrating the potentially transformative opportunities offered by ultra-high-energy neutrino beams.

\section{Discussion and conclusions}

Among cosmic colliders, the L$\nu$C offers distinct advantages over those using other cosmic rays, such as electrons or hadrons. Cosmic-ray electrons suffer significant radiative losses that limit their attainable energies, with the highest measured all-electron spectrum reaching only $\sim 40$ TeV, as observed by H.E.S.S.~\cite{HESS:2024etj}. Cosmic-ray hadrons can reach energies well beyond the EeV scale and possess larger cross sections, but they experience strong-interaction losses that restrict their propagation horizon and cosmological reach~\cite{Anchordoqui:2018qom, AlvesBatista:2019tlv, Globus:2025ftu}. Moreover, the large QCD elastic and inelastic cross sections in hadron--hadron collisions create a much higher background for new-physics searches compared to the L$\nu$C, where the dominant background arises only from weak interactions.

In this work, we have taken the first systematic step toward exploring new physics at large-volume neutrino telescopes, focusing on starting and throughgoing track-like events. A natural next step is to incorporate shower-like events in order to maximize the L$\nu$C reach for new physics. Moreover, since many new physics models can produce distinctive double-shower or double-track topologies, dedicated searches based on such event signatures may further suppress SM backgrounds and enhance sensitivity to new physics~\cite{Zhou:2021xuh}. In this study, we have focused on a few representative models to illustrate the projected coverage of L$\nu$Cs, though other exotic scenarios, including long-lived particles such as those considered in Ref.~\cite{Albuquerque:2003mi}, may also be probed more effectively at L$\nu$Cs.

The HE and UHE neutrinos may originate from particular astrophysical sources. Once their origins are identified, the directional information can be used to enhance sensitivity to new physics, rather than averaging over the entire sky. This makes the L$\nu$C even more analogous to traditional colliders, where the beam direction is known and the beam can be ``turned on".

In conclusion, we have demonstrated the feasibility of using high-energy cosmic neutrinos as beams to search for BSM physics.
The HE and UHE neutrino fluxes measured by IceCube and KM3NeT are sufficient to produce multi-TeV particles at the L$\nu$C and probe a wide range of BSM models.
Because the UHE intensity is not yet established, we presented both the KM3NeT-normalized UHE flux and a lower, theory-motivated conservative flux. The former can yield collider-competitive or stronger reach in selected benchmarks, while the latter provides a slightly weaker reach.
Future measurements of the UHE spectrum will therefore directly sharpen L$\nu$C sensitivity.
Next-generation neutrino telescopes with effective volumes of tens of km$^3$ will not only address key astrophysical questions but also probe fundamental particles and interactions in particle physics. Our results highlight that the L$\nu$C offers rich opportunities for discovery and a promising new frontier in both astrophysics and particle physics.

\section{Acknowledgement}

We would like to thank Benjamin Fuks and J\"urgen Reuter for help with some calculations. YB is supported by the U.S. Department of Energy under the contract DE-SC-0017647 and DEAC02-06CH11357 at Argonne National Laboratory. KX is supported by the U.S. National Science Foundation under Grant No. PHY-2310291 and PHY-2310497. BZ is supported by Fermi Forward Discovery Group, LLC under Contract No. 89243024CSC000002 with the U.S. Department of Energy, Office of Science, Office of High Energy Physics. This work was performed in part at the Aspen Center for Physics, which is supported by National Science Foundation grant PHY-2210452.

\vspace{1em}
\noindent\textbf{Corresponding authors:}
{\\
{
Yang Bai: yangbai@physics.wisc.edu \\
Keping Xie: kepingxie@sjtu.edu.cn \\
Bei Zhou: beizhou@fnal.gov
}
}

\bibliographystyle{utphys}
\bibliography{References.bib}

@article{Reno:2023sdm,
    author = "Reno, Mary Hall",
    title = "{High-Energy to Ultrahigh-Energy Neutrino Interactions}",
    doi = "10.1146/annurev-nucl-111422-040200",
    journal = "Ann. Rev. Nucl. Part. Sci.",
    volume = "73",
    number = "1",
    pages = "181--204",
    year = "2023"
}

@article{IceCube:2017roe,
    author = "Aartsen, M. G. and others",
    collaboration = "IceCube",
    title = "{Measurement of the multi-TeV neutrino cross section with IceCube using Earth absorption}",
    eprint = "1711.08119",
    archivePrefix = "arXiv",
    primaryClass = "hep-ex",
    doi = "10.1038/nature24459",
    journal = "Nature",
    volume = "551",
    pages = "596--600",
    year = "2017"
}

@article{Gandhi:1995tf,
    author = "Gandhi, Raj and Quigg, Chris and Reno, Mary Hall and Sarcevic, Ina",
    title = "{Ultrahigh-energy neutrino interactions}",
    eprint = "hep-ph/9512364",
    archivePrefix = "arXiv",
    reportNumber = "FERMILAB-PUB-95-221-T, CLNS-95-1357, MRI-PHY-16-95, UIOWA-95-06, AZPH-TH-95-15",
    doi = "10.1016/0927-6505(96)00008-4",
    journal = "Astropart. Phys.",
    volume = "5",
    pages = "81--110",
    year = "1996"
}

@article{Dziewonski:1981xy, author = "Dziewonski, Adam M. and Anderson, Don L.", title = "{Preliminary reference Earth model}", doi = "10.1016/0031-9201(81)90046-7", journal = "Phys. Earth Planet. Interiors", volume = "25", pages = "297--356", year = "1981"}

@article{MICE:2019jkl,
    author = "Bogomilov, M. and others",
    collaboration = "MICE",
    title = "{Demonstration of cooling by the Muon Ionization Cooling Experiment}",
    eprint = "1907.08562",
    archivePrefix = "arXiv",
    primaryClass = "physics.acc-ph",
    reportNumber = "RAL-P-2019-003",
    doi = "10.1038/s41586-020-1958-9",
    journal = "Nature",
    volume = "578",
    number = "7793",
    pages = "53--59",
    year = "2020"
}

@article{Narain:2022qud,
    author = "Narain, Meenakshi and others",
    title = "{The Future of US Particle Physics - The Snowmass 2021 Energy Frontier Report}",
    eprint = "2211.11084",
    archivePrefix = "arXiv",
    primaryClass = "hep-ex",
    reportNumber = "FERMILAB-FN-1219-PPD-T",
    doi = "10.2172/1908199",
    month = "11",
    year = "2022"
}

@article{Adolphsen:2022ibf,
    author = "Adolphsen, C. and others",
    editor = "Mounet, N.",
    title = "{European Strategy for Particle Physics -- Accelerator R{\&}D Roadmap}",
    eprint = "2201.07895",
    archivePrefix = "arXiv",
    primaryClass = "physics.acc-ph",
    reportNumber = "CERN-2022-001",
    doi = "10.23731/CYRM-2022-001",
    journal = "CERN Yellow Rep. Monogr.",
    volume = "1",
    pages = "1--270",
    year = "2022"
}

@article{Roser:2022sht,
    author = "Roser, Thomas and others",
    title = "{On the feasibility of future colliders: report of the Snowmass'21 Implementation Task Force}",
    eprint = "2208.06030",
    archivePrefix = "arXiv",
    primaryClass = "physics.acc-ph",
    reportNumber = "FERMILAB-FN-1184-AD-PPD",
    doi = "10.1088/1748-0221/18/05/P05018",
    journal = "JINST",
    volume = "18",
    number = "05",
    pages = "P05018",
    year = "2023"
}

@book{Evans:2645935,
      author        = "Evans, Lyndon and Evans, Lyn",
      title         = "{The Large Hadron Collider}",
      publisher     = "EPFL Press",
      address       = "Lausanne",
      series        = "Physics (EPFL Press)",
      year          = "2018",
      url           = "https://cds.cern.ch/record/2645935",
      note          = "On the cover : Including the discovery of the higgs boson",
}

@article{KM3NeT:2009xxi,
    author = "Bagley, P. and others",
    collaboration = "KM3NeT",
    title = "{KM3NeT:
Technical Design Report for a Deep-Sea
Research Infrastructure in the
Mediterranean Sea Incorporating a
Very Large Volume Neutrino Telescope}",
    year = "2009"
}

@article{Ishihara:2019aao,
    author = "Ishihara, Aya",
    collaboration = "IceCube",
    title = "{The IceCube Upgrade - Design and Science Goals}",
    eprint = "1908.09441",
    archivePrefix = "arXiv",
    primaryClass = "astro-ph.HE",
    reportNumber = "PoS-ICRC2019-1031",
    doi = "10.22323/1.358.1031",
    journal = "PoS",
    volume = "ICRC2019",
    pages = "1031",
    year = "2021"
}

@article{Babu:2022fje,
    author = "Babu, K. S. and Dev, P. S. Bhupal and Jana, Sudip",
    title = "{Probing neutrino mass models through resonances at neutrino telescopes}",
    eprint = "2202.06975",
    archivePrefix = "arXiv",
    primaryClass = "hep-ph",
    doi = "10.1142/S0217751X22300034",
    journal = "Int. J. Mod. Phys. A",
    volume = "37",
    number = "11n12",
    pages = "2230003",
    year = "2022"
}

@article{Dev:2016uxj,
    author = "Dev, P. S. Bhupal and Ghosh, Dilip Kumar and Rodejohann, Werner",
    title = "{R-parity Violating Supersymmetry at IceCube}",
    eprint = "1605.09743",
    archivePrefix = "arXiv",
    primaryClass = "hep-ph",
    doi = "10.1016/j.physletb.2016.08.066",
    journal = "Phys. Lett. B",
    volume = "762",
    pages = "116--123",
    year = "2016"
}

@article{Bhaskar:2023ftn,
    author = "Bhaskar, Arvind and Das, Arijit and Mandal, Tanumoy and Mitra, Subhadip and Sharma, Rachit",
    title = "{Fresh look at the LHC limits on scalar leptoquarks}",
    eprint = "2312.09855",
    archivePrefix = "arXiv",
    primaryClass = "hep-ph",
    doi = "10.1103/PhysRevD.109.055018",
    journal = "Phys. Rev. D",
    volume = "109",
    number = "5",
    pages = "055018",
    year = "2024"
}

@article{Das:2025osr,
    author = "Das, Arijit and Mandal, Tanumoy and Mitra, Subhadip and Sharma, Rachit",
    title = "{Fresh look at the LHC limits on vector leptoquarks}",
    eprint = "2507.18295",
    archivePrefix = "arXiv",
    primaryClass = "hep-ph",
    month = "7",
    year = "2025"
}

@article{Han:2021lnp,
    author = {Han, Tao and Kilian, Wolfgang and Kreher, Nils and Ma, Yang and Reuter, J{\"u}rgen and Striegl, Tobias and Xie, Keping},
    title = "{Precision test of the muon-Higgs coupling at a high-energy muon collider}",
    eprint = "2108.05362",
    archivePrefix = "arXiv",
    primaryClass = "hep-ph",
    reportNumber = "DESY 21-086, PITT-PACC-2110",
    doi = "10.1007/JHEP12(2021)162",
    journal = "JHEP",
    volume = "12",
    pages = "162",
    year = "2021"
}

@article{AbdulKhalek:2021gbh,
    author = "Abdul Khalek, R. and others",
    title = "{Science Requirements and Detector Concepts for the Electron-Ion Collider}: {EIC Yellow Report}",
    eprint = "2103.05419",
    archivePrefix = "arXiv",
    primaryClass = "physics.ins-det",
    reportNumber = "BNL-220990-2021-FORE, JLAB-PHY-21-3198, LA-UR-21-20953",
    doi = "10.1016/j.nuclphysa.2022.122447",
    journal = "Nucl. Phys. A",
    volume = "1026",
    pages = "122447",
    year = "2022"
}

@article{Alvarez-Muniz:2002snq,
    author = "Alvarez-Muniz, Jaime and Feng, Jonathan L. and Halzen, Francis and Han, Tao and Hooper, Dan",
    title = "{Detecting microscopic black holes with neutrino telescopes}",
    eprint = "hep-ph/0202081",
    archivePrefix = "arXiv",
    reportNumber = "MIT-CTP-3221, UCI-TR-2001-43, MADPH-02-1255",
    doi = "10.1103/PhysRevD.65.124015",
    journal = "Phys. Rev. D",
    volume = "65",
    pages = "124015",
    year = "2002"
}

@article{BookMotzkin:2024qgd,
    author = "Book Motzkin, Julia",
    title = "{Simulating and searching for Heavy Neutral Leptons in IceCube}",
    doi = "10.22323/1.441.0338",
    journal = "PoS",
    volume = "TAUP2023",
    pages = "338",
    year = "2024"
}

@article{IceCube:2025kve,
    author = "Abbasi, R. and others",
    collaboration = "IceCube",
    title = "{Search for Heavy Neutral Leptons with IceCube DeepCore}",
    eprint = "2502.09454",
    archivePrefix = "arXiv",
    primaryClass = "hep-ex",
    month = "2",
    year = "2025"
}

@article{Anchordoqui:2006wc,
    author = "Anchordoqui, Luis A. and Garcia Canal, Carlos A. and Goldberg, Haim and Gomez Dumm, Daniel and Halzen, Francis",
    title = "{Probing leptoquark production at IceCube}",
    eprint = "hep-ph/0609214",
    archivePrefix = "arXiv",
    doi = "10.1103/PhysRevD.74.125021",
    journal = "Phys. Rev. D",
    volume = "74",
    pages = "125021",
    year = "2006"
}

@article{Romero:2009vu,
    author = "Romero, Ismael and Sampayo, O. A.",
    title = "{Leptoquarks signals in KM**3 neutrino telescopes}",
    eprint = "0906.5245",
    archivePrefix = "arXiv",
    primaryClass = "hep-ph",
    doi = "10.1088/1126-6708/2009/05/111",
    journal = "JHEP",
    volume = "05",
    pages = "111",
    year = "2009"
}

@article{Alvarez-Muniz:2001efi,
    author = "Alvarez-Muniz, J. and Halzen, F. and Han, Tao and Hooper, D.",
    title = "{Phenomenology of high-energy neutrinos in low scale quantum gravity models}",
    eprint = "hep-ph/0107057",
    archivePrefix = "arXiv",
    reportNumber = "MADPH-01-1236",
    doi = "10.1103/PhysRevLett.88.021301",
    journal = "Phys. Rev. Lett.",
    volume = "88",
    pages = "021301",
    year = "2002"
}

@article{Albuquerque:2003mi,
    author = "Albuquerque, Ivone and Burdman, Gustavo and Chacko, Z.",
    title = "{Neutrino telescopes as a direct probe of supersymmetry breaking}",
    eprint = "hep-ph/0312197",
    archivePrefix = "arXiv",
    doi = "10.1103/PhysRevLett.92.221802",
    journal = "Phys. Rev. Lett.",
    volume = "92",
    pages = "221802",
    year = "2004"
}

@article{CEPCStudyGroup:2025kmw,
    collaboration = "CEPC Study Group",
    title = "{CEPC Technical Design Report -- Reference Detector}",
    eprint = "2510.05260",
    archivePrefix = "arXiv",
    primaryClass = "hep-ex",
    reportNumber = "IHEP-CEPC-DR-2025-01, IHEP-EP-2025-01",
    month = "10",
    year = "2025"
}

@article{IceCube:2021rpz,
    author = "Aartsen, M. G. and others",
    collaboration = "IceCube",
    title = "{Detection of a particle shower at the Glashow resonance with IceCube}",
    eprint = "2110.15051",
    archivePrefix = "arXiv",
    primaryClass = "hep-ex",
    doi = "10.1038/s41586-021-03256-1",
    journal = "Nature",
    volume = "591",
    number = "7849",
    pages = "220--224",
    year = "2021",
    note = "[Erratum: Nature 592, E11 (2021)]"
}

@article{Gonzalez-Alonso:2016etj,
    author = "Gonz{\'a}lez-Alonso, Mart{\'\i}n and Martin Camalich, Jorge",
    title = "{Global Effective-Field-Theory analysis of New-Physics effects in (semi)leptonic kaon decays}",
    eprint = "1605.07114",
    archivePrefix = "arXiv",
    primaryClass = "hep-ph",
    reportNumber = "INT-PUB-16-014, MITP-16-047",
    doi = "10.1007/JHEP12(2016)052",
    journal = "JHEP",
    volume = "12",
    pages = "052",
    year = "2016"
}

@article{Cepeda:2019klc,
    author = "Cepeda, M. and others",
    editor = "Dainese, Andrea and Mangano, Michelangelo and Meyer, Andreas B. and Nisati, Aleandro and Salam, Gavin and Vesterinen, Mika Anton",
    title = "{Report from Working Group 2}: {Higgs Physics at the HL-LHC and HE-LHC}",
    eprint = "1902.00134",
    archivePrefix = "arXiv",
    primaryClass = "hep-ph",
    reportNumber = "CERN-LPCC-2018-04",
    doi = "10.23731/CYRM-2019-007.221",
    journal = "CERN Yellow Rep. Monogr.",
    volume = "7",
    pages = "221--584",
    year = "2019"
}

@article{ZurbanoFernandez:2020cco,
    author = "Zurbano Fernandez, I. and others",
    editor = {B{\'e}jar Alonso, I. and Br{\"u}ning, O. and Fessia, P. and Rossi, L. and Tavian, L. and Zerlauth, M.},
    title = "{High-Luminosity Large Hadron Collider (HL-LHC): Technical design report}",
    reportNumber = "CERN-2020-010",
    doi = "10.23731/CYRM-2020-0010",
    volume = "10/2020",
    month = "12",
    year = "2020"
}

@article{CEPCStudyGroup:2023quu,
    author = "Abdallah, Waleed and others",
    collaboration = "CEPC Study Group",
    title = "{CEPC Technical Design Report: Accelerator}",
    eprint = "2312.14363",
    archivePrefix = "arXiv",
    primaryClass = "physics.acc-ph",
    reportNumber = "IHEP-CEPC-DR-2023-01, IHEP-AC-2023-01",
    doi = "10.1007/s41605-024-00463-y",
    journal = "Radiat. Detect. Technol. Methods",
    volume = "8",
    number = "1",
    pages = "1--1105",
    year = "2024",
    note = "[Erratum: Radiat.Detect.Technol.Methods 9, 184--192 (2025)]"
}

@article{Shtabovenko:2020gxv,
    author = "Shtabovenko, Vladyslav and Mertig, Rolf and Orellana, Frederik",
    title = "{FeynCalc 9.3: New features and improvements}",
    eprint = "2001.04407",
    archivePrefix = "arXiv",
    primaryClass = "hep-ph",
    reportNumber = "P3H-20-002, TTP19-020, TUM-EFT 130/19",
    doi = "10.1016/j.cpc.2020.107478",
    journal = "Comput. Phys. Commun.",
    volume = "256",
    pages = "107478",
    year = "2020"
}

@article{Clark:2016jgm,
    author = "Clark, D. B. and Godat, E. and Olness, F. I.",
    title = "{ManeParse : A Mathematica  reader for Parton Distribution Functions}",
    eprint = "1605.08012",
    archivePrefix = "arXiv",
    primaryClass = "hep-ph",
    reportNumber = "NSF-KITP-16-032, SMU-HEP-16-05",
    doi = "10.1016/j.cpc.2017.03.004",
    journal = "Comput. Phys. Commun.",
    volume = "216",
    pages = "126--137",
    year = "2017"
}

@article{Kilian:2007gr,
    author = "Kilian, Wolfgang and Ohl, Thorsten and Reuter, Jurgen",
    title = "{WHIZARD: Simulating Multi-Particle Processes at LHC and ILC}",
    eprint = "0708.4233",
    archivePrefix = "arXiv",
    primaryClass = "hep-ph",
    reportNumber = "DESY-11-126, EDINBURGH-2010-36, FR-PHENO-2010-037, SI-HEP-2010-18",
    doi = "10.1140/epjc/s10052-011-1742-y",
    journal = "Eur. Phys. J. C",
    volume = "71",
    pages = "1742",
    year = "2011"
}

@phdthesis{Campana:2024,
    author = "Campana, Mattia",
    title = "{Search for leptoquarks coupling to muons in lepton-quark collisions at LHC}",
    url = "https://www.roma1.infn.it/exp/cms/tesiPHD/tesi_phd_completate/campana.pdf",
    school = "Sapienza University of Rome",
    year = "2024"
}

@article{Hahn:2016ebn,
    author = "Hahn, Thomas and Pa{\ss}ehr, Sebastian and Schappacher, Christian",
    editor = "Salinas, Luis and Torres, Claudio",
    title = "{FormCalc 9 and Extensions}",
    eprint = "1604.04611",
    archivePrefix = "arXiv",
    primaryClass = "hep-ph",
    reportNumber = "DESY-16-060, MPP-2016-63",
    doi = "10.1088/1742-6596/762/1/012065",
    journal = "PoS",
    volume = "LL2016",
    pages = "068",
    year = "2016"
}

@article{Zhang:2024slv,
    author = "Zhang, Huiming and Cui, Yudong and Huang, Yunlei and Lin, Sujie and Liu, Yihan and Qiu, Zijian and Shao, Chengyu and Shi, Yihan and Xie, Caijin and Yang, Lili",
    title = "{A proposed deep sea Neutrino Observatory in the Nanhai}",
    eprint = "2408.05122",
    archivePrefix = "arXiv",
    primaryClass = "astro-ph.HE",
    doi = "10.1016/j.astropartphys.2025.103123",
    journal = "Astropart. Phys.",
    volume = "171",
    pages = "103123",
    year = "2025"
}

@article{Altarelli:1989ff,
    author = "Altarelli, Guido and Mele, B. and Ruiz-Altaba, M.",
    title = "{Searching for New Heavy Vector Bosons in $p \bar{p}$ Colliders}",
    reportNumber = "CERN-TH-5323-89",
    doi = "10.1007/BF01556677",
    journal = "Z. Phys. C",
    volume = "45",
    pages = "109",
    year = "1989",
    note = "[Erratum: Z.Phys.C 47, 676 (1990)]"
}

@article{Georgi:1974sy,
    author = "Georgi, H. and Glashow, S. L.",
    title = "{Unity of All Elementary Particle Forces}",
    doi = "10.1103/PhysRevLett.32.438",
    journal = "Phys. Rev. Lett.",
    volume = "32",
    pages = "438--441",
    year = "1974"
}

@article{Pati:1974yy,
    author = "Pati, Jogesh C. and Salam, Abdus",
    title = "{Lepton Number as the Fourth Color}",
    reportNumber = "IC-74-7",
    doi = "10.1103/PhysRevD.10.275",
    journal = "Phys. Rev. D",
    volume = "10",
    pages = "275--289",
    year = "1974",
    note = "[Erratum: Phys.Rev.D 11, 703--703 (1975)]"
}

@article{Aynutdinov:2023ifk,
    author = "Aynutdinov, V. M. and others",
    title = "{The Baikal-GVD Neutrino Telescope: Current Status and Development Prospects}",
    doi = "10.1134/S1063778824010101",
    journal = "Phys. Atom. Nucl.",
    volume = "86",
    number = "6",
    pages = "989--994",
    year = "2023"
}

@misc{Baikal_web,
    title        = {},
    author       = {},
    year         = {},
    month        = {},
    note         = {},
    howpublished = {\url{https://tass.com/science/1265679}}
}

@article{Shao:2025gyx,
    author = "Shao, Hengbin and others",
    title = "{A Cost Effective Optimization of the hybrid-DOM Design for TRIDENT}",
    eprint = "2507.10256",
    archivePrefix = "arXiv",
    primaryClass = "hep-ex",
    month = "7",
    year = "2025"
}

@misc{PONE_EPPSU,
    title        = "{P-ONE: The Pacific Ocean Neutrino Experiment: A Next-Generation Deep-Sea Neutrino Detector for Fundamental Physics}",
    author       = {},
    year         = {},
    month        = {},
    note         = {},
    howpublished = {\url{https://indico.cern.ch/event/1439855/contributions/6461460/}}
}

@misc{IceCube_web,
    title        = {},
    author       = {},
    year         = {},
    month        = {},
    note         = {},
    howpublished = {\url{https://icecube.wisc.edu/}}
}

@article{IceCube-Gen2:2020qha,
    title        = {{IceCube-Gen2: the window to the extreme Universe}},
    author       = {Aartsen, M. G. and others},
    year         = 2021,
    journal      = {J. Phys. G},
    volume       = 48,
    number       = 6,
    pages        = {060501},
    doi          = {10.1088/1361-6471/abbd48},
    collaboration = {IceCube-Gen2},
    eprint       = {2008.04323},
    archiveprefix = {arXiv},
    primaryclass = {astro-ph.HE}
}

@article{Huang:2023mzt,
    author = "Huang, Tian-Qi and Cao, Zhen and Chen, Mingjun and Liu, Jiali and Wang, Zike and You, Xiaohao and Qi, Ying",
    title = "{Proposal for the High Energy Neutrino Telescope}",
    doi = "10.22323/1.444.1080",
    journal = "PoS",
    volume = "ICRC2023",
    pages = "1080",
    year = "2023"
}

@article{KM3Net:2016zxf,
    author = "Adrian-Martinez, S. and others",
    collaboration = "KM3Net",
    title = "{Letter of intent for KM3NeT 2.0}",
    eprint = "1601.07459",
    archivePrefix = "arXiv",
    primaryClass = "astro-ph.IM",
    doi = "10.1088/0954-3899/43/8/084001",
    journal = "J. Phys. G",
    volume = "43",
    number = "8",
    pages = "084001",
    year = "2016"
}

@article{Allakhverdyan:2021vkk,
    author = "Allakhverdyan, V. A. and others",
    title = "{Deep-Water Neutrino Telescope in Lake Baikal}",
    doi = "10.1134/S1063778821090064",
    journal = "Phys. Atom. Nucl.",
    volume = "84",
    number = "9",
    pages = "1600--1609",
    year = "2021"
}

@article{FCC:2018evy,
    author = "Abada, A. and others",
    collaboration = "FCC",
    title = "{FCC-ee: The Lepton Collider}: {Future Circular Collider Conceptual Design Report Volume 2}",
    reportNumber = "CERN-ACC-2018-0057",
    doi = "10.1140/epjst/e2019-900045-4",
    journal = "Eur. Phys. J. ST",
    volume = "228",
    number = "2",
    pages = "261--623",
    year = "2019"
}

@article{FCC:2018vvp,
    author = "Abada, A. and others",
    collaboration = "FCC",
    title = "{FCC-hh: The Hadron Collider}: {Future Circular Collider Conceptual Design Report Volume 3}",
    reportNumber = "CERN-ACC-2018-0058",
    doi = "10.1140/epjst/e2019-900087-0",
    journal = "Eur. Phys. J. ST",
    volume = "228",
    number = "4",
    pages = "755--1107",
    year = "2019"
}

@article{ATLAS:2024lda,
    author = "Aad, Georges and others",
    collaboration = "ATLAS",
    title = "{The quest to discover supersymmetry at the ATLAS experiment}",
    eprint = "2403.02455",
    archivePrefix = "arXiv",
    primaryClass = "hep-ex",
    reportNumber = "CERN-EP-2024-056",
    doi = "10.1016/j.physrep.2024.09.010",
    journal = "Phys. Rept.",
    volume = "1116",
    pages = "261--300",
    year = "2025"
}

@article{Weinberg:2018apv,
    author = "Weinberg, Steven",
    title = "{Essay: Half a Century of the Standard Model}",
    doi = "10.1103/PhysRevLett.121.220001",
    journal = "Phys. Rev. Lett.",
    volume = "121",
    number = "22",
    pages = "220001",
    year = "2018"
}

@article{Anderson:1933mb,
    author = "Anderson, C. D.",
    title = "{The Positive Electron}",
    doi = "10.1103/PhysRev.43.491",
    journal = "Phys. Rev.",
    volume = "43",
    pages = "491--494",
    year = "1933"
}

@article{ParticleDataGroup:2024cfk,
    author = "Navas, S. and others",
    collaboration = "Particle Data Group",
    title = "{Review of particle physics}",
    doi = "10.1103/PhysRevD.110.030001",
    journal = "Phys. Rev. D",
    volume = "110",
    number = "3",
    pages = "030001",
    year = "2024"
}

@article{Cowan:2010js,
  author = "Cowan, Glen and Cranmer, Kyle and Gross, Eilam and Vitells, Ofer",
  title = "Asymptotic formulae for likelihood-based tests of new physics",
  journal = "Eur. Phys. J. C",
  volume = "71",
  pages = "1554",
  year = "2011",
  eprint = "1007.1727",
  archivePrefix = "arXiv"
}

@phdthesis{Sarkar:2023dvr,
    author = "Sarkar, Sourav",
    title = "{Search for TeV-Scale Neutrino Dimuon Events with 10.7 Years of IceCube Data}",
    doi = "10.7939/r3-y8x0-mn08",
    school = "Alberta U.",
    year = "2023"
}

@article{IceCube:2018pgc,
    author = "Aartsen, M. G. and others",
    collaboration = "IceCube",
    title = "{Measurements using the inelasticity distribution of multi-TeV neutrino interactions in IceCube}",
    eprint = "1808.07629",
    archivePrefix = "arXiv",
    primaryClass = "hep-ex",
    doi = "10.1103/PhysRevD.99.032004",
    journal = "Phys. Rev. D",
    volume = "99",
    number = "3",
    pages = "032004",
    year = "2019"
}

@article{IceCubeCollaborationSS:2025zgz,
    author = "Abbasi, R. and others",
    collaboration = "(IceCube Collaboration){\textsection}, IceCube",
    title = "{Measurement of the inelasticity distribution of neutrino-nucleon interactions for 80{\,}{\,}GeV{\ensuremath{<}}E{\ensuremath{\nu}}{\ensuremath{<}}560{\,}{\,}GeV with IceCube DeepCore}",
    eprint = "2502.13299",
    archivePrefix = "arXiv",
    primaryClass = "hep-ex",
    doi = "10.1103/PhysRevD.111.112001",
    journal = "Phys. Rev. D",
    volume = "111",
    number = "11",
    pages = "112001",
    year = "2025"
}

@article{Koehne:2013gpa,
    author = "Koehne, J. H. and Frantzen, K. and Schmitz, M. and Fuchs, T. and Rhode, W. and Chirkin, D. and Becker Tjus, J.",
    title = "{PROPOSAL: A tool for propagation of charged leptons}",
    doi = "10.1016/j.cpc.2013.04.001",
    journal = "Comput. Phys. Commun.",
    volume = "184",
    pages = "2070--2090",
    year = "2013"
}

@article{Gaisser:2002jj,
    author = "Gaisser, T. K. and Honda, M.",
    title = "{Flux of atmospheric neutrinos}",
    eprint = "hep-ph/0203272",
    archivePrefix = "arXiv",
    doi = "10.1146/annurev.nucl.52.050102.090645",
    journal = "Ann. Rev. Nucl. Part. Sci.",
    volume = "52",
    pages = "153--199",
    year = "2002"
}

@article{Borschensky:2021hbo,
    author = {Borschensky, Christoph and Fuks, Benjamin and Kulesza, Anna and Schwartl\"ander, Daniel},
    title = "{Scalar leptoquark pair production at the LHC: precision predictions in the era of flavour anomalies}",
    eprint = "2108.11404",
    archivePrefix = "arXiv",
    primaryClass = "hep-ph",
    doi = "10.1007/JHEP02(2022)157",
    journal = "JHEP",
    volume = "02",
    pages = "157",
    year = "2022"
}

@article{Borschensky:2022xsa,
    author = "Borschensky, Christoph and Fuks, Benjamin and Jueid, Adil and Kulesza, Anna",
    title = "{Scalar leptoquarks at the LHC and flavour anomalies: a comparison of pair-production modes at NLO-QCD}",
    eprint = "2207.02879",
    archivePrefix = "arXiv",
    primaryClass = "hep-ph",
    reportNumber = "KIAS-Q22010, CTPU-PTC-22-24, CERN-TH-2022-106, KA-TP-17-2022, MS-TP-22-18",
    doi = "10.1007/JHEP11(2022)006",
    journal = "JHEP",
    volume = "11",
    pages = "006",
    year = "2022"
}

@article{Falkowski:2017pss,
    author = "Falkowski, Adam and Gonz\'alez-Alonso, Mart\'\i{}n and Mimouni, Kin",
    title = "{Compilation of low-energy constraints on 4-fermion operators in the SMEFT}",
    eprint = "1706.03783",
    archivePrefix = "arXiv",
    primaryClass = "hep-ph",
    doi = "10.1007/JHEP08(2017)123",
    journal = "JHEP",
    volume = "08",
    pages = "123",
    year = "2017"
}

@article{Baker:2019sli,
    author = {Baker, Michael J. and Fuentes-Mart\'\i{}n, Javier and Isidori, Gino and K\"onig, Matthias},
    title = "{High- $p_T$ signatures in vector\textendash{}leptoquark models}",
    eprint = "1901.10480",
    archivePrefix = "arXiv",
    primaryClass = "hep-ph",
    reportNumber = "ZU-TH-19/05",
    doi = "10.1140/epjc/s10052-019-6853-x",
    journal = "Eur. Phys. J. C",
    volume = "79",
    number = "4",
    pages = "334",
    year = "2019"
}

@article{Zhou:2019frk,
    author = "Zhou, Bei and Beacom, John F.",
    title = "{W-boson and trident production in TeV\textendash{}PeV neutrino observatories}",
    eprint = "1910.10720",
    archivePrefix = "arXiv",
    primaryClass = "hep-ph",
    doi = "10.1103/PhysRevD.101.036010",
    journal = "Phys. Rev. D",
    volume = "101",
    number = "3",
    pages = "036010",
    year = "2020"
}

@article{Zhou:2019vxt,
    author = "Zhou, Bei and Beacom, John F.",
    title = "{Neutrino-nucleus cross sections for W-boson and trident production}",
    eprint = "1910.08090",
    archivePrefix = "arXiv",
    primaryClass = "hep-ph",
    doi = "10.1103/PhysRevD.101.036011",
    journal = "Phys. Rev. D",
    volume = "101",
    number = "3",
    pages = "036011",
    year = "2020"
}

@article{DiLuzio:2018zxy,
    author = "Di Luzio, Luca and Fuentes-Martin, Javier and Greljo, Admir and Nardecchia, Marco and Renner, Sophie",
    title = "{Maximal Flavour Violation: a Cabibbo mechanism for leptoquarks}",
    eprint = "1808.00942",
    archivePrefix = "arXiv",
    primaryClass = "hep-ph",
    reportNumber = "IPPP/18/69, MITP/18-072, CERN-TH-2018-182",
    doi = "10.1007/JHEP11(2018)081",
    journal = "JHEP",
    volume = "11",
    pages = "081",
    year = "2018"
}

@article{Cornella:2021sby,
    author = "Cornella, Claudia and Faroughy, Darius A. and Fuentes-Martin, Javier and Isidori, Gino and Neubert, Matthias",
    title = "{Reading the footprints of the B-meson flavor anomalies}",
    eprint = "2103.16558",
    archivePrefix = "arXiv",
    primaryClass = "hep-ph",
    doi = "10.1007/JHEP08(2021)050",
    journal = "JHEP",
    volume = "08",
    pages = "050",
    year = "2021"
}

@article{Diaz:2017lit,
    author = "Diaz, Bastian and Schmaltz, Martin and Zhong, Yi-Ming",
    title = "{The leptoquark Hunter\textquoteright{}s guide: Pair production}",
    eprint = "1706.05033",
    archivePrefix = "arXiv",
    primaryClass = "hep-ph",
    doi = "10.1007/JHEP10(2017)097",
    journal = "JHEP",
    volume = "10",
    pages = "097",
    year = "2017"
}

@article{Schmaltz:2018nls,
    author = "Schmaltz, Martin and Zhong, Yi-Ming",
    title = "{The leptoquark Hunter\textquoteright{}s guide: large coupling}",
    eprint = "1810.10017",
    archivePrefix = "arXiv",
    primaryClass = "hep-ph",
    doi = "10.1007/JHEP01(2019)132",
    journal = "JHEP",
    volume = "01",
    pages = "132",
    year = "2019"
}

@inproceedings{deBlas:2022ofj,
    author = "de Blas, Jorge and Du, Yong and Grojean, Christophe and Gu, Jiayin and Miralles, Victor and Peskin, Michael E. and Tian, Junping and Vos, Marcel and Vryonidou, Eleni",
    title = "{Global SMEFT Fits at Future Colliders}",
    booktitle = "{Snowmass 2021}",
    eprint = "2206.08326",
    archivePrefix = "arXiv",
    primaryClass = "hep-ph",
    month = "6",
    year = "2022"
}

@article{Dorsner:2018ynv,
    author = "Dor\v{s}ner, Ilja and Greljo, Admir",
    title = "{Leptoquark toolbox for precision collider studies}",
    eprint = "1801.07641",
    archivePrefix = "arXiv",
    primaryClass = "hep-ph",
    reportNumber = "MITP-18-005",
    doi = "10.1007/JHEP05(2018)126",
    journal = "JHEP",
    volume = "05",
    pages = "126",
    year = "2018"
}

@article{Alwall:2014hca,
    author = "Alwall, J. and Frederix, R. and Frixione, S. and Hirschi, V. and Maltoni, F. and Mattelaer, O. and Shao, H. -S. and Stelzer, T. and Torrielli, P. and Zaro, M.",
    title = "{The automated computation of tree-level and next-to-leading order differential cross sections, and their matching to parton shower simulations}",
    eprint = "1405.0301",
    archivePrefix = "arXiv",
    primaryClass = "hep-ph",
    reportNumber = "CERN-PH-TH-2014-064, CP3-14-18, LPN14-066, MCNET-14-09, ZU-TH-14-14",
    doi = "10.1007/JHEP07(2014)079",
    journal = "JHEP",
    volume = "07",
    pages = "079",
    year = "2014"
}

@article{Anchordoqui:2005ey,
    author = "Anchordoqui, Luis and Han, Tao and Hooper, Dan and Sarkar, Subir",
    title = "{Exotic neutrino interactions at the Pierre Auger Observatory}",
    eprint = "hep-ph/0508312",
    archivePrefix = "arXiv",
    reportNumber = "NUB-3257-TH-05, MADPH-05-1435, FERMILAB-PUB-05-373-A",
    doi = "10.1016/j.astropartphys.2005.10.006",
    journal = "Astropart. Phys.",
    volume = "25",
    pages = "14--32",
    year = "2006"
}

@article{Kirk:2023fin,
    author = "Kirk, Matthew and Okawa, Shohei and Wu, Keyun",
    title = "{A \ensuremath{\nu} window onto leptoquarks?}",
    eprint = "2307.11152",
    archivePrefix = "arXiv",
    primaryClass = "hep-ph",
    doi = "10.1007/JHEP12(2023)093",
    journal = "JHEP",
    volume = "12",
    pages = "093",
    year = "2023"
}

@article{Becirevic:2018uab,
    author = "Be\v{c}irevi\'c, Damir and Panes, Boris and Sumensari, Olcyr and Zukanovich Funchal, Renata",
    title = "{Seeking leptoquarks in IceCube}",
    eprint = "1803.10112",
    archivePrefix = "arXiv",
    primaryClass = "hep-ph",
    doi = "10.1007/JHEP06(2018)032",
    journal = "JHEP",
    volume = "06",
    pages = "032",
    year = "2018"
}

@article{King:1984qk,
    author = "King, Stephen F. and Sharpe, Stephen R.",
    title = "{Exotic {CERN} Events From Exotic Color States}",
    reportNumber = "HUTP-84/A062",
    doi = "10.1016/0550-3213(85)90517-6",
    journal = "Nucl. Phys. B",
    volume = "253",
    pages = "1--13",
    year = "1985"
}

@article{Mandal:2016csb,
    author = "Mandal, Tanumoy and Mitra, Subhadip and Seth, Satyajit",
    title = "{Probing Compositeness with the CMS $eejj$ \& $eej$ Data}",
    eprint = "1602.01273",
    archivePrefix = "arXiv",
    primaryClass = "hep-ph",
    reportNumber = "MITP/16-012, MITP-16-012",
    doi = "10.1016/j.physletb.2016.05.020",
    journal = "Phys. Lett. B",
    volume = "758",
    pages = "219--225",
    year = "2016"
}

@article{Almeida:2022udp,
    author = "Almeida, Eduardo da Silva and Alves, Alexandre and \'Eboli, Oscar J. P. and Queiroz, F. S.",
    title = "{Resonant lepton-gluon collisions at the Large Hadron Collider}",
    eprint = "2212.06178",
    archivePrefix = "arXiv",
    primaryClass = "hep-ph",
    doi = "10.1103/PhysRevD.107.055024",
    journal = "Phys. Rev. D",
    volume = "107",
    number = "5",
    pages = "055024",
    year = "2023"
}

@article{Goncalves-Netto:2013nla,
    author = "Goncalves-Netto, Dorival and Lopez-Val, David and Mawatari, Kentarou and Wigmore, Ioan and Plehn, Tilman",
    title = "{Looking for leptogluons}",
    eprint = "1303.0845",
    archivePrefix = "arXiv",
    primaryClass = "hep-ph",
    doi = "10.1103/PhysRevD.87.094023",
    journal = "Phys. Rev. D",
    volume = "87",
    pages = "094023",
    year = "2013"
}

@article{H1:1993vsn,
    author = "Abt, I. and others",
    collaboration = "H1",
    title = "{A Search for leptoquarks, leptogluons and excited leptons in H1 at HERA}",
    reportNumber = "DESY-93-029",
    doi = "10.1016/0550-3213(93)90255-N",
    journal = "Nucl. Phys. B",
    volume = "396",
    pages = "3--26",
    year = "1993"
}

@article{Dobado:1987pj,
    author = "Dobado, A. and Herrero, M. J. and Munoz, C.",
    title = "{Production of Leptoquarks From Superstring Models in $e p$ Colliders}",
    reportNumber = "FTUAM/87-4",
    doi = "10.1016/0370-2693(87)90638-1",
    journal = "Phys. Lett. B",
    volume = "191",
    pages = "449--455",
    year = "1987"
}

@article{Dobrescu:2021fny,
    author = "Dobrescu, Bogdan A.",
    title = "{Quark and Lepton Compositeness: A Renormalizable Model}",
    eprint = "2112.15132",
    archivePrefix = "arXiv",
    primaryClass = "hep-ph",
    reportNumber = "FERMILAB-PUB-21-771-T",
    doi = "10.1103/PhysRevLett.128.241804",
    journal = "Phys. Rev. Lett.",
    volume = "128",
    number = "24",
    pages = "241804",
    year = "2022"
}

@article{Hou:2019efy,
    author = "Hou, Tie-Jiun and others",
    title = "{New CTEQ global analysis of quantum chromodynamics with high-precision data from the LHC}",
    eprint = "1912.10053",
    archivePrefix = "arXiv",
    primaryClass = "hep-ph",
    reportNumber = "MSUHEP-19-025, PITT-PACC-1911, SMU-HEP-19-03",
    doi = "10.1103/PhysRevD.103.014013",
    journal = "Phys. Rev. D",
    volume = "103",
    number = "1",
    pages = "014013",
    year = "2021"
}

@article{Assi:2025rjx,
    author = "Assi, Beno\^\i{}t and Dobrescu, Bogdan",
    title = "{Composite quarks and leptons with embedded QCD}",
    eprint = "2501.11607",
    archivePrefix = "arXiv",
    primaryClass = "hep-ph",
    reportNumber = "FERMILAB-PUB-25-0021-T",
    month = "1",
    year = "2025"
}

@article{Shupe:1979fv,
    author = "Shupe, M. A.",
    title = "{A Composite Model of Leptons and Quarks}",
    doi = "10.1016/0370-2693(79)90627-0",
    journal = "Phys. Lett. B",
    volume = "86",
    pages = "87--92",
    year = "1979"
}

@article{Harari:1979gi,
    author = "Harari, Haim",
    title = "{A Schematic Model of Quarks and Leptons}",
    reportNumber = "SLAC-PUB-2310",
    doi = "10.1016/0370-2693(79)90626-9",
    journal = "Phys. Lett. B",
    volume = "86",
    pages = "83--86",
    year = "1979"
}

@article{Harari:1981uh,
    author = "Harari, Haim and Seiberg, Nathan",
    title = "{The Rishon Model}",
    reportNumber = "WIS-81/38-Ph",
    doi = "10.1016/0550-3213(82)90426-6",
    journal = "Nucl. Phys. B",
    volume = "204",
    pages = "141--167",
    year = "1982"
}

@article{Han:2025wdy,
    author = "Han, Tao and Low, Matthew and Wu, Tong Arthur and Xie, Keping",
    title = "{Colorful Particle Production at High-Energy Muon Colliders}",
    eprint = "2502.20443",
    archivePrefix = "arXiv",
    primaryClass = "hep-ph",
    month = "2",
    year = "2025"
}

@article{Han:2003wu,
    author = "Han, Tao and Logan, Heather E. and McElrath, Bob and Wang, Lian-Tao",
    title = "{Phenomenology of the little Higgs model}",
    eprint = "hep-ph/0301040",
    archivePrefix = "arXiv",
    reportNumber = "MADPH-02-1317",
    doi = "10.1103/PhysRevD.67.095004",
    journal = "Phys. Rev. D",
    volume = "67",
    pages = "095004",
    year = "2003"
}

@article{Fritzsch:1981zh,
    author = "Fritzsch, H. and Mandelbaum, G.",
    title = "{Weak Interactions as Manifestations of the Substructure of Leptons and Quarks}",
    reportNumber = "MPI-PAE/PTh 22/81",
    doi = "10.1016/0370-2693(81)90626-2",
    journal = "Phys. Lett. B",
    volume = "102",
    pages = "319--322",
    year = "1981"
}

@misc{nuflux:2024,
author = "IceCube Collaboration",
title  = "NuFlux: A library for calculating atmospheric neutrino fluxes",
doi = "10.5281/zenodo.5874708",
url = "https://zenodo.org/records/8180337",
version = "2.0.7",
}

@misc{MCEq,
	title        = {MCEq},
	howpublished = {\url{https://github.com/mceq-project/MCEq}},
}

@article{Riehn:2017mfm,
    author = "Riehn, Felix and Dembinski, Hans P. and Engel, Ralph and Fedynitch, Anatoli and Gaisser, Thomas K. and Stanev, Todor",
    title = "{The hadronic interaction model SIBYLL 2.3c and Feynman scaling}",
    eprint = "1709.07227",
    archivePrefix = "arXiv",
    primaryClass = "hep-ph",
    doi = "10.22323/1.301.0301",
    journal = "PoS",
    volume = "ICRC2017",
    pages = "301",
    year = "2018"
}

@article{Gaisser:2011klf,
    author = "Gaisser, Thomas K.",
    title = "{Spectrum of cosmic-ray nucleons, kaon production, and the atmospheric muon charge ratio}",
    eprint = "1111.6675",
    archivePrefix = "arXiv",
    primaryClass = "astro-ph.HE",
    doi = "10.1016/j.astropartphys.2012.02.010",
    journal = "Astropart. Phys.",
    volume = "35",
    pages = "801--806",
    year = "2012"
}

@article{Torre:2024edq,
    author = "Torre, S. Dalla",
    title = "{The Electron-Ion Collider and the ePIC experiment}",
    doi = "10.1393/ncc/i2024-24144-6",
    journal = "Nuovo Cim. C",
    volume = "47",
    number = "4",
    pages = "144",
    year = "2024"
}

@article{Xie:2023suk,
    author = "Xie, Keping and Gao, Jun and Hobbs, T. J. and Stump, Daniel R. and Yuan, C. -P.",
    collaboration = "CTEQ-TEA",
    title = "{High-energy neutrino deep inelastic scattering cross sections}",
    eprint = "2303.13607",
    archivePrefix = "arXiv",
    primaryClass = "hep-ph",
    reportNumber = "ANL-181430, MSUHEP-23-003, PITT-PACC-2302",
    doi = "10.1103/PhysRevD.109.113001",
    journal = "Phys. Rev. D",
    volume = "109",
    number = "11",
    pages = "113001",
    year = "2024"
}

@article{IceCube:2024fxo,
    author = "Abbasi, R. and others",
    collaboration = "IceCube",
    title = "{Characterization of the astrophysical diffuse neutrino flux using starting track events in IceCube}",
    eprint = "2402.18026",
    archivePrefix = "arXiv",
    primaryClass = "astro-ph.HE",
    doi = "10.1103/PhysRevD.110.022001",
    journal = "Phys. Rev. D",
    volume = "110",
    number = "2",
    pages = "022001",
    year = "2024"
}

@article{Abbasi:2021qfz,
    author = "Abbasi, R. and others",
    title = "{Improved Characterization of the Astrophysical Muon\textendash{}neutrino Flux with 9.5 Years of IceCube Data}",
    eprint = "2111.10299",
    archivePrefix = "arXiv",
    primaryClass = "astro-ph.HE",
    doi = "10.3847/1538-4357/ac4d29",
    journal = "Astrophys. J.",
    volume = "928",
    number = "1",
    pages = "50",
    year = "2022"
}

@article{Greisen:1966jv,
    author = "Greisen, Kenneth",
    title = "{End to the cosmic ray spectrum?}",
    doi = "10.1103/PhysRevLett.16.748",
    journal = "Phys. Rev. Lett.",
    volume = "16",
    pages = "748--750",
    year = "1966"
}

@article{Zatsepin:1966jv,
    author = "Zatsepin, G. T. and Kuzmin, V. A.",
    title = "{Upper limit of the spectrum of cosmic rays}",
    journal = "JETP Lett.",
    volume = "4",
    pages = "78--80",
    year = "1966"
}

@article{IceCube:2025ezc,
    author = "Abbasi, R. and others",
    collaboration = "IceCube",
    title = "{A search for extremely-high-energy neutrinos and first constraints on the ultra-high-energy cosmic-ray proton fraction with IceCube}",
    eprint = "2502.01963",
    archivePrefix = "arXiv",
    primaryClass = "astro-ph.HE",
    month = "2",
    year = "2025"
}

@misc{Gen2_TDR_web,
    howpublished = {\url{https://icecube-gen2.wisc.edu/science/publications/TDR/}}
}

@article{P-ONE:2020ljt,
    title        = {{The Pacific Ocean Neutrino Experiment}},
    author       = {Agostini, Matteo and others},
    year         = 2020,
    journal      = {Nature Astron.},
    volume       = 4,
    number       = 10,
    pages        = {913--915},
    doi          = {10.1038/s41550-020-1182-4},
    collaboration = {P-ONE},
    eprint       = {2005.09493},
    archiveprefix = {arXiv},
    primaryclass = {astro-ph.HE}
}

@article{Seckel:1997kk,
    author = "Seckel, D.",
    title = "{Neutrino photon reactions in astrophysics and cosmology}",
    eprint = "hep-ph/9709290",
    archivePrefix = "arXiv",
    reportNumber = "BA-97-32",
    doi = "10.1103/PhysRevLett.80.900",
    journal = "Phys. Rev. Lett.",
    volume = "80",
    pages = "900--903",
    year = "1998"
}

@article{Plestid:2024bva,
    author = "Plestid, Ryan and Zhou, Bei",
    title = "{Final state radiation from high and ultrahigh energy neutrino interactions}",
    eprint = "2403.07984",
    archivePrefix = "arXiv",
    primaryClass = "hep-ph",
    reportNumber = "FERMILAB-PUB-24-0087-T",
    doi = "10.1103/PhysRevD.111.043007",
    journal = "Phys. Rev. D",
    volume = "111",
    number = "4",
    pages = "043007",
    year = "2025"
}

@article{Xie:2023qbn,
    author = "Xie, Keping and Zhou, Bei and Hobbs, T. J.",
    collaboration = "CTEQ-TEA",
    title = "{The photon content of the neutron}",
    eprint = "2305.10497",
    archivePrefix = "arXiv",
    primaryClass = "hep-ph",
    reportNumber = "ANL-182626, MSUHEP-24-004, PITT-PACC-2314",
    doi = "10.1007/JHEP04(2024)022",
    journal = "JHEP",
    volume = "04",
    pages = "022",
    year = "2024"
}

@article{Alikhanov:2015kla,
    author = "Alikhanov, I.",
    title = "{Hidden Glashow resonance in neutrino\textendash{}nucleus collisions}",
    eprint = "1503.08817",
    archivePrefix = "arXiv",
    primaryClass = "hep-ph",
    doi = "10.1016/j.physletb.2016.03.009",
    journal = "Phys. Lett. B",
    volume = "756",
    pages = "247--253",
    year = "2016"
}

@article{Ye:2023dch,
    author = "Ye, Z. P. and others",
    title = "{A multi-cubic-kilometre neutrino telescope in the western Pacific Ocean}",
    doi = "10.1038/s41550-023-02087-6",
    journal = "Nature Astron.",
    volume = "7",
    number = "12",
    pages = "1497--1505",
    year = "2023"
}

@article{PierreAuger:2023pjg,
    author = "Abdul Halim, Adila and others",
    collaboration = "Pierre Auger",
    title = "{Latest results from the searches for ultra-high-energy photons and neutrinos at the Pierre Auger Observatory}",
    doi = "10.22323/1.444.1488",
    journal = "PoS",
    volume = "ICRC2023",
    pages = "1488",
    year = "2023"
}

@article{KM3NeT:2025npi,
    author = "Aiello, S. and others",
    collaboration = "KM3NeT",
    title = "{Observation of an ultra-high-energy cosmic neutrino with KM3NeT}",
    doi = "10.1038/s41586-024-08543-1",
    journal = "Nature",
    volume = "638",
    number = "8050",
    pages = "376--382",
    year = "2025",
    note = "[Erratum: Nature 640, E3 (2025)]"
}

@article{Ackermann:2022rqc,
    author = "Ackermann, Markus and others",
    title = "{High-energy and ultra-high-energy neutrinos: A Snowmass white paper}",
    eprint = "2203.08096",
    archivePrefix = "arXiv",
    primaryClass = "hep-ph",
    doi = "10.1016/j.jheap.2022.08.001",
    journal = "JHEAp",
    volume = "36",
    pages = "55--110",
    year = "2022"
}

@article{Anchordoqui:2018qom,
    author = "Anchordoqui, Luis A.",
    title = "{Ultra-High-Energy Cosmic Rays}",
    eprint = "1807.09645",
    archivePrefix = "arXiv",
    primaryClass = "astro-ph.HE",
    doi = "10.1016/j.physrep.2019.01.002",
    journal = "Phys. Rept.",
    volume = "801",
    pages = "1--93",
    year = "2019"
}

@article{AlvesBatista:2019tlv,
    author = "Alves Batista, Rafael and others",
    title = "{Open Questions in Cosmic-Ray Research at Ultrahigh Energies}",
    eprint = "1903.06714",
    archivePrefix = "arXiv",
    primaryClass = "astro-ph.HE",
    doi = "10.3389/fspas.2019.00023",
    journal = "Front. Astron. Space Sci.",
    volume = "6",
    pages = "23",
    year = "2019"
}

@article{Globus:2025ftu,
    author = "Globus, No{\'e}mie and Blandford, Roger D.",
    title = "{Ultrahigh-Energy Cosmic Rays}",
    doi = "10.1146/annurev-astro-052622-033150",
    journal = "Ann. Rev. Astron. Astrophys.",
    volume = "63",
    number = "1",
    pages = "339--377",
    year = "2025"
}

@article{Amoroso:2022eow,
    author = "Amoroso, S. and others",
    title = "{Snowmass 2021 Whitepaper: Proton Structure at the Precision Frontier}",
    eprint = "2203.13923",
    archivePrefix = "arXiv",
    primaryClass = "hep-ph",
    reportNumber = "Edinburgh 2022/08, FERMILAB-PUB-22-222-QIS-SCD-T, MPP-2022-32,
  SLAC-PUB-17652, SMU-HEP-22-02, TIF-UNIMI-2022-6",
    doi = "10.5506/APhysPolB.53.12-A1",
    journal = "Acta Phys. Polon. B",
    volume = "53",
    number = "12",
    pages = "12-A1",
    year = "2022"
}

@article{Ariga:2025qup,
    author = "Ariga, Akitaka and Boyd, Jamie and Kling, Felix and De Roeck, Albert",
    title = "{Neutrino Experiments at the Large Hadron Collider}",
    eprint = "2501.10078",
    archivePrefix = "arXiv",
    primaryClass = "hep-ex",
    doi = "10.1146/annurev-nucl-121423-101000",
    month = "1",
    year = "2025"
}

@article{Altmannshofer:2024hqd,
    author = {Altmannshofer, Wolfgang and M\"akel\"a, Toni and Sarkar, Subir and Trojanowski, Sebastian and Xie, Keping and Zhou, Bei},
    title = "{Discovering neutrino tridents at the Large Hadron Collider}",
    eprint = "2406.16803",
    archivePrefix = "arXiv",
    primaryClass = "hep-ph",
    reportNumber = "FERMILAB-PUB-24-0294-T, MSUHEP-24-007",
    doi = "10.1103/PhysRevD.110.072018",
    journal = "Phys. Rev. D",
    volume = "110",
    number = "7",
    pages = "072018",
    year = "2024"
}

@article{Bigaran:2024zxk,
    author = "Bigaran, Innes and Dev, P. S. Bhupal and Lopez Gutierrez, Diego and Machado, Pedro A. N.",
    title = "{Tau Tridents at Accelerator Neutrino Facilities}",
    eprint = "2406.20067",
    archivePrefix = "arXiv",
    primaryClass = "hep-ph",
    reportNumber = "FERMILAB-PUB-24-0318-T",
    month = "6",
    year = "2024"
}

@article{Francener:2024wul,
    author = "Francener, Reinaldo and Goncalves, Victor P. and Gratieri, Diego R.",
    title = "{Neutrino trident scattering at the LHC energy regime}",
    eprint = "2406.13593",
    archivePrefix = "arXiv",
    primaryClass = "hep-ph",
    doi = "10.1140/epjc/s10052-024-13323-2",
    journal = "Eur. Phys. J. C",
    volume = "84",
    number = "9",
    pages = "923",
    year = "2024"
}

@article{Wang:2025qap,
    author = "Wang, Isaac R. and Xu, Xun-Jie and Zhou, Bei",
    title = "{Widen the Resonance: Probing a New Regime of Neutrino Self-Interactions with Astrophysical Neutrinos}",
    eprint = "2501.07624",
    archivePrefix = "arXiv",
    primaryClass = "hep-ph",
    reportNumber = "FERMILAB-PUB-25-0016-T",
    month = "1",
    year = "2025"
}

@article{DeGouvea:2020ang,
    author = "De Gouv{\^e}a, Andr{\'e} and Martinez-Soler, Ivan and Perez-Gonzalez, Yuber F. and Sen, Manibrata",
    title = "{Fundamental physics with the diffuse supernova background neutrinos}",
    eprint = "2007.13748",
    archivePrefix = "arXiv",
    primaryClass = "hep-ph",
    reportNumber = "NUHEP-TH/20-08, FERMILAB-PUB-20-353-T",
    doi = "10.1103/PhysRevD.102.123012",
    journal = "Phys. Rev. D",
    volume = "102",
    pages = "123012",
    year = "2020"
}

@article{deGouvea:2022dtw,
    author = "de Gouv{\^e}a, Andr{\'e} and Martinez-Soler, Ivan and Perez-Gonzalez, Yuber F. and Sen, Manibrata",
    title = "{Diffuse supernova neutrino background as a probe of late-time neutrino mass generation}",
    eprint = "2205.01102",
    archivePrefix = "arXiv",
    primaryClass = "hep-ph",
    reportNumber = "IPPP/22/29",
    doi = "10.1103/PhysRevD.106.103026",
    journal = "Phys. Rev. D",
    volume = "106",
    number = "10",
    pages = "103026",
    year = "2022"
}

@article{Perez-Gonzalez:2025qjh,
    author = "Perez-Gonzalez, Yuber F. and Sen, Manibrata",
    title = "{Dynamic Neutrino Mass Ordering and Its Imprint on the Diffuse Supernova Neutrino Background}",
    eprint = "2501.16412",
    archivePrefix = "arXiv",
    primaryClass = "hep-ph",
    reportNumber = "IFT-UAM/CSIC-25-4",
    month = "1",
    year = "2025"
}

@article{IceCube:2024ayt,
    author = "Abbasi, R. and others",
    collaboration = "IceCube",
    title = "{Search for Neutrino Emission from Hard X-Ray AGN with IceCube}",
    eprint = "2406.06684",
    archivePrefix = "arXiv",
    primaryClass = "astro-ph.HE",
    doi = "10.3847/1538-4357/ada94b",
    journal = "Astrophys. J.",
    volume = "981",
    number = "2",
    pages = "131",
    year = "2025"
}

@article{IceCube:2024dou,
    author = "Abbasi, R. and others",
    collaboration = "IceCube",
    title = "{IceCube Search for Neutrino Emission from X-Ray Bright Seyfert Galaxies}",
    eprint = "2406.07601",
    archivePrefix = "arXiv",
    primaryClass = "astro-ph.HE",
    doi = "10.3847/1538-4357/addd05",
    journal = "Astrophys. J.",
    volume = "988",
    number = "1",
    pages = "141",
    year = "2025"
}

@article{IceCube:2022der,
    author = "Abbasi, R. and others",
    collaboration = "IceCube",
    title = "{Evidence for neutrino emission from the nearby active galaxy NGC 1068}",
    eprint = "2211.09972",
    archivePrefix = "arXiv",
    primaryClass = "astro-ph.HE",
    doi = "10.1126/science.abg3395",
    journal = "Science",
    volume = "378",
    number = "6619",
    pages = "538--543",
    year = "2022"
}

@article{IceCube:2019cia,
    author = "Aartsen, M. G. and others",
    collaboration = "IceCube",
    title = "{Time-Integrated Neutrino Source Searches with 10 Years of IceCube Data}",
    eprint = "1910.08488",
    archivePrefix = "arXiv",
    primaryClass = "astro-ph.HE",
    doi = "10.1103/PhysRevLett.124.051103",
    journal = "Phys. Rev. Lett.",
    volume = "124",
    number = "5",
    pages = "051103",
    year = "2020"
}

@article{IceCube:2018cha,
    author = "Aartsen, M. G. and others",
    collaboration = "IceCube",
    title = "{Neutrino emission from the direction of the blazar TXS 0506+056 prior to the IceCube-170922A alert}",
    eprint = "1807.08794",
    archivePrefix = "arXiv",
    primaryClass = "astro-ph.HE",
    doi = "10.1126/science.aat2890",
    journal = "Science",
    volume = "361",
    number = "6398",
    pages = "147--151",
    year = "2018"
}

@article{IceCube:2018dnn,
    author = "Aartsen, M. G. and others",
    collaboration = "IceCube, Fermi-LAT, MAGIC, AGILE, ASAS-SN, HAWC, H.E.S.S., INTEGRAL, Kanata, Kiso, Kapteyn, Liverpool Telescope, Subaru, Swift NuSTAR, VERITAS, VLA/17B-403",
    title = "{Multimessenger observations of a flaring blazar coincident with high-energy neutrino IceCube-170922A}",
    eprint = "1807.08816",
    archivePrefix = "arXiv",
    primaryClass = "astro-ph.HE",
    doi = "10.1126/science.aat1378",
    journal = "Science",
    volume = "361",
    number = "6398",
    pages = "eaat1378",
    year = "2018"
}

@article{Chang:2022hqj,
    author = "Chang, Po-Wen and Zhou, Bei and Murase, Kohta and Kamionkowski, Marc",
    title = "{High-energy neutrinos from choked-jet supernovae: Searches and implications}",
    eprint = "2210.03088",
    archivePrefix = "arXiv",
    primaryClass = "astro-ph.HE",
    doi = "10.1103/PhysRevD.109.103041",
    journal = "Phys. Rev. D",
    volume = "109",
    number = "10",
    pages = "103041",
    year = "2024"
}

@article{Zhou:2021rhl,
    author = "Zhou, Bei and Kamionkowski, Marc and Liang, Yun-feng",
    title = "{Search for High-Energy Neutrino Emission from Radio-Bright AGN}",
    eprint = "2103.12813",
    archivePrefix = "arXiv",
    primaryClass = "astro-ph.HE",
    doi = "10.1103/PhysRevD.103.123018",
    journal = "Phys. Rev. D",
    volume = "103",
    number = "12",
    pages = "123018",
    year = "2021"
}

@article{Neronov:2023aks,
    author = "Neronov, A. and Savchenko, D. and Semikoz, D. V.",
    title = "{Neutrino Signal from a Population of Seyfert Galaxies}",
    eprint = "2306.09018",
    archivePrefix = "arXiv",
    primaryClass = "astro-ph.HE",
    doi = "10.1103/PhysRevLett.132.101002",
    journal = "Phys. Rev. Lett.",
    volume = "132",
    number = "10",
    pages = "101002",
    year = "2024"
}

@article{IceCube:2023woj,
    author = "Abbasi, R. and others",
    collaboration = "IceCube",
    title = "{Search for 10{\textendash}1000 GeV Neutrinos from Gamma-Ray Bursts with IceCube}",
    eprint = "2312.11515",
    archivePrefix = "arXiv",
    primaryClass = "astro-ph.HE",
    doi = "10.3847/1538-4357/ad220b",
    journal = "Astrophys. J.",
    volume = "964",
    number = "2",
    pages = "126",
    year = "2024"
}

@article{IceCube:2016qvd,
    author = "Aartsen, M. G. and others",
    collaboration = "IceCube",
    title = "{The contribution of Fermi-2LAC blazars to the diffuse TeV-PeV neutrino flux}",
    eprint = "1611.03874",
    archivePrefix = "arXiv",
    primaryClass = "astro-ph.HE",
    doi = "10.3847/1538-4357/835/1/45",
    journal = "Astrophys. J.",
    volume = "835",
    number = "1",
    pages = "45",
    year = "2017"
}

@article{Bouri:2024ctc,
    author = "Bouri, Subhadip and Parashari, Priyank and Das, Mousumi and Laha, Ranjan",
    title = "{First search for high-energy neutrino emission from galaxy mergers}",
    eprint = "2404.06539",
    archivePrefix = "arXiv",
    primaryClass = "astro-ph.HE",
    doi = "10.1103/PhysRevD.111.063059",
    journal = "Phys. Rev. D",
    volume = "111",
    number = "6",
    pages = "063059",
    year = "2025"
}

@article{Balantekin:2023jlg,
    author = "Balantekin, A. Baha and Fuller, George M. and Ray, Anupam and Suliga, Anna M.",
    title = "{Probing self-interacting sterile neutrino dark matter with the diffuse supernova neutrino background}",
    eprint = "2310.07145",
    archivePrefix = "arXiv",
    primaryClass = "hep-ph",
    reportNumber = "N3AS-23-031, INT-PUB-23-042, CETUP-2023-010",
    doi = "10.1103/PhysRevD.108.123011",
    journal = "Phys. Rev. D",
    volume = "108",
    number = "12",
    pages = "123011",
    year = "2023"
}

@article{Beacom:2010kk,
    author = "Beacom, John F.",
    title = "{The Diffuse Supernova Neutrino Background}",
    eprint = "1004.3311",
    archivePrefix = "arXiv",
    primaryClass = "astro-ph.HE",
    doi = "10.1146/annurev.nucl.010909.083331",
    journal = "Ann. Rev. Nucl. Part. Sci.",
    volume = "60",
    pages = "439--462",
    year = "2010"
}

@article{Feng:2022inv,
    author = "Feng, Jonathan L. and others",
    title = "{The Forward Physics Facility at the High-Luminosity LHC}",
    eprint = "2203.05090",
    archivePrefix = "arXiv",
    primaryClass = "hep-ex",
    reportNumber = "UCI-TR-2022-01, CERN-PBC-Notes-2022-001, INT-PUB-22-006, BONN-TH-2022-04, FERMILAB-PUB-22-094-ND-SCD-T",
    doi = "10.1088/1361-6471/ac865e",
    journal = "J. Phys. G",
    volume = "50",
    number = "3",
    pages = "030501",
    year = "2023"
}

@article{Machado:2025ktc,
    author = "Machado, Pedro and Zhou, Bei",
    title = "{Neutrino Physics and Astrophysics at Colliders}",
    eprint = "2506.20855",
    archivePrefix = "arXiv",
    primaryClass = "hep-ph",
    reportNumber = "FERMILAB-PUB-25-0421-T",
    month = "6",
    year = "2025"
}

@article{HESS:2024etj,
    author = "Aharonian, F. and others",
    collaboration = "H.E.S.S.",
    title = "{High-Statistics Measurement of the Cosmic-Ray Electron Spectrum with H.E.S.S.}",
    eprint = "2411.08189",
    archivePrefix = "arXiv",
    primaryClass = "astro-ph.HE",
    doi = "10.1103/PhysRevLett.133.221001",
    journal = "Phys. Rev. Lett.",
    volume = "133",
    number = "22",
    pages = "221001",
    year = "2024"
}

@article{Taboada:2019lfc,
    author = "Taboada, Ignacio",
    editor = "Spiering, C.",
    collaboration = "IceCube",
    title = "{IceCube: An overview of physics results}",
    doi = "10.1051/epjconf/201920701002",
    journal = "EPJ Web Conf.",
    volume = "207",
    pages = "01002",
    year = "2019"
}

@article{Blum:2014ewa,
    author = "Blum, Kfir and Hook, Anson and Murase, Kohta",
    title = "{High energy neutrino telescopes as a probe of the neutrino mass mechanism}",
    eprint = "1408.3799",
    archivePrefix = "arXiv",
    primaryClass = "hep-ph",
    month = "8",
    year = "2014"
}

@article{Bustamante:2016ciw,
    author = "Bustamante, Mauricio and Beacom, John F. and Murase, Kohta",
    title = "{Testing decay of astrophysical neutrinos with incomplete information}",
    eprint = "1610.02096",
    archivePrefix = "arXiv",
    primaryClass = "astro-ph.HE",
    doi = "10.1103/PhysRevD.95.063013",
    journal = "Phys. Rev. D",
    volume = "95",
    number = "6",
    pages = "063013",
    year = "2017"
}

@article{Laha:2013lka,
    author = "Laha, Ranjan and Beacom, John F. and Dasgupta, Basudeb and Horiuchi, Shunsaku and Murase, Kohta",
    title = "{Demystifying the PeV Cascades in IceCube: Less (Energy) is More (Events)}",
    eprint = "1306.2309",
    archivePrefix = "arXiv",
    primaryClass = "astro-ph.HE",
    doi = "10.1103/PhysRevD.88.043009",
    journal = "Phys. Rev. D",
    volume = "88",
    pages = "043009",
    year = "2013"
}

@inproceedings{Muzio:2025xen,
    author = "Muzio, Marco Stein",
    title = "{Ultrahigh energy cosmic rays and neutrino flux models}",
    eprint = "2502.11834",
    archivePrefix = "arXiv",
    primaryClass = "astro-ph.HE",
    doi = "10.1140/epjs/s11734-025-01502-5",
    month = "2",
    year = "2025"
}

@article{HiRes:2007lra,
    author = "Abbasi, R. U. and others",
    collaboration = "HiRes",
    title = "{First observation of the Greisen-Zatsepin-Kuzmin suppression}",
    eprint = "astro-ph/0703099",
    archivePrefix = "arXiv",
    reportNumber = "RU-PNA-002",
    doi = "10.1103/PhysRevLett.100.101101",
    journal = "Phys. Rev. Lett.",
    volume = "100",
    pages = "101101",
    year = "2008"
}

@article{IceCube:2020acn,
    author = "Aartsen, M. G. and others",
    collaboration = "IceCube",
    title = "{Characteristics of the diffuse astrophysical electron and tau neutrino flux with six years of IceCube high energy cascade data}",
    eprint = "2001.09520",
    archivePrefix = "arXiv",
    primaryClass = "astro-ph.HE",
    doi = "10.1103/PhysRevLett.125.121104",
    journal = "Phys. Rev. Lett.",
    volume = "125",
    number = "12",
    pages = "121104",
    year = "2020"
}

@article{IceCube:2013low,
    author = "Aartsen, M. G. and others",
    collaboration = "IceCube",
    title = "{Evidence for High-Energy Extraterrestrial Neutrinos at the IceCube Detector}",
    eprint = "1311.5238",
    archivePrefix = "arXiv",
    primaryClass = "astro-ph.HE",
    doi = "10.1126/science.1242856",
    journal = "Science",
    volume = "342",
    pages = "1242856",
    year = "2013"
}

@article{Harari:1982xy,
    author = "Harari, Haim",
    editor = "Mosher, Anne and Feldman, Gary J. and Gilman, Frederick J. and Leith, David W. G. S.",
    title = "{Composite Models for Quarks and Leptons}",
    reportNumber = "WIS-82/60-Ph",
    doi = "10.1016/0370-1573(84)90207-2",
    journal = "Phys. Rept.",
    volume = "104",
    pages = "159",
    year = "1984"
}

@article{Zhou:2021xuh,
    author = "Zhou, Bei and Beacom, John F.",
    title = "{Dimuons in neutrino telescopes: New predictions and first search in IceCube}",
    eprint = "2110.02974",
    archivePrefix = "arXiv",
    primaryClass = "hep-ph",
    doi = "10.1103/PhysRevD.105.093005",
    journal = "Phys. Rev. D",
    volume = "105",
    number = "9",
    pages = "093005",
    year = "2022"
}

@article{IceCube:2018fhm,
    author = "Aartsen, M. G. and others",
    collaboration = "IceCube",
    title = "{Differential limit on the extremely-high-energy cosmic neutrino flux in the presence of astrophysical background from nine years of IceCube data}",
    eprint = "1807.01820",
    archivePrefix = "arXiv",
    primaryClass = "astro-ph.HE",
    doi = "10.1103/PhysRevD.98.062003",
    journal = "Phys. Rev. D",
    volume = "98",
    number = "6",
    pages = "062003",
    year = "2018"
}

@article{Leal:2025eou,
    author = "Leal, Luighi P. S. and Naredo-Tuero, Daniel and Funchal, Renata Zukanovich",
    title = "{Cosmogenic neutrinos as probes of new physics}",
    eprint = "2504.10576",
    archivePrefix = "arXiv",
    primaryClass = "hep-ph",
    doi = "10.1007/JHEP08(2025)057",
    journal = "JHEP",
    volume = "08",
    pages = "057",
    year = "2025"
}

@article{Machado:2025uhe,
    author = "Machado, Pedro A. N. and Wang, Isaac R. and Xu, Xun-Jie and Zhou, Bei",
    title = "{Widen the Resonance at Ultra-High Energies: Novel Probes of Neutrino Self-interactions in the High-Mass Regime}",
    eprint = "2512.00165",
    archivePrefix = "arXiv",
    primaryClass = "hep-ph",
    year = "2025"
}

\clearpage
\onecolumngrid
\setcounter{section}{0}
\setcounter{subsection}{0}
\setcounter{subsubsection}{0}
\setcounter{equation}{0}
\setcounter{figure}{0}
\setcounter{table}{0}
\renewcommand{\theequation}{S\arabic{equation}}
\renewcommand{\thefigure}{S\arabic{figure}}
\renewcommand{\thetable}{S\arabic{table}}
\makeatletter
\renewcommand{\theHsection}{supp.\arabic{section}}
\renewcommand{\theHsubsection}{supp.\arabic{section}.\arabic{subsection}}
\renewcommand{\theHsubsubsection}{supp.\arabic{section}.\arabic{subsection}.\arabic{subsubsection}}
\renewcommand{\theHequation}{supp.\arabic{equation}}
\renewcommand{\theHfigure}{supp.\arabic{figure}}
\renewcommand{\theHtable}{supp.\arabic{table}}
\makeatother

\begin{center}
{\large\bfseries Supplemental Material for Large Neutrino ``Colliders"}\\[0.8em]
\begin{minipage}{0.92\textwidth}
\small
In this Supplemental Material, we present the relevant technical details of our sensitivity estimates for several representative new-physics models considered in the main text. In Sec.~\ref{app:flux}, we discuss high-energy and ultra-high-energy neutrino fluxes. The Standard Model neutrino interactions are presented in Sec.~\ref{app:SM-cross-section}. The event rates and statistical treatment are discussed in Sec.~\ref{app_EventRatesStat}. Finally, in Sec.~\ref{app:models}, we provide the details of all the new-physics models---contact interactions, leptogluons, leptoquarks, and $W'/H^-$---as well as the corresponding parameter-space reaches at the Large Neutrino Colliders.
\end{minipage}
\end{center}
\vspace{1em}

\section{HE and UHE neutrino fluxes}
\label{app:flux}

The atmospheric neutrinos originate from cosmic rays that bombard the Earth's atmosphere, producing mesons that subsequently decay into neutrinos~\cite{Gaisser:2002jj}.
They dominate the flux at $E_\nu \lesssim 10^5$GeV.
For the atmospheric neutrino fluxes at production, we use \texttt{NuFlux}~\cite{nuflux:2024} with the widely-used {\tt H3a\_SIBYLL23C} model, which is calculated with {\tt {MCEq~1.2.1}}~\cite{MCEq} assuming cosmic ray model {\tt {H3a}}~\cite{Gaisser:2011klf} and the hadronic interaction model {\tt{Sibyll\,2.3C}}~\cite{Riehn:2017mfm}. The effects of solar modulation and geomagnetic cutoff on cosmic rays and neutrino oscillations during propagation are only important at lower energies and are thus negligible in our work.
To estimate uncertainties, we compare {\tt H3a\_SIBYLL23C} to other recent flux models in \texttt{NuFlux}, such as {\tt H3a\_SIBYLL21} and {\tt IPhonda2014}, finding that the uncertainty is about 30\% at $E_\nu \sim 10^5$~GeV.

The astrophysical neutrino flux dominates at $E_\nu \gtrsim 10^5$~GeV, which has been measured by the IceCube Neutrino Observatory since 2013~\cite{IceCube:2013low}.
While the origins of these neutrinos remain uncertain, significant progress has been made over the past decade~\cite{IceCube:2016qvd, IceCube:2018dnn, IceCube:2018cha, IceCube:2019cia, Zhou:2021rhl, IceCube:2022der, Chang:2022hqj, IceCube:2023woj, Neronov:2023aks, IceCube:2024ayt, IceCube:2024dou, Bouri:2024ctc}, and the all-sky diffuse flux relevant for our calculation is now well measured~\cite{IceCube:2020acn, Abbasi:2021qfz, IceCube:2024fxo}. We use the most recent measurements from IceCube using starting track events~\cite{IceCube:2024fxo}, which gives
\begin{equation}
\begin{aligned}
\phi_\nu^{\text {Astro}}(E_\nu)=1.68_{-0.22}^{+0.19} \times 10^{-18} \left(\frac{E_\nu}{100 \, \mathrm{TeV}}\right)^{-2.58_{-0.09}^{+0.10}}
\label{eq_phi_astro}
\end{aligned}
\end{equation}
in the unit of $\mathrm{GeV}^{-1} \mathrm{~cm}^{-2} \mathrm{~s}^{-1} \mathrm{sr}^{-1}$ and per flavor (e.g., $\nu_\mu+\bar{\nu}_\mu$).

At the highest energies, beyond $\sim$ PeV, there is an UHE neutrino component, with two main possible contributions within the SM (see, e.g., Refs.~\cite{Ackermann:2022rqc, Muzio:2025xen} for recent reviews).
The first is cosmogenic, resulting from the interaction of UHE cosmic rays (UHECRs) with the cosmic microwave background, a process responsible for the Greisen--Zatsepin--Kuzmin (GZK) cutoff~\cite{Greisen:1966jv, Zatsepin:1966jv, HiRes:2007lra}. The resulting neutrinos are also referred to as GZK neutrinos. The other possible contribution comes from astrophysical sources. However, there are huge uncertainties in both contributions, the former primarily due to the UHECR composition and source evolution, and the latter primarily due to the huge variety of astrophysical environments.
The $E_\nu^2 \phi_\nu$ from different predictions varies between $\sim 10^{-11}$ and $\sim \rm 10^{-7}\, GeV\, cm^{-2}\, s^{-1}\, sr^{-1}$, with different spectral shape, though most of them roughly follow $E_\nu^{-2}$.
Excitingly, the recent observations of KM3NeT have identified an UHE neutrino event of 220~PeV (median) and 72~PeV -- 2.6~EeV (90\% confidence level)~\cite{KM3NeT:2025npi}, which implies an UHE neutrino flux of
\begin{equation}
\begin{aligned}
E_\nu^2 \phi^{\rm UHE}_\nu(E_\nu) =
5.8_{-3.7}^{+10.1} \times 10^{-8} \, \mathrm{GeV} \mathrm{~cm}^{-2} \mathrm{~s}^{-1} \mathrm{sr}^{-1} \,.
\label{eq_nuflux_KM3NeT}
\end{aligned}
\end{equation}

{
To test the flux dependence with a lower physical benchmark, we also use a conservative cosmogenic flux~\cite{Leal:2025eou,Machado:2025uhe}. The UHECR sources inject protons with spectral index $\Gamma=2.5$ up to $E_p^{\rm max}=250$~EeV, and their redshift evolution is parameterized by
\begin{equation}
S_E(z;m)=
\begin{cases}
(1+z)^m, & z<1,\\
2^m, & 1\leq z<4,\\
2^m[(1+z)/5]^{-3.5}, & 4\leq z<7.
\end{cases}
\label{eq:source_evolution_uhe}
\end{equation}
Our conservative central flux is the physical moderate-evolution case $m=1.5$. We take the pessimistic $m=0$ and optimistic $m=3$ cases as the lower and upper boundaries of a $2\sigma$ envelope.

The atmospheric and astrophysical components and the two different UHE fluxes are shown in Fig.~2 of the main text. They provide the estimated ``luminosity" of the L$\nu$C and expose the dominant uncertainty in the UHE region.
An important effect for all three components is the absorption when these neutrinos pass through the Earth due to neutrino interactions with matter, which we discuss next.
}

As discussed in the main text, HE and UHE neutrinos may originate from specific astrophysical sources. Identifying these sources would allow one to exploit their directional information to enhance sensitivity to new physics, rather than relying on sky-averaged fluxes. This would make the L$\nu$C even more analogous to conventional colliders, where the beam direction is controlled and well defined.
Moreover, once the redshift distribution of the sources becomes known, new physics can be tested across a range of cosmological baselines, offering a capability that terrestrial colliders cannot achieve.
For instance, the diffuse supernova neutrino background at MeV energies~\cite{Beacom:2010kk}, with its well-understood redshift distribution, exemplifies how cosmological-distance information can be used to probe new physics~\cite{DeGouvea:2020ang, deGouvea:2022dtw, Balantekin:2023jlg, Wang:2025qap, Perez-Gonzalez:2025qjh}.

\section{Neutrino interactions in the SM}
\label{app:SM-cross-section}

Above TeV, the dominant neutrino interaction in the SM are neutrino-nucleus deep-inelastic scattering (DIS)~\cite{Xie:2023suk}, including charged current (CC) and neutral current (NC) DIS.
The cross sections per nucleon at 100~GeV are $\sim 5 \times 10^{-37}$~cm$^2$ for CCDIS and $\sim 2 \times 10^{-37}$~cm$^2$ for NCDIS. At lower energies, they grow linearly on $E_\nu$, and then the growth slows down and eventually saturates at $E_\nu^{0.3}$ above $\sim 10^7$~GeV, due to the masses of the weak bosons ($\simeq 80$~GeV).
For this work, we adopt the DIS cross sections and uncertainties calculated to next-to-next-to-next-leading-order (N$^3$LO) in Ref.~\cite{Xie:2023suk}, which is based CT18 parton-distribution functions (PDFs) of proton~\cite{Hou:2019efy} and neutron~\cite{Xie:2023qbn}, with effects including heavy-quark mass, small-$x$ resummation, and nuclear effects.

Another important interaction is the Glashow resonance, $\bar{\nu}_e + e^- \to W^-$, and the cross section is characterized by a large peak around 6.3~PeV. However, this is only relevant for $\bar{\nu}_e$ and is therefore irrelevant to our work, which focuses on $\nu_\mu + \bar{\nu}_\mu$ or CC-like events.

In addition to the dominant interactions, a few other interactions become increasingly important as the statistics keep increasing, especially when the next-generation HE neutrino detectors start operation with $\sim$10 times larger volume~\cite{IceCube-Gen2:2020qha, Gen2_TDR_web, P-ONE:2020ljt, Ye:2023dch, Huang:2023mzt, Zhang:2024slv}.
The second largest interaction is the neutrino-nucleus W-boson production (WBP), $\nu + A \to l^\pm + W^\mp + A'$, where $A$ and $A'$ denote the initial and final state nucleus, respectively~\cite{Seckel:1997kk, Alikhanov:2015kla, Zhou:2019vxt, Zhou:2019frk}.
WBP contributes $\simeq 10\%$ of the total neutrino-nucleus cross section~\cite{Zhou:2019vxt, Zhou:2019frk}.
We use the calculations initially developed in Refs.~\cite{Zhou:2019vxt, Zhou:2019frk} and updated in Ref.~\cite{Xie:2023qbn}, and the cross sections and uncertainties can be found on \href{https://github.com/beizhouphys/neutrino-W-boson-and-trident-production}{GitHub} \github{beizhouphys/neutrino-W-boson-and-trident-production}. Note that at the relevant energies, neutrino trident production~\cite{Altmannshofer:2024hqd, Bigaran:2024zxk, Francener:2024wul} is a part of WBP~\cite{Zhou:2019vxt}, so we do not include it separately.

On the other hand, final-state radiation of photons (FSR) introduces non-negligible corrections at high energies~\cite{Plestid:2024bva}.
FSR mainly impacts the differential cross sections, as the emitted photon takes away energy from the final-state charged lepton and is typically indistinguishable from the hadronic cascade.
Take CCDIS for example [$\nu_{\ell} (\bar{\nu}_{\ell}) +A \rightarrow \ell^{-} (\ell^{+}) + X + \gamma$].
This can reduce the charged-lepton energy ($E_\ell$) by as much as 5\% and increase the cascade energy by as much as 25\%.
The corrections can be applied to the $\dd \sigma/ \dd E_\ell$ by~\cite{Plestid:2024bva}
\begin{equation}
\frac{\mathrm{d} \sigma^{(1)}}{\mathrm{d} E_{\ell}}=\frac{\alpha}{2 \pi} \int \mathrm{~d} y \int \mathrm{~d} z \frac{\mathrm{~d} \sigma^{(0)}}{\mathrm{d} y} \delta\left[E_{\ell}-(1-y)\, z\,E_\nu\right]
\times \log \left(\frac{s}{m_{\ell}^2}\right)\left[\frac{1+z^2}{[1-z]_{+}}+\frac{3}{2} \delta(1-z)\right] \,,
\end{equation}
where the superscripts ``(1)'' and ``(0)'' denote the $\dd \sigma/ \dd E_\ell$ with and without the FSR correction, respectively, $y \equiv (1-E_\ell)/E_\nu$, $(1-z)$ the fraction of the charged lepton's momentum carried away by the photon, $s$ the center of mass energy, and $m_\ell$ the charged-lepton mass.

\section{Event rates and statistics for Large Neutrino Colliders}
\label{app_EventRatesStat}
In this section, we first overview of the currently running and future neutrino telescopes as Large Neutrino Colliders (L$\nu$Cs), including their volumes, operational timelines, and budget costs in Sec.~\ref {app:time&cost}.
Meanwhile, we also provide a comparison with traditional terrestrial colliders, highlighting the L$\nu$Cs's cost-effectiveness and promptness. Afterwards, we will provide the details to estimate the signal and background event rates and statistical analysis at L$\nu$Cs in Sec.~\ref{app_EventRatesStat_ER} and Sec.~\ref{app:statistics}, respectively, based on the typical L$\nu$C volumes, and operational time.

\subsection{The cost and operational time of L$\nu$Cs}
\label{app:time&cost}

\begin{figure}
    \centering
    \includegraphics[width=0.99\linewidth]{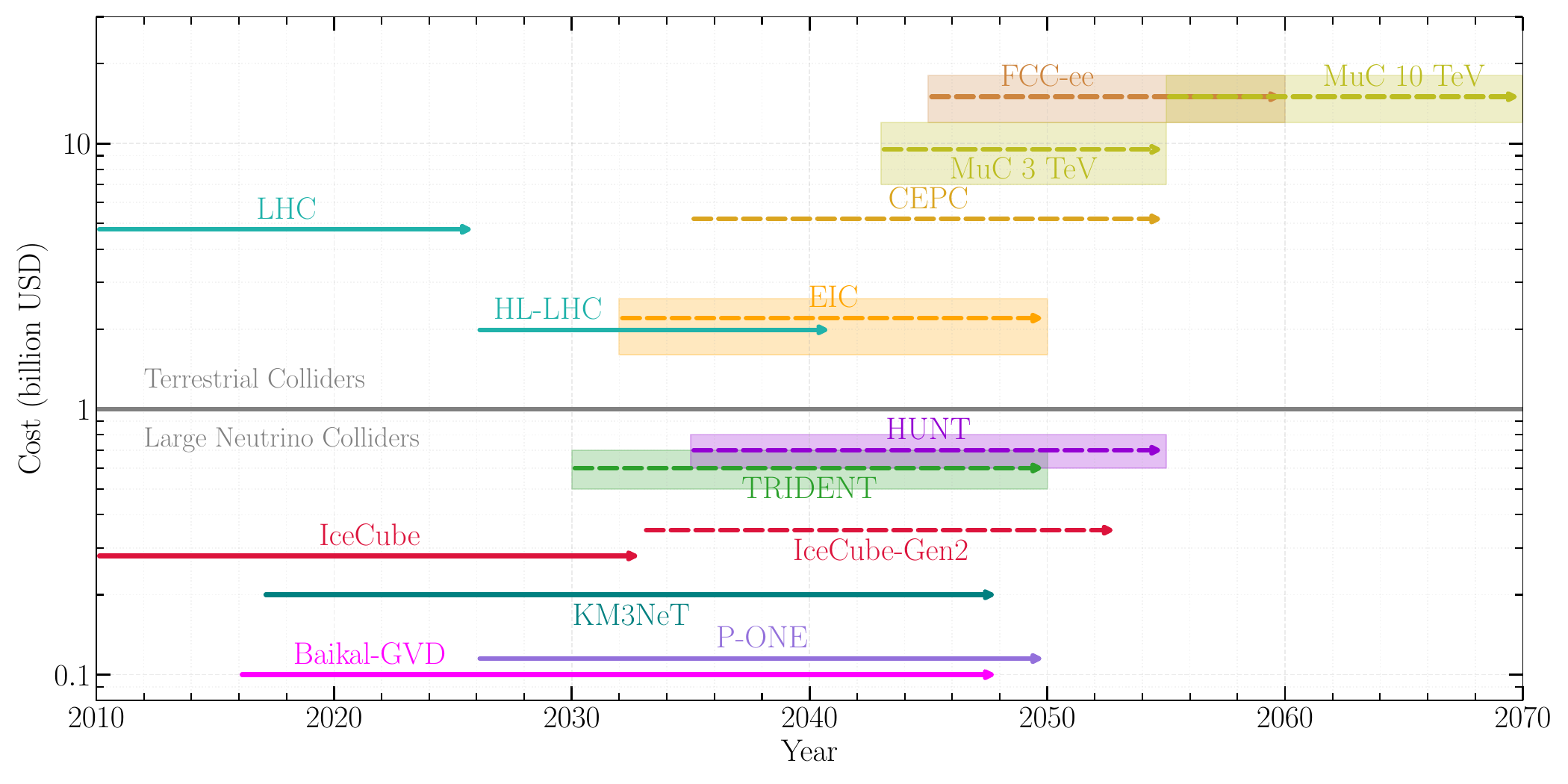}
    \caption{The projected timelines and costs of Neutrino Telescopes as Large Neutrino Colliders (L$\nu$Cs) in comparison with ongoing and future Terrestrial Colliders.}
    \label{fig:LnuCvsFCC}
\end{figure}

As briefly outlined in the main text, a few neutrino telescopes as L$\nu$Cs are fully or partially operational. The IceCube Neutrino Observatory~\cite{IceCube_web} has instrumented about 1~km$^3$ volume, and has been operational for over 17 years. Together with its upgrade phase~\cite{Ishihara:2019aao}, IceCube is expected to run until 2035, as illustrated in Fig.~\ref{fig:LnuCvsFCC}. Meanwhile, we also present the total cost of \$279 million~\cite{IceCube_web}, as the vertical axis, to be compared with other L$\nu$Cs, as well as terrestrial colliders. KM3NeT~\cite{KM3Net:2016zxf}, with a total cost around \texteuro200 million~\cite{KM3NeT:2009xxi}, began installing its detector around 2017, and is expected to complete its full $\simeq1$~km$^3$ volume by 2027-2028. Roughly around the same time, Baikal-GVD~\cite{Allakhverdyan:2021vkk,Aynutdinov:2023ifk} started around 2016 and completed its first phase with a total cost of around \$34 million in 2021~\cite{Baikal_web}, which is planned to expand to a similar 1~km$^3$ volume. In terms of its three phases, we take its total cost as around \$1 million in Fig.~\ref{fig:LnuCvsFCC}.

Towards the future, as a successor, IceCube-Gen2~\cite{IceCube-Gen2:2020qha} is designed to have an instrumented volume up to 8-km$^3$. The construction is expected to take eight years with a cost of about \$350 million~\cite{Gen2_TDR_web}, which is shown in the red dashed arrow in Fig.~\ref{fig:LnuCvsFCC}. P-ONE~\cite{P-ONE:2020ljt} with more than 1 km$^3$ volume, is expected to start operation and data analysis around 2026, even though the full construction is expected to be completed around 2028, with a total investment of about \texteuro100 million~\cite{PONE_EPPSU}, benefiting from the existing oceanographic infrastructure. TRIDENT~\cite{Ye:2023dch} is expected to be completed in 2030 with a 7.5 km$^3$ volume. Currently, no official cost has been made for TRIDENT, yet, which we scale from KM3NeT and P-ONE based on the volume and the density of Digital Optical Module (DOM)~\cite{Shao:2025gyx}. The NEON~\cite{Zhang:2024slv} is designed to have a similar volume, with a comparable cost, which is not shown here.  Finally, HUNT~\cite{Huang:2023mzt}, which is designed for the largest instrumented volume up to 30 km$^3$, might come later in the next decade.

In Fig.~\ref{fig:LnuCvsFCC}, we also include the (expected) cost and operational time of current and future terrestrial colliders for comparison. A possible timeline for future colliders has been investigated during the US Snowmass 2021 Community Study~\cite{Narain:2022qud} and the European Strategy~\cite{Adolphsen:2022ibf}, with a cost estimated by the Snowmass Implementation Task Force~\cite{Roser:2022sht}. Here, we include the LHC with extension to its high-luminosity (HL) phase until the 2040s, with a total cost of about \$7 billion~\cite{Evans:2645935,ZurbanoFernandez:2020cco}. The EIC is approved by the US DOE with a cost between \$1.6 and \$2.6 billion, which is expected to operate at the beginning of the 2030s~\cite{Torre:2024edq}.
If ideally, the CEPC can start operation in the middle of the 2030s, which will cost about \$5.2 billion~\cite{CEPCStudyGroup:2023quu} in total. Alternatively, the FCC-ee might become on shell in the mid-2040s, with the cost about \$12 billion~\cite{Roser:2022sht}. If the muon cooling technology~\cite{MICE:2019jkl} becomes mature in time, a 3 TeV muon collider might come out in the mid-2040s~\cite{Narain:2022qud,Adolphsen:2022ibf}, at the cost of \$7 billion, and a 10 TeV one can come after in the 2050s~\cite{Narain:2022qud} with the cost of \$12 billion~\cite{Roser:2022sht}. The 100 TeV hadron-hadron collider, such as FCC-hh, will cost above \$30 billion~\cite{Roser:2022sht}, and the potential operation time might go beyond 2070~\cite{Narain:2022qud}, which is not shown here.

As compared in Fig.~\ref{fig:LnuCvsFCC}, the currently running and proposed L$\nu$Cs have clear advantages over traditional terrestrial colliders in terms of the costs and timelines. In general, the cost of the L$\nu$Cs can be one magnitude smaller, and the operational time can come earlier by 5 to 10 years with respect to future high-energy colliders. Also note that the main drive of building the large-volume neutrino telescopes is for neutrino astrophysics and for neutrino BSM in general. Using neutrino telescopes to search for TeV-scale new physics is just a plus to the main drive, which is almost cost-free, up to manpower for data analysis without any new experimental add-ons. In addition, as to be carefully examined in Sec.~\ref{app:models}, some new physics potential of L$\nu$Cs can be comparable, or even outperform with respect to traditional colliders, highlighting the importance of the L$\nu$Cs.

\subsection{Event rates}
\label{app_EventRatesStat_ER}

For the detectability of the BSM scenarios, this work focuses on the starting and through-going muons from $\nu_\mu$CC-like interactions that produce a muon-like track, due to its large effective volume and low backgrounds compared to shower-like events.

The event rate of starting muons can be calculated by
\begin{equation}
\begin{aligned}
\frac{\dd N^{\rm st}}{\dd E_\mu} (E_\mu)
&=
T\,  V\, \rho N_A
\int_{E_\mu}^\infty \dd E_\nu
\frac{\dd \sigma}{\dd E_\mu} (E_\mu, E_\nu) \,
2\pi\int^1_{-1} \dd \cos \theta_z\, \frac{\dd \phi_\nu}{\dd E_\nu}\left(E_\nu, \, \cos \theta_z\right) e^{-\tau\left(E_\nu, \, \cos \theta_z\right)}  \,.
\label{eq_dNdEnuST}
\end{aligned}
\end{equation}
The event rate of throughgoing muons can be calculated by
\begin{equation}
\begin{aligned}
\frac{\dd N^{\mathrm{thr}}}{\dd E_\mu}\left(E_\mu\right) & =T\, \frac{N_A V^{2/3} }{\alpha+\beta E_\mu} \int_{E_\mu}^{\infty} \dd E_\nu \int_{E_\mu}^{E_\nu} \dd E_\mu^{\prime} \frac{\dd \sigma}{\dd E_\mu^{\prime}}\left(E_\mu^{\prime}, E_\nu\right)
\, 2\pi
\int^0_{-1} \dd \cos \theta_z\, \frac{\dd \phi_\nu}{\dd E_\nu}\left(E_\nu, \, \cos \theta_z\right) e^{-\tau\left(E_\nu, \, \cos \theta_z\right)}\, .
\label{eq_dNdEnuThrgo}
\end{aligned}
\end{equation}

In the above equations, $T$ is the exposure time, which we set to 20 years, as these telescopes can operate for a long time (e.g., IceCube has been operating for 17 years already), and
$\theta_z$ is the zenith angle.
$V$ is the detection volume. For 1~km$^3$ detectors, we use the fiducial volume of IceCube, which is $\simeq (0.8\, \rm km)^3 = 0.512 \, \rm km^3$~\cite{IceCube:2013low, Taboada:2019lfc}.
For 8~km$^3$ and 30~km$^3$ detectors, we estimate their fiducial volumes following the same idea as IceCube's, which gives $\simeq (2-0.2\, \rm km)^3 = 0.728\times8 \, \rm km^3$ and $\simeq (30^{1/3}-0.2\, \rm km)^3 \simeq 0.819\times30 \, \rm km^3$, respectively.
$\rho$ is the density of the detection material of the detector, which is either ice or water.
$N_A$ is the Avogadro constant.
$\alpha$ and $\beta$ are the parameters for muon energy losses while traveling in matter, and we use
$\alpha = 2.54\times10^{-3} \, \rm GeV\, cm^2\, g^{-1}$
and $\beta  = 3.93\times10^{-6} \, \rm cm{^2}\, g^{-1}$, which are the averaged values of Tables~3 and 4 in Ref.~\cite{Koehne:2013gpa}.

This continuous-energy-loss treatment is standard for inclusive through-going-muon event-rate calculations and describes the ensemble-averaged propagation range relevant for our statistical analysis. At multi-PeV energies, discrete radiative processes produce fluctuations and occasional catastrophic losses for individual muons; these effects primarily broaden the relation between the initial muon energy and its reconstructed energy in a specific detector. Their detailed treatment requires detector-specific stochastic propagation and reconstruction simulations, whereas the present calculation concerns the expected inclusive event spectrum.
Moreover, we do not include detector-specific trigger and reconstruction efficiencies or analysis cuts, which are expected to have only a small impact in the high-energy range relevant here. Since energy-dependent effective masses have not yet been published for the proposed detectors, we characterize each detector by its fiducial volume for starting events and its projected area for through-going events, using the same treatment for all configurations.

The $\dd \sigma/\dd E_\mu$ in Eqs.~(\ref{eq_dNdEnuST},~\ref{eq_dNdEnuThrgo}) is the neutrino differential cross section, and it can be approximated as
\begin{equation}\label{eq:Eavg}
\frac{\dd \sigma}{\dd E'_\mu}(E'_\mu, E_\nu) \simeq
\sigma(E_\nu) \delta\Big(E_\mu^\prime - \langle E_\mu^\prime \rangle(E_\nu)\Big),
\end{equation}
as commonly adopted in, e.g., Refs.~\cite{Laha:2013lka, Blum:2014ewa, Bustamante:2016ciw, Zhou:2019frk}.

The last term in Eqs.~(\ref{eq_dNdEnuST},~\ref{eq_dNdEnuThrgo}) accounts for neutrino absorption in the Earth through the survival probability $\exp[-\tau(E_\nu,\cos\theta_z)]$~\cite{Gandhi:1995tf,IceCube:2017roe}. The optical depth is calculated as
\begin{equation}
\tau_{\nu(\bar\nu)}(E_\nu,\cos\theta_z)=N_A\,\sigma^{\rm tot}_{\nu(\bar\nu)}(E_\nu)\int_{\rm chord}\dd \ell\,\rho_{\rm PREM}\!\left[r(\ell,\cos\theta_z)\right],
\label{eq:earth_absorption}
\end{equation}
where the mass density along the Earth chord is taken from the Preliminary Reference Earth Model (PREM)~\cite{Dziewonski:1981xy}. Neutrinos and antineutrinos are evaluated separately using their respective cross sections. For each BSM hypothesis, the corresponding BSM contribution is included in $\sigma^{\rm tot}$, so that its enhancement of the interaction rate and of Earth absorption is treated consistently. The uncertainty in the attenuation is governed mainly by the neutrino-nucleus cross-section uncertainty already included in the event-rate analysis; the smaller uncertainty associated with the Earth density profile is neglected in this forecast. For signal events, the cross sections and differential cross sections include both SM and BSM components, while for background events, they only include the SM component.

The SM component is detailed in App.~\ref{app:SM-cross-section}, while the BSM component contains both model-independent and model-dependent frameworks in App.~\ref{app:models}.

We also note that there are other event classes that can be used to search for BSM physics.
For the $\nu_\mu$CC-like events, those produced by BSM and SM would have different distributions of inelasticity ($y=1-E_\mu/E_\nu$), which can be measured by neutrino telescopes using starting events~\cite{IceCube:2018pgc, IceCubeCollaborationSS:2025zgz}.
In addition, some BSM models like leptogluon (App.~\ref{app:LG}) also produce dimuon-like signatures, which have much lower SM backgrounds~\cite{Zhou:2021xuh, Sarkar:2023dvr}.
On the other hand, the BSM scenarios we consider also produce pure-shower events. However, the SM background (including $\nu_e$CC, $\nu_\tau$CC, and NC; see, e.g., left panel of Fig.~7 in Ref.~\cite{Zhou:2019frk}) is expected to be higher than that of $\nu_\mu$CC-like events.
We leave the consideration of these event classes for future work.

\subsection{Statistics}
\label{app:statistics}

To quantify the expected sensitivity of each BSM scenario, we use the profile likelihood ratio test statistic, incorporating systematic uncertainties on both the signal and background predictions.

We divide the events into five bins in $E_\mu$ per decade.
In each energy bin $i$, we calculate the expected BSM signal ($s_{i,j}$) and SM background ($b_{i,j}$) event counts, where $j$ denotes the event type, i.e., start muons (Eq.~\eqref{eq_dNdEnuST}) and throughgoing muons (Eq.~\eqref{eq_dNdEnuThrgo}).
In addition, we introduce nuisance parameters $\theta_{i,j}^s$ and $\theta_{i,j}^b$ to account for systematic uncertainties for the signal and background, respectively. Thus, the effective event counts become
$s_{i,j}^\text{eff} = s_{i,j} \times 10^{\theta_{i,j}^s}$ and
$b_{i,j}^\text{eff} = b_{i,j} \times 10^{\theta_{i,j}^b}$.
The nuisance parameters $\theta_{i,j}^s$ and $\theta_{i,j}^b$ have Gaussian priors:
$\theta_{i,j}^s \sim \mathcal{N}(0, \log_{10} \delta_{i,j}^s)$ and
$\theta_{i,j}^b \sim \mathcal{N}(0, \log_{10} \delta_{i,j}^b)$,
where $\delta_{i,j}^s$ and $\delta_{i,j}^b$ are the fractional uncertainties of the signal and background event counts, respectively.

Given the observed event counts $n_{i,j}$ in each bin,
the full likelihood function is~\cite{Cowan:2010js, ParticleDataGroup:2024cfk}
\begin{align}
\mathcal{L}(\{n_{i,j}\} \,|\, s, b, \theta^s, \theta^b)
&=
\prod_{i,j} \mathrm{Poisson}\left(n_{i,j} \,\middle|\, s_{i,j}^\text{eff} + b_{i,j}^\text{eff}\right)
\times \mathcal{N}(\theta_{i,j}^s \,|\, 0, \log_{10} \delta_{i,j}^s) \times \mathcal{N}(\theta_{i,j}^s \,|\, 0, \log_{10} \delta_{i,j}^b).
\end{align}
To estimate the expected sensitivity to each BSM scenario, we compute the profile likelihood ratio
\begin{align}
q_0 = -2 \log \left( \frac{
\mathcal{L}(n_{i,j} = s_{i,j} + b_{i,j} \,|\, s_{i,j} = 0,\, \hat{\hat{\theta}}_{i,j}^b)
}{
\mathcal{L}(n_{i,j} = s_{i,j} + b_{i,j} \,|\, s_{i,j},\, \hat{\theta}_{i,j}^s,\, \hat{\theta}_{i,j}^b)
} \right),
\end{align}
where $\hat{\hat{\theta}}_{i,j}^b$ denotes the conditional maximum likelihood estimates under the null (background-only) hypothesis, and $(\hat{\theta}_{i,j}^s, \hat{\theta}_{i,j}^b)$ are the profiled maximum likelihood estimates under the signal-plus-background hypothesis.

The sensitivities at 95\% confidence level can be set at $\sqrt{q_0} \simeq 1.960$.
To obtain the expected significance without statistical fluctuations, we use the ``Asimov dataset''~\cite{Cowan:2010js,ParticleDataGroup:2024cfk}, in which the observed number of events is set equal to its expectation under the signal-plus-background hypothesis:
$n_{i,j} = s_{i,j} + b_{i,j}$.
This allows efficient computation of the median significance without Monte Carlo sampling.

{
The baseline forecast treats different energy bins as independent, setting all bin-to-bin correlations to zero.
To test bin-to-bin correlations directly, we repeat the analysis with a single bin-to-bin correlation coefficient $\rho$.
For $x=s,b$, the systematic covariance is taken to be

$ C^{x}_{ik}=\Delta_i^x\Delta_k^x\left[(1-\rho)\delta_{ik}+\rho\right],
 \label{eq:correlation_stress_test}
$

where $\Delta_i^x$ is the systematic uncertainty in bin $i$. We evaluate $Z^2=\boldsymbol{s}^{T}V^{-1}\boldsymbol{s}$ with $V=\operatorname{diag}(\boldsymbol{b})+C^s+C^b$.
$\rho=0$ means no bin-to-bin correlations and reproduces the baseline result; $\rho=0.99$ represents nearly complete positive bin-to-bin correlations. The latter is an intentionally extreme test, not a detector-specific model.
These tests use the 8~km$^3$, 20-year configuration and the contact-interaction and leptogluon benchmarks, which represent smooth and resonance-induced spectral distortions, respectively.

The contact-interaction limits remain within the same scale range under these variations.
For the leptogluon scenario, bin-to-bin correlations alone lower the mass reach by about 10\% for both the KM3NeT flux and the conservative flux.
Thus, neglecting bin-to-bin correlations can overestimate the reach, but not significant.
Because bin-to-bin correlations and detector responses depend on the source model and detector, the quoted sensitivities are forecast-level estimates. A precision projection requires collaboration-provided bin-to-bin correlation and detector response matrices.
}

\section{New physics models}
\label{app:models}

In this section, we explain the details of some representative new physics models that can be probed at L$\nu$Cs. We also note that L$\nu$Cs can probe not only BSM physics but also SM physics, such as quantum chromodynamics~\cite{Amoroso:2022eow}, although we will not cover the latter case here. We start with the contact four-fermion interactions among the leptons and quarks in the SMEFT framework in App.~\ref{app:4f}. Afterwards, we build a vector-like leptogluon model in App.~\ref{app:LG}, where the neutrinogluon component can be directly produced in the neutrino-nucleon scattering. The leptoquark models, including both scalar and vector types, are explored in App.~\ref{app:LQ}. Finally, we include the heavy vector boson $W'$ and charged scalar $H^\pm$ models to represent the neutrino-electron collision at the L$\nu$C in App.~\ref{app:Wprime}, and conclude then.

\subsection{Contact interactions}
\label{app:4f}

For the neutrino-nucleon scattering at L$\nu$Cs, the dominant parton-level processes come from the neutrino scattering with quark and gluon partons inside the nucleon. Limiting ourselves to the CC-like events, the relevant four-fermion contact operators are
\begin{equation}\label{eq:op}
	\begin{aligned}
		&O_{\ell q}^{(3)}=(\bar{\ell}\bar{\sigma}_\mu\sigma^i\ell)(\bar{q}\bar{\sigma}^\mu\sigma^i q),~&
		&O_{\ell edq}=(\bar{\ell}\bar{e}^c)(d^cq),    \\
		&O_{\ell equ}^{(3)}=(\bar{\ell}\bar{\sigma}_{\mu\nu}\bar{e}^c)\epsilon(\bar{q}\bar{\sigma}^{\mu\nu}\bar{u}^c),~&
		&O_{\ell equ}^{(1)}=(\bar{\ell}\bar{e}^c)\epsilon(\bar{q}\bar{u}),
	\end{aligned}
\end{equation}
with the $\epsilon$ tensor to contract $SU(2)_W$ gauge indices. The corresponding Wilson coefficients are labeled correspondingly, e.g., $C_{\ell q}^{(3)}$.
Here, $O_{\ell q}^{(3)}$ conserves the chirality, which can be induced by a heavy $W'$ boson.
Other operators violate chirality and match with heavy-scalar models.
As we only consider experimental signal from the charged muon (second-generation lepton), i.e., $\nu_\mu$CC-like events, in this work, and the nucleon is mainly composed of first-generation quarks, we will focus on the lepton and quark flavor index 2211 of the operators in Eq.~(\ref{eq:op}).

\begin{figure}
\centering
\includegraphics[width=0.32\textwidth]{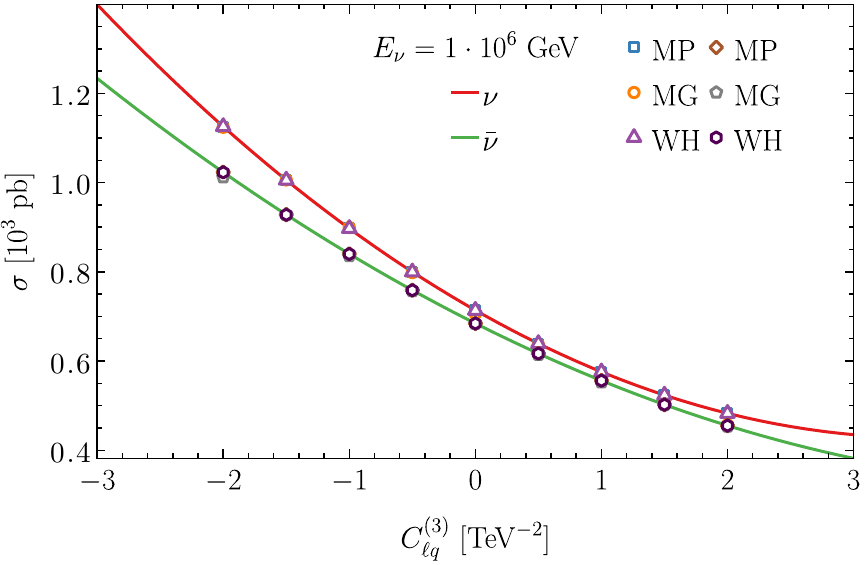}
\includegraphics[width=0.32\textwidth]{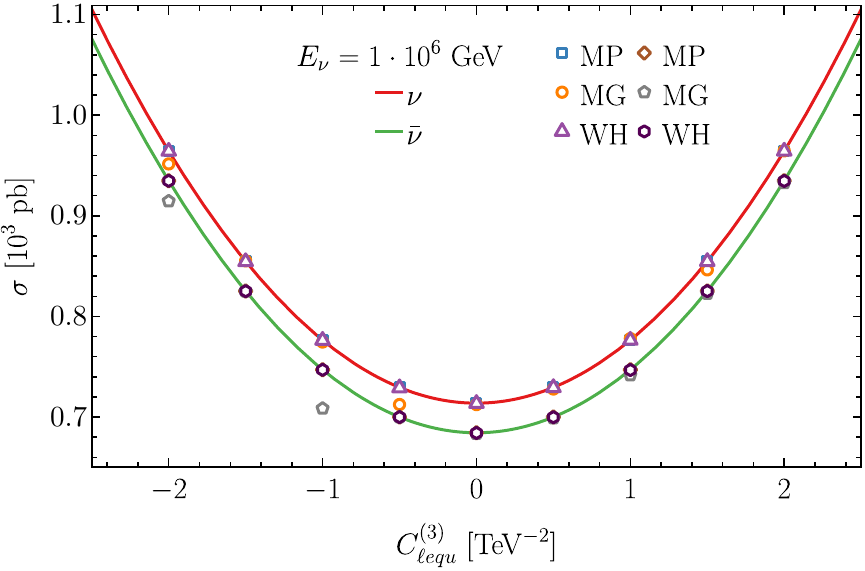}
\includegraphics[width=0.32\textwidth]{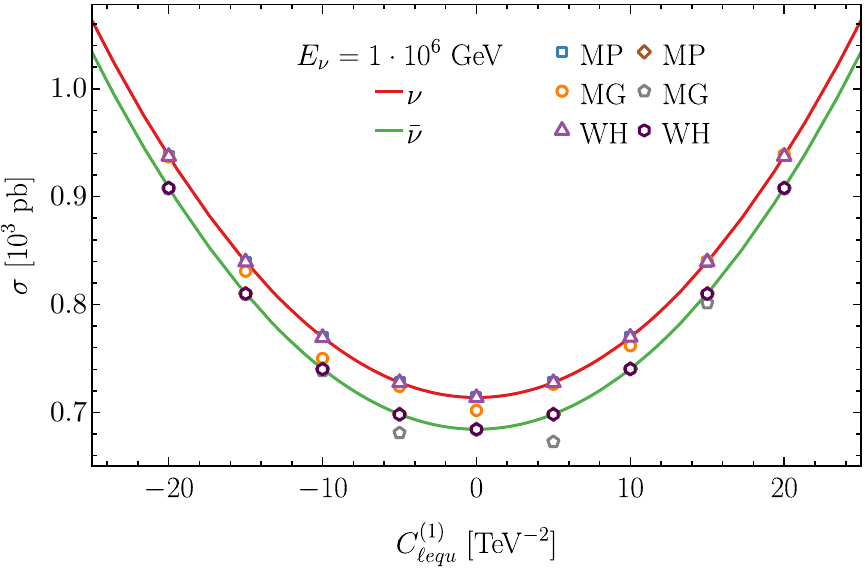}
\includegraphics[width=0.32\textwidth]{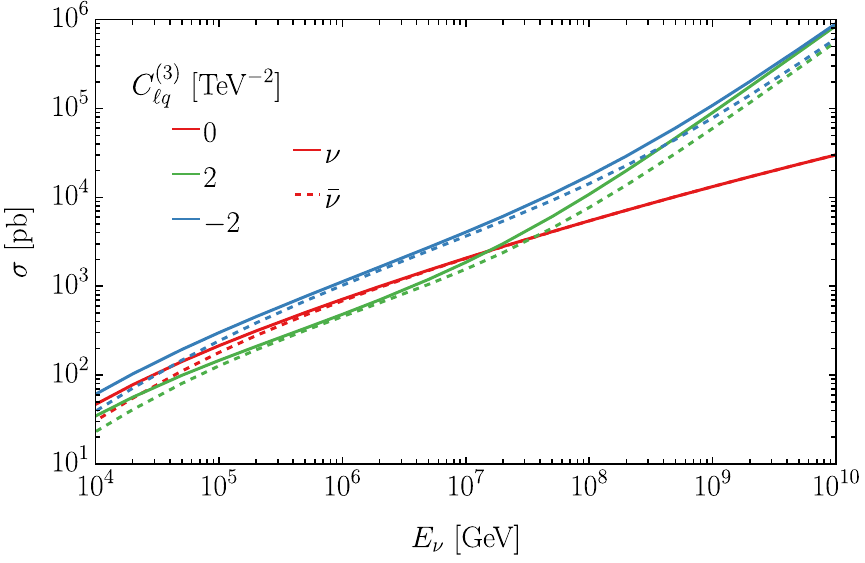}
\includegraphics[width=0.32\textwidth]{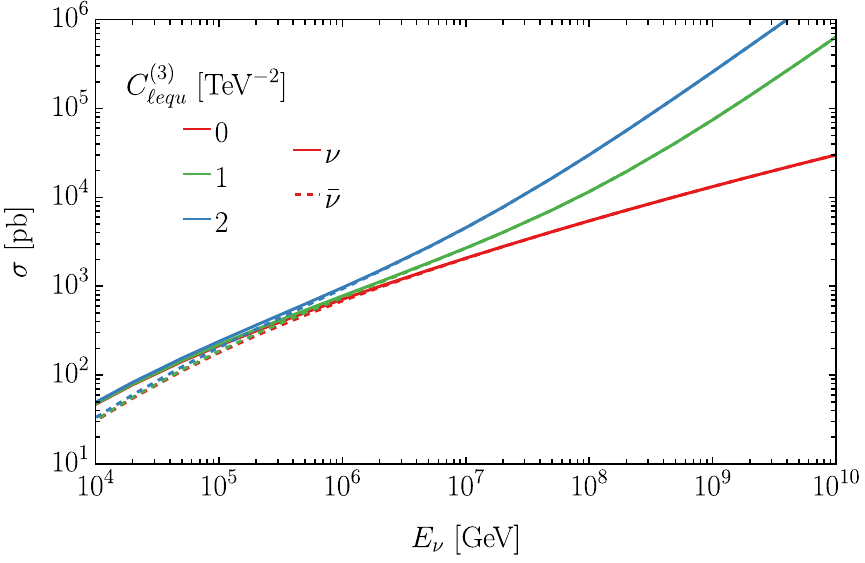}
\includegraphics[width=0.32\textwidth]{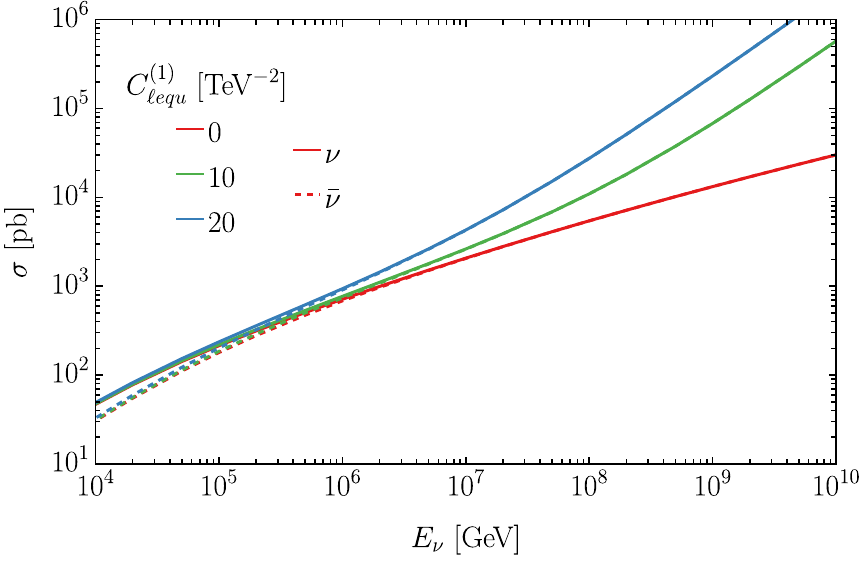}
\caption{Upper: The dependence of neutrino-nucleon scattering cross sections on the four-fermion contact operators with a fixed neutrino incident energy $E_\nu=1\cdot10^{6}~\GeV$. Here, the dots with labels are calculated with \texttt{FeynCalc}~\cite{Shtabovenko:2020gxv}/\texttt{FormCalc}~\cite{Hahn:2016ebn} interfacing \texttt{ManeParse} (MP)~\cite{Clark:2016jgm}, \texttt{MadGraph} (MG)~\cite{Alwall:2014hca}, and \texttt{Whizard} (WH)~\cite{Kilian:2007gr}, respectively. The lines are interpolated or extrapolated with Eq.~(\ref{eq:WCxs}). Lower: The cross-section dependence on the neutrino energy, by fixing SMEFT operators with a few representative coefficient values. The growing features of the lower panels are due to the violation of the EFT, but in our actual limits we did ensure that the parton-level COM energy is smaller than around the cutoff scale or $\Lambda\gtrsim\sqrt{\hat{s}}/3(4)$ (see Fig.~\ref{fig:XSRatio}).}
\label{fig:4f}
\end{figure}

For the neutrino-nucleon scattering, the 4-fermion interaction can potentially interfere with the SM CC scattering mediated by the $W$ boson. The parton-level scattering amplitude squared for $\nu d\to \ell^- u$ is
\begin{equation}\label{eq:neu}
|\mathcal{M}_{\nu d\to\ell^-u}|^2=\left(\frac{g_W^2}{2(\hat{t}-M_W^2)}+2C^{(3)}_{\ell q}\right)^2\hat{s}^2+
\left[\left(\frac{1}{2}C^{(1)}_{\ell equ}+2C^{(3)}_{\ell equ}(1+2\hat{s}/\hat{t})\right)^2+
\frac{1}{4}C_{\ell edq}^2\right]\hat{t}^2.
\end{equation}
Here, the $g_W$ is the SM weak gauge coupling, and $\hat{s},\hat{t},\hat{u}$ are the partonic Mandelstam variables. Similarly, the antineutrino scattering is
\begin{equation}\label{eq:ant}
|\mathcal{M}_{\bar{\nu}u\to\ell^+d}|^2=\left(\frac{g_W^2}{2(\hat{t}-M_W^2)}+2C^{(3)}_{\ell q}\right)^2\hat{u}^2+
\left[\left(\frac{1}{2}C^{(1)}_{\ell equ}+2C^{(3)}_{\ell equ}(1+2\hat{u}/\hat{t})\right)^2+
\frac{1}{4}C_{\ell edq}^2\right]\hat{t}^2,
\end{equation}
which shares a cross symmetry with Eq.~(\ref{eq:neu}) by a swap $\hat{s}\leftrightarrow\hat{u}$. For the anti-quark scattering, we have the relation that
\begin{equation}
\mathcal{M}_{\nu\bar{u}\to\ell^-\bar{d}}=\mathcal{M}_{\bar{\nu}u\to\ell^+d},~ \qquad ~
\mathcal{M}_{\bar{\nu}\bar{d}\to\ell^+\bar{u}}=\mathcal{M}_{\nu d\to\ell^-u}.
\end{equation}
We see that only the $\calO_{\ell q}^{(3)}$ operator interferes with the SM $W$-boson diagram, while other operators have no interference, as a consequence of the chirality-conserving or violating features. Also note that $\hat{t} < 0$, so there is destructive interference between the $C^{(3)}_{\ell q}$ contribution and the SM one. By integrating out the kinematical angle and the corresponding PDFs, the dependence of the neutrino-nucleon scattering cross sections on the Wilson coefficients is demonstrated in Fig.~\ref{fig:4f} (upper), with a representative neutrino incident energy $E_\nu=10^6~\GeV$.
The numerical evaluations are performed with \texttt{FeynCalc}~\cite{Shtabovenko:2020gxv}/\texttt{FormCalc}~\cite{Hahn:2016ebn} with PDFs interpolated with \texttt{ManeParse} (MP)~\cite{Clark:2016jgm}, and independently with general-purpose event generators, \texttt{MadGraph} (MG)~\cite{Alwall:2014hca} and \texttt{Whizard} (WH)~\cite{Kilian:2007gr}. The overall agreement is quite good among these three methods, except for a few random MG points, which suffer from a larger Monte-Carlo uncertainty.
The SMEFT operators are added one by one to examine the corresponding impact. We have left out the $\calO_{\ell edq}$ case, as the cross section is identical to the $\calO^{(1)}_{\ell equ}$, which can be inferred from Eqs.~(\ref{eq:neu}) and (\ref{eq:ant}).

For a general Wilson coefficient, the cross-section dependence can be parameterized as
\begin{equation}\label{eq:WCxs}
\sigma(C)=\sigma^{(0)}+\sigma^{(1)}C+\sigma^{(2)}C^2,
\end{equation}
where $\sigma^{(0)}$ is the SM cross section. The coefficients can be determined as
\begin{equation}
\begin{aligned}
\sigma^{(0)}=\sigma(0),\quad
\sigma^{(1)}=\frac{\sigma(1)-\sigma(-1)}{2},\quad
\sigma^{(2)}=\frac{\sigma(1)-2\sigma(0)+\sigma(-1)}{2},
\end{aligned}
\end{equation}
where $\sigma(\pm1)$ corresponds to the benchmark Wilson coefficient choice $\pm C_0$. In Fig.~\ref{fig:4f} (upper), we compare the interpolated and extrapolated results with the corresponding numerical evaluations for a few representative Wilson coefficients, which show quite good agreement. Moreover, the linear term $\sigma^{(1)}C$ only shows up for the $\calO_{\ell q}^{(3)}$ scenario, while other scenarios start from the quadratic term $\sigma^{(2)}C^2$, due to the lack of interference.

With the MP method, we present the cross-section dependence on the neutrino energy in the lower panel of Fig.~\ref{fig:4f}, where we fix the SMEFT operator coefficients to a few representative values. We observe only in the $\calO^{(3)}_{\ell q}$ case that the low-$E_\nu$ cross sections undergo a sizable modification due to interference with the SM $W$ mediation. As the neutrino energy increases, the quadratic terms gradually take over and enhance the neutrino cross section with respect to the SM ones.
Among these contact operators, the relative impact of $\calO_{\ell q}^{(3)}$ in the entire $E_\nu$ region is relatively more significant than the rest. Thus, we will mainly focus on this operator below for the BSM sensitivity exploration. For this dimension-six operator, the asymptotic partonic cross sections in Eqs.~(\ref{eq:neu}-\ref{eq:ant}) grow linear in $\hat{s}$ and are
\begin{equation}
\hat{\sigma}(\nu d\to\ell^-u)\sim\frac{|C_{\ell q}^{(3)}|^2}{2\pi}\hat{s},\quad
\hat{\sigma}(\bar{\nu}u\to\ell^+d)\sim\frac{|C_{\ell q}^{(3)}|^2}{6\pi}\hat{s}.
\end{equation}
With the perturbative unitarity bound $\hat{\sigma}\leq4\pi/\hat{s}$~\cite{Han:2021lnp}, the parton-level COM energy has an upper bound of  $\sqrt{\hat{s}}/3(4) \lesssim \Lambda\sim1/\sqrt{C_{\ell q}^{(3)}}$.

\begin{figure}
    \centering
    \includegraphics[width=0.32\linewidth]{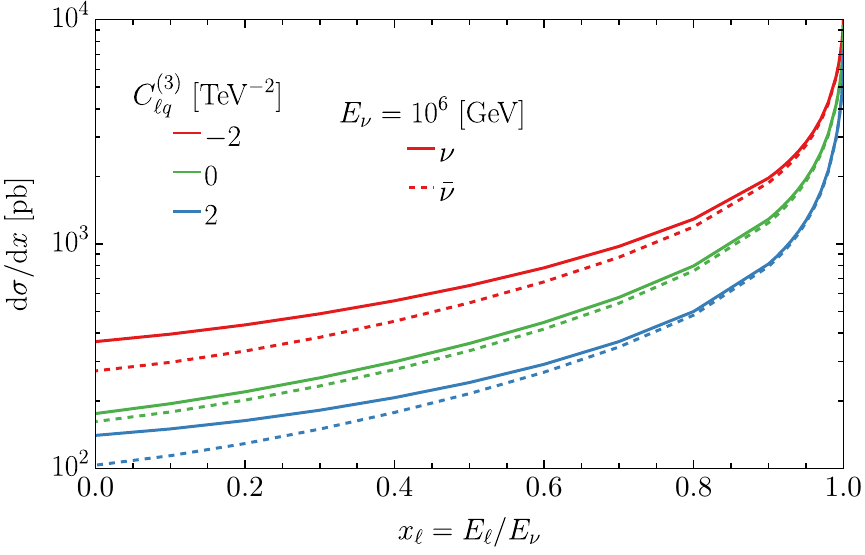}
    \includegraphics[width=0.32\linewidth]{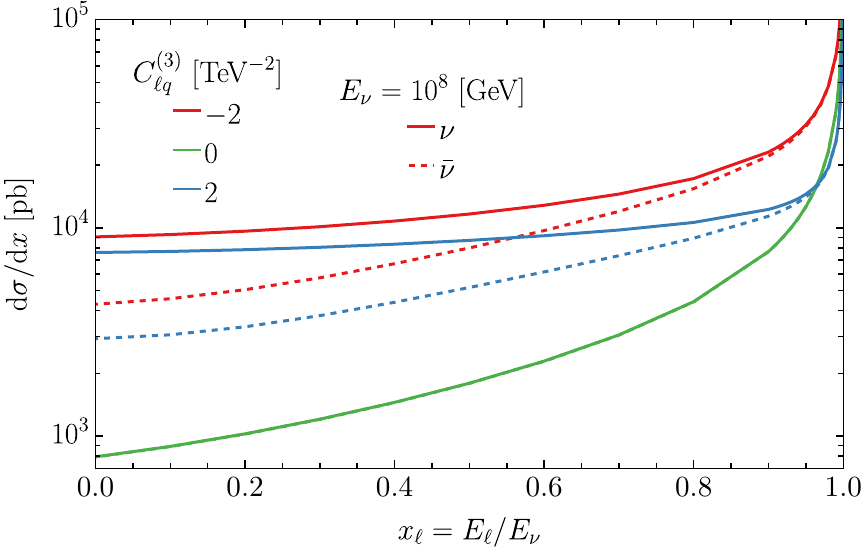}
    \includegraphics[width=0.32\textwidth]{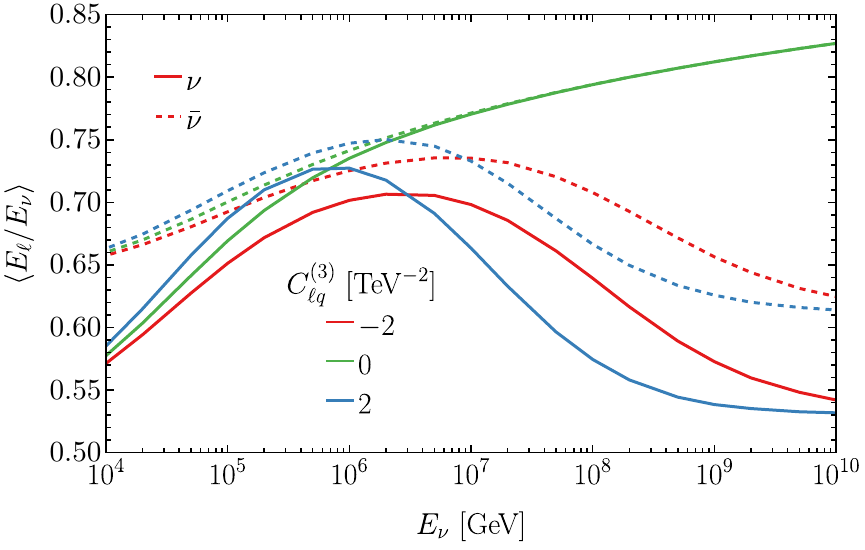}
    \caption{Left and Middle: the energy fraction $x_\ell=E_\ell/E_\nu$ distribution of the final-state charged lepton in the SM+$\calO_{\ell q}^{(3)}$ with $E_\nu=10^{6}~\GeV$ and $10^{8}~\GeV$. Right: the corresponding dependence of the average energy fraction $\langle E_\ell/E_\nu\rangle$ on the incident neutrino energy.}
    \label{fig:distClq3}
\end{figure}

In Fig.~\ref{fig:distClq3} (left and middle), we present the energy fraction $x_\ell=E_\ell/E_\nu$ distribution of the final-state charged lepton in the SM and together with the $\calO_{\ell q}^{(3)}$ operator, with $E_\nu=10^{6}~\GeV$ and $10^{8}~\GeV$ to represent HE and UHE scenarios, respectively. In the HE case, the $x_\ell$ distribution largely remains similar to the SM case, while the overall rate gets modified due to constructive or destructive interference. In contrast, the UHE $x_\ell$ distribution is noticeably different from the SM one, as the quadratic $\sigma^{(2)}C^2$ term in Eq.~(\ref{eq:WCxs}) begins to take over (for our projected sensitivity, we will check to ensure that perturbative unitarity holds).
In Fig.~\ref{fig:distClq3} (right), we present the dependence of the averaged energy fraction $\langle x_\ell\rangle$ on the incident neutrino energy $E_\nu$, which explicitly shows the transition from the SM baseline to the $\calO_{\ell q}^{(3)}$ operator dominant regime. In the $E_\nu\to\infty$ limit, the average energy fraction approaches constants  $\langle x_\ell\rangle\simeq0.53$ and 0.61, for neutrino and anti-neutrino, respectively, independent of the Wilson coefficients.

\begin{figure}[htb!]
\centering
\includegraphics[width=0.32\linewidth]{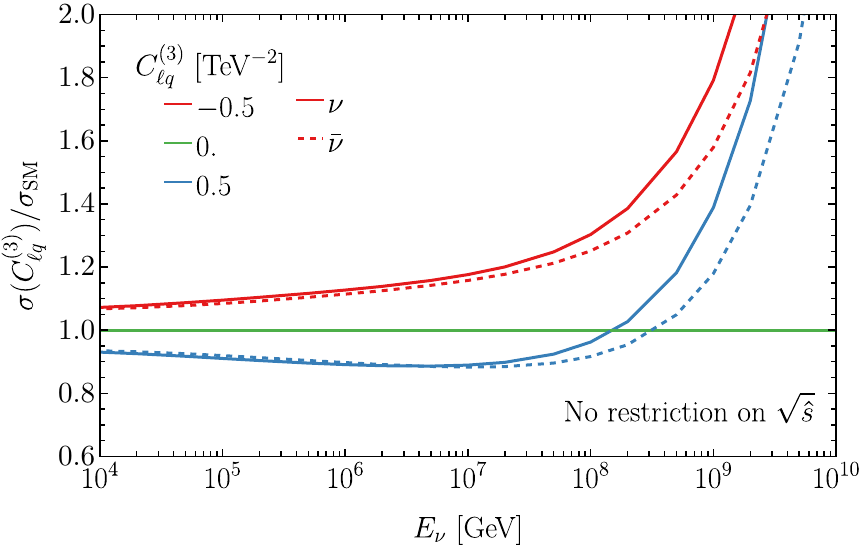}
\includegraphics[width=0.32\textwidth]{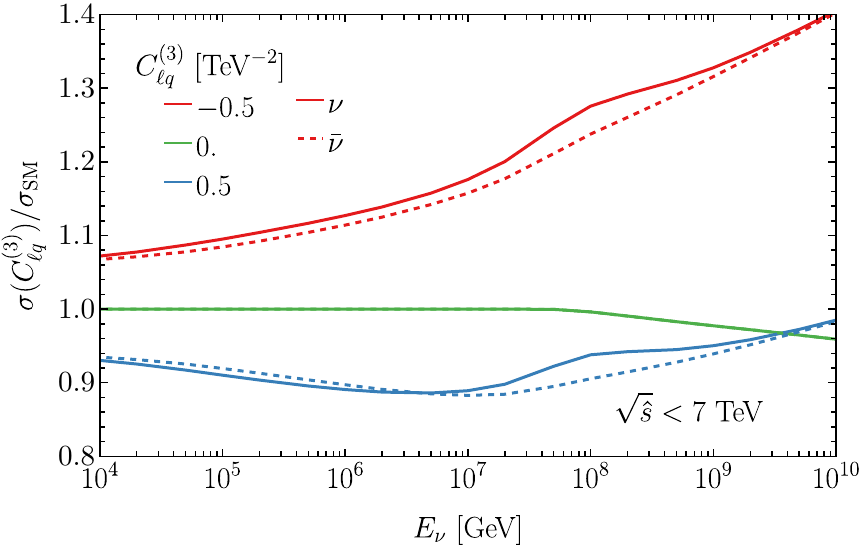}
\includegraphics[width=0.32\textwidth]{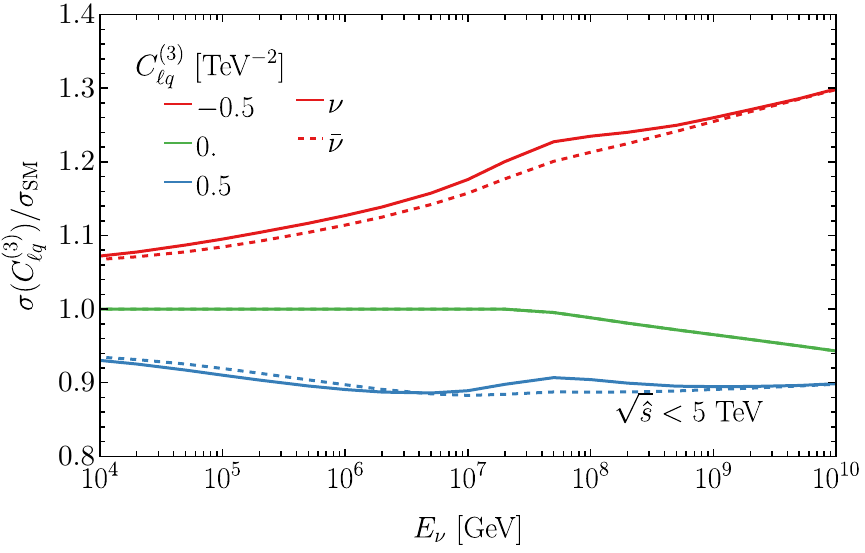}
\caption{The cross-section ratios of the SM+$\calO_{\ell q}^{(3)}$ contact operators with respect to the SM, without (left) and with the restriction of $\sqrt{\hat{s}}<7~\TeV$ (middle) and $\sqrt{\hat{s}}<5~\TeV$ (right).}
\label{fig:XSRatio}
\end{figure}

To check the effects of the perturbative bounds of $\sqrt{\hat{s}}/3(4) \lesssim \Lambda$, we show the cross-sections ratio to the SM one in Fig.~\ref{fig:XSRatio} for without the restriction in the left panel, with $\sqrt{\hat{s}} < 7$~TeV in the middle panel and $\sqrt{\hat{s}} < 5$~TeV in the right panel. The last choice corresponds to $\sqrt{\hat{s}}/3(4) < \Lambda= 1/\sqrt{C_{\ell q}^{(3)}}$ with $C_{\ell q}^{(3)} = 0.4\sim0.6~\TeV^{-2}$.
One can see that the requirement of perturbative unitarity makes the cross section grow reasonably slower.

\begin{figure}[t]
\includegraphics[width=0.48\columnwidth]{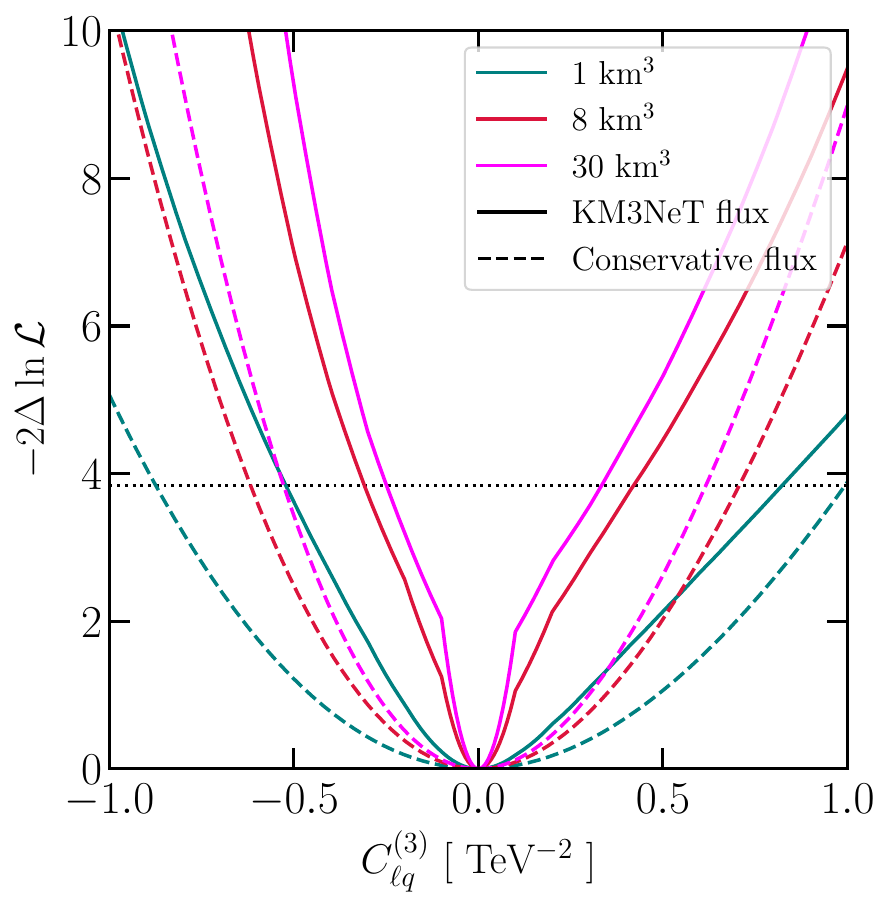}
\caption{
Profile likelihoods for the Wilson coefficient of the contact operator $\calO_{\ell q}^{(3)}$ at different L$\nu$Cs, with the restriction $\sqrt{\hat{s}}<5$~TeV.
}
\label{Fig_SMEFT_sensitivity}
\end{figure}

{
By substituting the average energy fraction $\langle E_\ell/E_\nu\rangle$ in Fig.~\ref{fig:distClq3} (where $\ell=\mu$) into Eq.~(\ref{eq:Eavg}), we estimate the event rate of starting and thoroughgoing muons with Eqs.~(\ref{eq_dNdEnuST}) and (\ref{eq_dNdEnuThrgo}). Figure~\ref{Fig_SMEFT_sensitivity} presents the profile likelihood for $\calO_{\ell q}^{(3)}$ using $\nu_\mu$ CC-like events, detector volumes of 1, 8, and 30~km$^3$, 20 years of exposure, and both UHE-flux scenarios.
The corresponding 95\% CL constraints are listed in Table~\ref{tab:Clq3_bounds}. The conservative flux weakens the constraints, while the asymmetric boundaries in both cases reflect constructive and destructive interference. In the main text, we take the larger absolute value of the two-sided limits.
}

\begin{table}[hb!]
\centering
\renewcommand{\arraystretch}{1.5}
\begin{tabular}{c c @{\hspace{0.8cm}} c}
\hline
\text{Volume [km$^{3}$]} & \text{KM3NeT flux} & \text{Conservative flux} \\
\hline
1   & $-0.52<C_{\ell q}^{(3)}<+0.82$ & $-0.87<C_{\ell q}^{(3)}<+0.99$ \\
8   & $-0.31<C_{\ell q}^{(3)}<+0.42$ & $-0.62<C_{\ell q}^{(3)}<+0.71$ \\
30  & $-0.25<C_{\ell q}^{(3)}<+0.33$ & $-0.53<C_{\ell q}^{(3)}<+0.62$ \\
\hline
\end{tabular}
\caption{Constraints on $C_{\ell q}^{(3)}$ in TeV$^{-2}$ for different L$\nu$C detector volumes and UHE-flux scenarios.}
\label{tab:Clq3_bounds}
\end{table}

Besides the L$\nu$Cs discussed here, many existing experiments can constrain four-fermion contact interactions. Global analyses from collider measurements~\cite{deBlas:2022ofj} and low-energy flavor observables~\cite{Falkowski:2017pss} give bounds
\begin{equation}
    |C_{\ell q}^{(3)}|<0.59/0.46/0.02~\TeV^{-2},
\end{equation}
based on existing colliders/HL-LHC/Low-Energy observables, respectively.
Here, the first number comes from a global SMEFT fit~\cite{deBlas:2022ofj} of the collider observables from LEP, SLC, SLD, Tevatron, and LHC, while the second one is the HL-LHC projection with 14 TeV energy and 3 ab$^{-1}$ integrated luminosity~\cite{Cepeda:2019klc}. The third number is from the low-energy flavor observables~\cite{Falkowski:2017pss}, especially the (semi)leptonic hadron decays $d_j\to u_i\ell\bar{\nu}_\ell$~\cite{Gonzalez-Alonso:2016etj}. We have converted the Wilson coefficient unit $1/v^2$~\cite{Falkowski:2017pss,deBlas:2022ofj} to TeV$^{-2}$ for the convenience of calculation in this work, where $v=246~\GeV$ is the SM vacuum expectation value. We see that the L$\nu$Cs can impose a comparable or even stronger constraint on operator $\calO_{\ell q}^{(3)}$ than the collider experiments, even though the low-energy flavor observable remains to be the strongest one.
Finally, with a conversion $\Lambda=1/\sqrt{C_{\ell q}^{(3)}}$, we translate the bounds on the contact operator $\calO_{\ell q}^{(3)}$ to be the corresponding new-physics scale, and summarize in Fig.~1 of the main text.

If a resonance exists within the L$\nu$C energy reach, the SMEFT contact operator description will fail. A better strategy is to search for the resonance directly. In the rest of this work, we will discuss the potential cases one by one, including a vector-like leptogluon in Sec.~\ref{app:LG}, scalar and vector leptoquarks in Sec.~\ref{app:LQ}, and the resonance of vector boson $W'$ and charged scalar $H^\pm$ in Sec.~\ref{app:Wprime} and conclude there.

\subsection{Leptogluon}
\label{app:LG}
In many BSM unification scenarios, quarks and leptons can be composite particles, \emph{e.g.}, the ``Rishon model"~\cite{Harari:1981uh} or some schematic constructions~\cite{Harari:1979gi,Shupe:1979fv}. More recent studies can be found in Refs.~\cite{Dobrescu:2021fny,Assi:2025rjx}.
As the constituents of leptons can also be charged under QCD, one may anticipate the excited states of leptons to carry QCD charges, including the adjoint representation~\cite{Fritzsch:1981zh,Harari:1982xy}. A pioneer phenomenological study can be found in Ref.~\cite{King:1984qk}, followed by Refs.~\cite{Goncalves-Netto:2013nla,Mandal:2016csb,Almeida:2022udp,Han:2025wdy}. Some existing searches at HERA can be found in Ref.~\cite{H1:1993vsn}.

\subsubsection{A renormalizable vector-like leptogluon model}
Leaving aside the underlying model, we introduce vector-like fermions to guide our discussion for color-octet leptogluons (LGs) at the L$\nu$C. Under the SM $SU(3)_C\times SU(2)_W \times U(1)_Y$ gauge group, we can have doublet and singlet for LGs as
\begin{equation}
	(8, 2)_{-1/2}:~L_8=\begin{pmatrix}
		\nu_8 \\ \lambda_8
	\end{pmatrix},~\textrm{and}~
	(8, 1)_{-1}: E_8,
\end{equation}
respectively.
In order to avoid a gauge anomaly, we need to construct them as vector-like fermions. The corresponding mass term and Yukawa interaction can be written as
\begin{eqnarray}
	-\mathcal{L} \supset M_{L_8}\bar{L}_{8L}L_{8R} + M_{E_8}\bar{E}_{8L}E_{8R} + y_8\bar{L}_{8L}\Phi E_{8R}+\textrm{h.c.},
\end{eqnarray}
where $\Phi$ is the SM doublet Higgs with the vacuum expectation value (VEV) as $\langle\Phi^T\rangle=(0,v)/\sqrt{2}$, and
$v=246~\GeV$.

The mass eigenstate of the neutral LG (neutrinogluon, NG) is the same as the gauge eigenstate with a mass
\begin{eqnarray}
	M_{\nu_8} =  M_{L_8},
\end{eqnarray}
while the charged states $(\lambda_8,E_8)$ have a mass matrix of
\begin{equation}
	\mathcal{M} = \begin{pmatrix}
		M_{L_8} & m_L \\
		0 &   M_{E_8}
	\end{pmatrix},~\textrm{where}~
	m_L=\frac{y_8v}{\sqrt{2}}.
\end{equation}
We can diagonalize this mass matrix as
\begin{equation}
	U_L\mathcal{M}U_R^\dagger=\begin{pmatrix}
		M_{\ell_8} & 0\\
		0 & M_{e_8}
	\end{pmatrix},
\end{equation}
where the eigenvalues are
\begin{equation}
	\begin{aligned}
		M^2_{\ell_8,e_8}=\frac{1}{2}\left(M^2_{L_8}+M^2_{E_8}+m_L^2\pm\Delta\right),~
		\Delta=M^2_{\ell_8}-M^2_{e_8}=\sqrt{(M^2_{L_8}+M^2_{E_8}+m^2_L)^2-4M^2_{L_8}M^2_{E_8}}.
	\end{aligned}
\end{equation}
Here, we choose the convention for the mass hierarchy as
\begin{equation}\label{eq:hierachy}
	M_{\ell_8}>M_{\nu_8}=M_{L_8}>M_{E_8}>M_{e_8}.
\end{equation}
In the small $m_L$ limit, i.e., $m_L\ll M_{L_8,E_8}$, we have the approximation as
\begin{equation}
	M_{\ell_8}^2\simeq M_{L_8}^2+\frac{M^2_{L_8}m_L^2}{M^2_{L_8}-M^2_{E_8}}, \quad
	M_{e_8}^2\simeq M_{E_8}^2-\frac{M^2_{E_8}m_L^2}{M^2_{L_8}-M^2_{E_8}}.~
\end{equation}
The rotational matrices are defined as
\begin{equation}
	U_L=\begin{pmatrix}
		\cos\theta_L & \sin\theta_L\\
		-\sin\theta_L & \cos\theta_L
	\end{pmatrix},\quad
	U_R=\begin{pmatrix}
		\cos\theta_R & \sin\theta_R\\
		-\sin\theta_R & \cos\theta_R
	\end{pmatrix},
\end{equation}
with mixing angles
\begin{equation}
	\sin(2\theta_L)=\frac{2M_{E_8}m_L}{\Delta},\quad
	\sin(2\theta_R)=\frac{2M_{L_8}m_L}{\Delta}.
\end{equation}
Consequently, we have the mass eigenstates as
\begin{equation}
	\begin{pmatrix}
		\ell_{8L} \\
		e_{8L}
	\end{pmatrix}=\begin{pmatrix}
		\cos\theta_L & -\sin\theta_L\\
		\sin\theta_L & \cos\theta_L
	\end{pmatrix}\begin{pmatrix}
		\lambda_{8L}\\ E_{8L}
	\end{pmatrix},\quad
	\begin{pmatrix}
		\ell_{8R} \\
		e_{8R}
	\end{pmatrix}=\begin{pmatrix}
		\cos\theta_R & -\sin\theta_R\\
		\sin\theta_R & \cos\theta_R
	\end{pmatrix}\begin{pmatrix}
		\lambda_{8R}\\ E_{8R}
	\end{pmatrix}.
\end{equation}

The charge-current interactions in terms of mass eigenstates are
\begin{eqnarray}
	\label{eq:W-coupling}
	-\mathcal{L} \supset \frac{ig_W}{\sqrt{2}}\,W^+_\mu\left(-\sin\theta_{L,R}\, \bar{\nu}_{8L,R}\gamma^\mu e_{8L,R} + \cos\theta_{L,R}\, \bar{\nu}_{8L,R}\gamma^\mu\ell_{8L,R}\right)
	+\textrm{h.c.}
\end{eqnarray}
Other than renormalizable operators from gauge interactions, the underlying composite dynamics could also generate a dimension-5 chromo-dipole operator
\begin{equation}
	\label{eq:dim-5-operator}
	-\mathcal{L} \supset \frac{g_s}{2\Lambda}	\left(a_L\bar{\ell}_L\sigma^{\mu\nu}G^a_{\mu\nu}L^a_{8R}+a_R\bar{e}_R\sigma^{\mu\nu}G^a_{\mu\nu}E_{8L}\right)+\textrm{h.c.},
\end{equation}
where $\sigma^{\mu\nu}=i(\gamma^\mu\gamma^\nu-\gamma^\nu\gamma^\mu)/2$. Similar to previous studies~\cite{Goncalves-Netto:2013nla,Mandal:2016csb,Almeida:2022udp,Han:2025wdy}, we choose $a_R=0$ with only focusing on $a_L=a$ throughout this work. To avoid a charged lepton flavor violation (CFLV)~\cite{ParticleDataGroup:2024cfk}, we need to introduce three flavors of LGs, aligning with the SM lepton flavors. However, only the muon flavor is relevant to this work, due to our choice of using CC-like muon events for the sensitivity study. In the following, we use $e_8$ with mass $M_{e_8}$ to represent all three-generation charged leptogluons.

In terms of the mass hierarchy in Eq.~(\ref{eq:hierachy}), we have two decay channels for the NG as
\begin{equation}
	\nu_8 \rightarrow \nu + g~\textrm{[from Eq.~\eqref{eq:dim-5-operator}]  and  }~ \nu_8 \rightarrow e_8 + W^+~\textrm{[from Eq.~\eqref{eq:W-coupling}]},
\end{equation}
respectively.
The LG interaction in Eq.~(\ref{eq:dim-5-operator}) gives the partial width as
\begin{eqnarray}
	\Gamma(\nu_8\to\nu g)=\left(\frac{ag_s}{2\Lambda}\right)^2\frac{M_{\nu_8}^3}{4\pi}.
\end{eqnarray}
In comparison, the gauge interactions in Eq.~(\ref{eq:W-coupling}) will induce an on-shell or off-shell decay.
In this study, we will focus on the on-shell $W$ case by requiring $M_{\nu_8} - M_{e_8} > M_W$, with the partial width as
\begin{equation}
	\Gamma(\nu_8 \rightarrow e_8 W) = \kappa^2\frac{\alpha}{16\,s_W^2}\, \frac{(M^2_{\nu_8} - M^2_{e_8})^2 + M^2_W(M^2_{\nu_8} + M^2_{e_8} - 2 M_W^2)}{M_W^2\, M_{\nu_8}}\, \beta ~,
\end{equation}
where $s_W = \sin{\theta_W}$ is the sine of the weak mixing angle, and the phase space factor is given by
\begin{eqnarray}
	\beta = \frac{1}{M^2_{\nu_8}}\lambda^{1/2}(M^2_{\nu_8}, M^2_{e_8}, M^2_W) ~,
	\label{eq:beta}
\end{eqnarray}
with the K\"allen function $\lambda(a, b, c) = a^2 + b^2 + c^2 - 2 (ab + bc + ca)$. The $\kappa$ coefficient is
\begin{equation}\label{eq:kappa}
	\begin{aligned}
		\kappa^2
		=\sin^2\theta_L+\sin^2\theta_R-\sin\theta_L\sin\theta_R
		\frac{12M_{\nu_8}M_{e_8}M_W^2}{(M_{\nu_8}^2-M_{e_8}^2)^2+(M_{\nu_8}^2+M_{e_8}^2)M_W^2-2M_W^4}
		\simeq\sin^2\theta_L+\sin^2\theta_R,
	\end{aligned}
\end{equation}
where the approximation applies when $M_{\nu_8}^2-M_{e_8}^2\gg M_W^2$.
In such a way, the parameter scales as $\kappa\sim y_8v/M_{\nu_8}$, which ensures the width scaling $\Gamma_{\nu_8}\sim M_{\nu_8}$ without a unitariy violation, as discussed in Ref.~\cite{Han:2003wu}.

\begin{figure}
	\centering
	\includegraphics[width=0.49\textwidth]{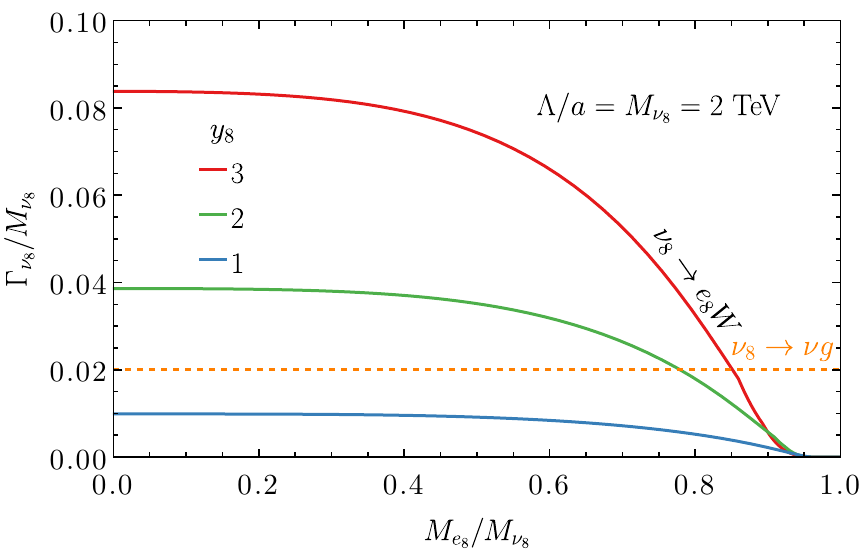}
	\includegraphics[width=0.49\textwidth]{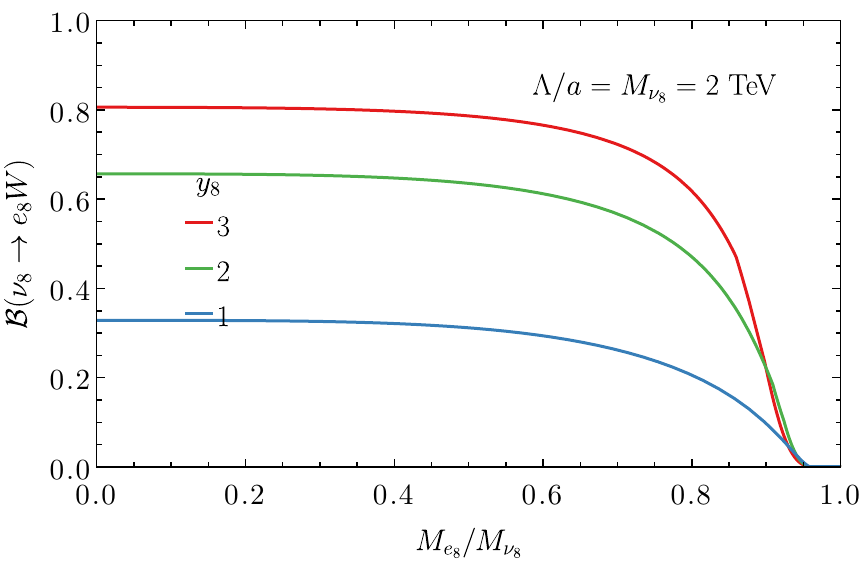}
	\includegraphics[width=0.49\textwidth]{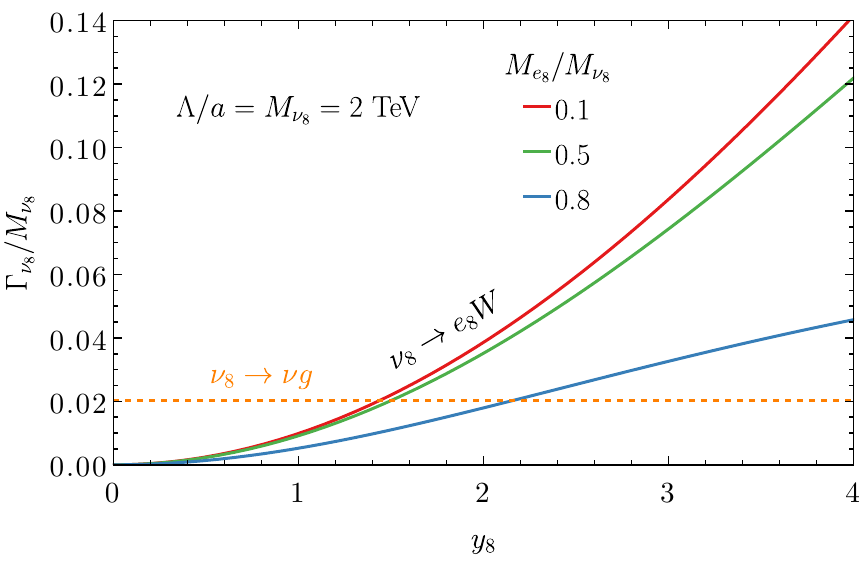}
	\includegraphics[width=0.49\textwidth]{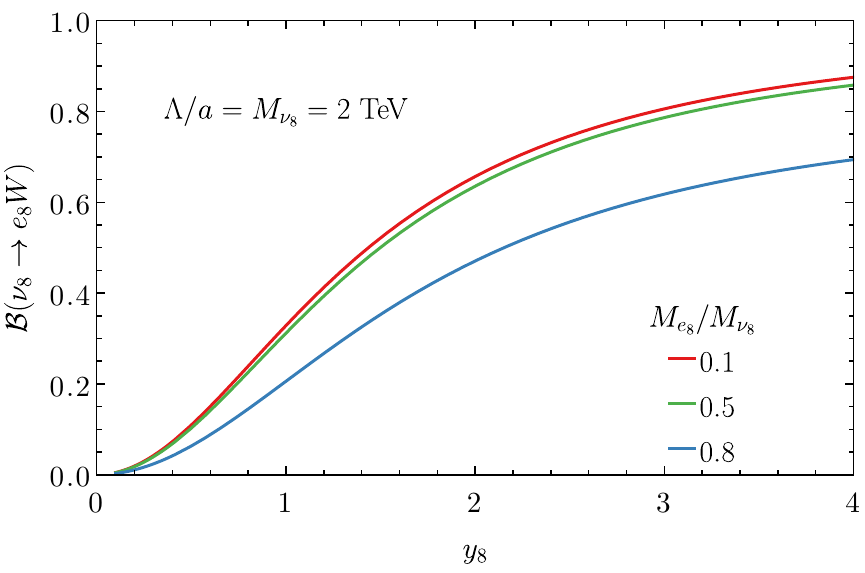}
	\caption{The partial-width-to-mass ratio $\Gamma(\nu_8\to X)/M_{\nu_8}$ (left) and branch fraction for $\nu_8\to e_8 \,W$ (right), as a function the mass ratio $M_{e_8}/M_{\nu_8}$ (upper) and the Yukawa coupling $y_8$ (lower) with fixing $\Lambda/a=M_{\nu_8}$, respectively.
    }
	\label{fig:Width}
\end{figure}
With benchmark Yukawa values $y_8=1/2/3$ or mass ratios $M_{e_8}/M_{\nu_8}=0.1/0.5/0.8$, we present the partial widths and branch fraction for $\nu_8\to e_8W$ in Fig.~\ref{fig:Width}.
In principle, the off-shell decay $\nu_8\to e_8(W^*\to jj/\ell\nu)$ in Fig~\ref{feyn:NG} (left) will emerge when $0<M_{\nu_8}-M_{e_8}<M_W$. However, we neglect this channel in this work for the sake of three-body phase-space suppression.
We see that with $y_8=3$, we can have the opportunity for the $\nu_8\to e_8W$ branch fraction as an order of a unit, which ensures a large fraction of NGs produced in the neutrino-nucleus scattering to decay into a charged lepton and be detected at L$\nu$C.

\subsubsection{Neutrinogluon ($\nu_8$) production at the L$\nu$C}

\begin{figure}[th!]
\centering
\includegraphics[width=0.3\linewidth]{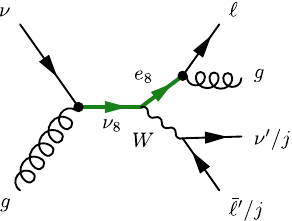}\quad
\includegraphics[width=0.27\linewidth]{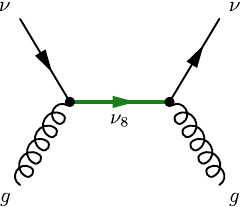}\quad
\includegraphics[width=0.29\textwidth]{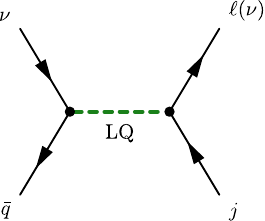}
\caption{Representative Feynman diagrams for the production of a neutrinogluon (left and middle) and leptoquark (right) from the neutrino-nucleus scattering and subsequent decays.
}
\label{feyn:NG}
\end{figure}

As implied above, NGs can be produced by the collision of one neutrino and the gluon parton in neutrino-nucleus scattering, as shown in the left and middle panels of Fig.~\ref{feyn:NG}.
Subsequently, the NG will decay into charged leptons as Fig.~\ref{feyn:NG} (left), or back into a neutrino and gluon as Fig.~\ref{feyn:NG} (middle).
Including the finite-width effect, the parton-level 2-to-2 scattering cross sections are calculated to be
\begin{equation}
\begin{aligned}\label{eq:partonic}
		\hat{\sigma}(\nu g\to\nu_8\to e_8W)&= \frac{a^2g_s^2\alpha\kappa^2(\hat{s})}{16\,\Lambda^2\,M_W^2\,s_W^2}\, \frac{\hat{s}\,[(\hat{s} - M_{e_8}^2)^2 + (\hat{s} + M_{e_8}^2 - 2\,M_W^2)\,M_W^2]}{(\hat{s} - M_{\nu_8}^2)^2 + \Gamma_{\nu_8}^2\,M_{\nu_8}^2}\,\beta(\hat{s}) ~, \\
		\hat{\sigma}(\nu g\to\nu_8\to\nu g) &=\frac{\pi a^4\alpha_s^2}{\Lambda^4}\frac{\hat{s}^3}{(\hat{s}-M_{\nu_8}^2)^2+M_{\nu_8}^2\Gamma_{\nu_8}^2}.
	\end{aligned}
\end{equation}
Here the phase space factor $\beta(\hat{s})$ shares a similar form as Eq.~\eqref{eq:beta} by replacing $M_{\nu_8}^2$ with $\hat{s}$. The mixing angle factor reads
\begin{equation}
	\kappa^2(\hat{s})=\sin^2\theta_L\frac{M_{\nu_8}^2}{\hat{s}}+\sin^2\theta_R
	-\sin\theta_L\sin\theta_R\frac{12M_{\nu_8}M_{e_8}M_W^2}{(\hat{s}-M_{e_8}^2)^2+(\hat{s}+M_{e_8}^2)M_W^2-2M_W^2},
\end{equation}
which returns to Eq.~(\ref{eq:kappa}) with a replacement $\hat{s}\to M_{\nu_8}^2$.

\begin{figure}
	\centering
	\includegraphics[width=0.49\textwidth]{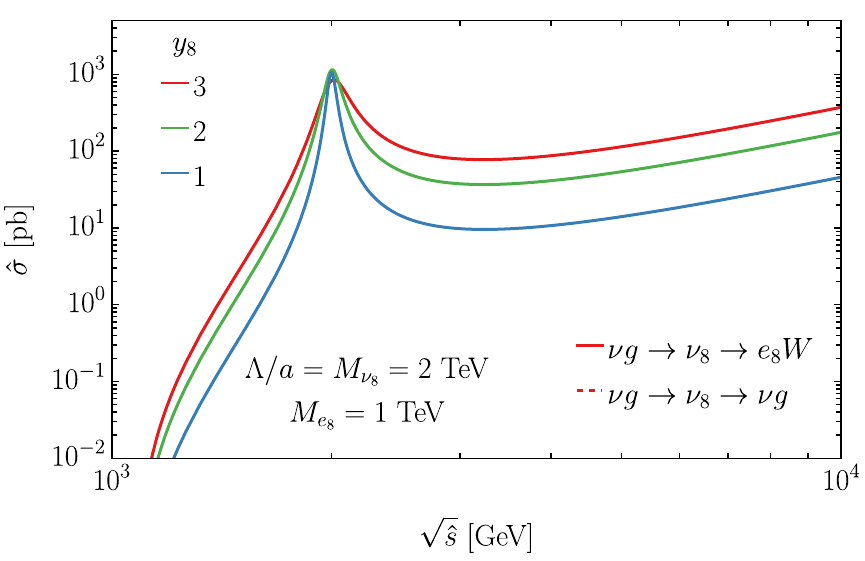}
	\caption{Benchmark partonic cross sections of the NG production and decays.}
	\label{fig:partonic}
\end{figure}

In Fig.~\ref{fig:partonic}, we present the dependence of the partonic cross section on the partonic collision energy $\sqrt{\hat{s}}$.
We see the resonance peak around the $\sqrt{\hat{s}}\sim M_{\nu_8}$, with a spread due to the finite width. However, when $\hat{s}\gtrsim M_{\nu_8}$, both charged- and neutral-current channels give a growing behavior, and will violate unitarity at some point. This is a natural consequence of the dimensional-5 operator in Eq.~(\ref{eq:dim-5-operator}), which gives an $\hat{s}/\Lambda^2$ dependence for the partonic cross section in Eq.~(\ref{eq:partonic}). It is expected to fail and give way to a UV complete model when $\hat{s}\gg M_{\nu_8}$. This ``flaw" can be cured with the narrow width approximation (NWA) as
\begin{equation}
	\frac{1}{(\hat{s}-M_{\nu_8}^2)^2+M_{\nu_8}^2\Gamma_{\nu_8}^2}\to \frac{\pi}{M_{\nu_8}\Gamma_{\nu_8}}\delta(\hat{s}-M_{\nu_8}^2).
\end{equation}
As a consequence, the partonic cross sections become
\begin{equation}
	\hat{\sigma}(\nu g\to\nu_8\to e_8W/\nu g)=
	16\pi^2\delta(\hat{s}-M_{\nu_8}^2)\frac{\Gamma(\nu_8\to\nu g)}{M_{\nu_8}}\mathcal{B}(\nu_8\to e_8W/\nu g),
\end{equation}
where the branch fraction $\mathcal{B}(\nu_8\to e_8W)$ corresponds to Fig.~\ref{fig:Width}.
Folded with the corresponding parton distribution function, the hadronic NG production cross section can be written as
\begin{equation}
	\begin{aligned}\label{eq:XS}
		\sigma(\nu N\to \nu_8\to e_8W/\nu g)
		=f_{g/N}(x,Q^2)\otimes \hat{\sigma}(\hat{s})
		=2\times\frac{4\pi^2}{s}(2J+1)\frac{\Gamma(\nu_8\to\nu g)}{M_{\nu_8}}\mathcal{B}(\nu_8\to e_8W/\nu g)f_{g/N}(M_{\nu_8}^2/s,M_{\nu_8}^2),
	\end{aligned}
\end{equation}
where $\hat{s}=xs$ and $J=1/2$ for its fermionic nature. The factor of 2 comes from the neutrino left-handed chirality, with respect to the electron case in Ref.~\cite{H1:1993vsn}. Here we choose the nominal scale as $Q^2=M_{\nu_8}^2$. In Fig.~3 of the main text, we compare the NG production cross sections of the neutrino-nucleus scattering for the NWA and the 2-to-2 partonic scattering with the explicit $\nu_8$ propagator. We see that the two approaches agree with each other very well, except at extremely high neutrino energies with a relatively small $M_{\nu_8}$. For example, when $E_\nu>10^{10}~\GeV$ for $M_{\nu_8}=2~\TeV$, the $\nu_8$ propagator approach gives larger predictions than NWA, which is exactly due to the high-energy partonic behavior $\hat{s}\gg M_{\nu_8}$ in Eq.~(\ref{eq:partonic}) and Fig.~\ref{fig:partonic}.

\begin{figure}
	\centering
	\includegraphics[width=0.49\textwidth]{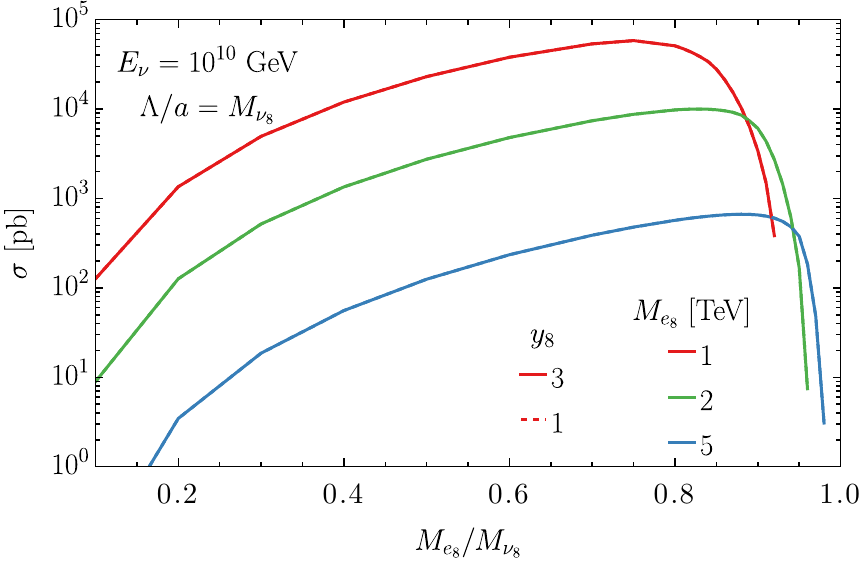}
	\includegraphics[width=0.49\textwidth]{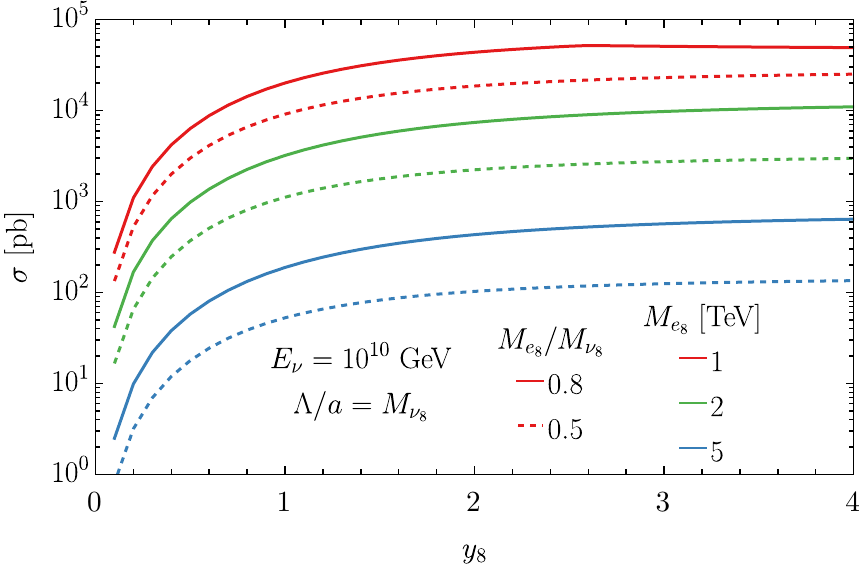}
	\caption{The neutrinogluon production cross section as a function of the mass ratio $M_{e_8}/M_{\nu_8}$ (left) and the Yukawa coupling $y_8$ (right) with fixed $M_{e_8}=1/2/5$~TeV, $E_\nu=10^{10}~\GeV$, and $\Lambda/a=M_{\nu_8}$.}
	\label{fig:app:xsLG}
\end{figure}

For a general LG model, direct searches can be done at colliders, such as HERA~\cite{H1:1993vsn}, LHC~\cite{Goncalves-Netto:2013nla,Mandal:2016csb,Almeida:2022udp} and muon collider~\cite{Han:2025wdy}. However, these searches only apply to the lighter charged component $e_8$. For the convenience of comparison later, we also present the dependence of the cross section on the mass ratio $= M_{\mu_8}/ M_{\nu_8}$, by fixing the LG mass $M_{e_8} = M_{\mu_8}$ in Fig.~\ref{fig:app:xsLG} (left). We see that with the increase of mass ratio, the NG production cross section normally increases, due to the decrease of the NG mass $M_{\nu_8}$. However, after a critical point approaching $M_{e_8}\sim M_{\nu_8}$, the cross sections decrease, resulting from the phase space suppression of the phase space $\beta$ factor in Eq.~(\ref{eq:beta}).

\subsubsection{Search for leptogluons and projected constraints}

\begin{figure}[th!]
	\centering
	\includegraphics[width=0.32\textwidth]{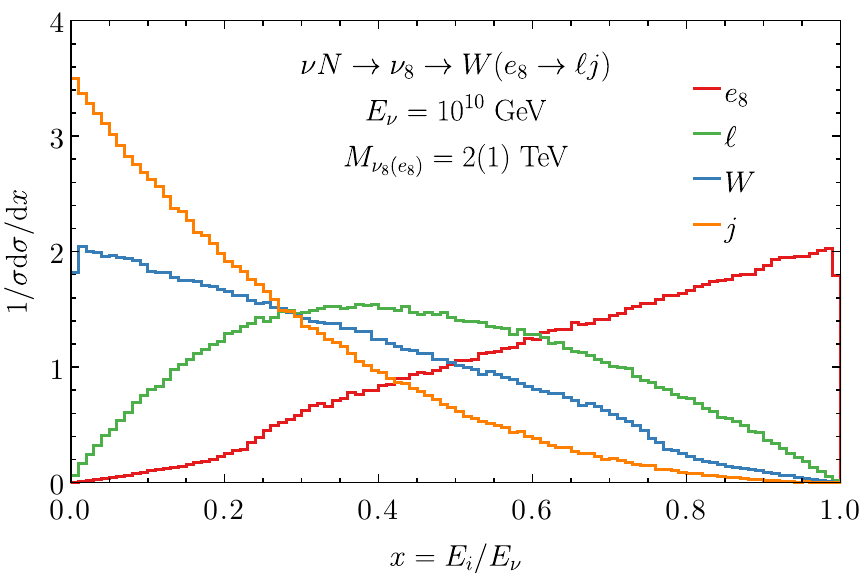}
    \includegraphics[width=0.32\textwidth]{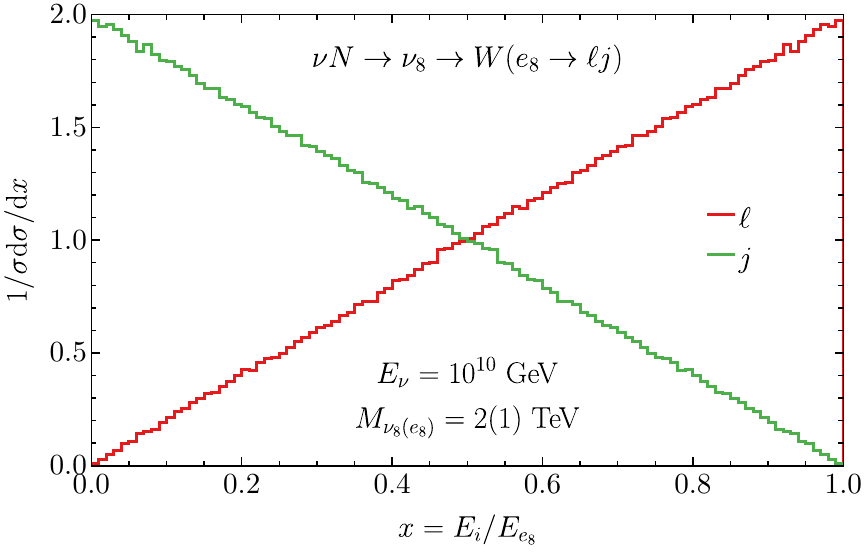}
	\includegraphics[width=0.32\textwidth]{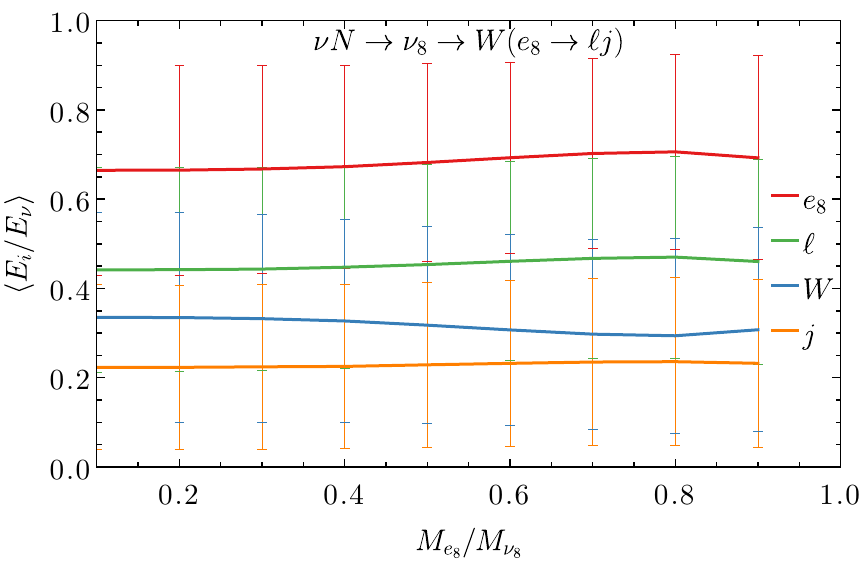}
	\caption{The energy distributions of final states and the average $\langle E_i/E_\nu\rangle$ with standard deviation as error bar.}
	\label{fig:dist}
\end{figure}

The leptogluon model has rich phenomena in large-volume neutrino telescopes, including
\begin{itemize}[noitemsep]
\item $\nu_\mu$CC-like events, from $\nu N \to\nu_8 \to (\mu_8\to\mu g) W$;
\item dimuons, if the $W$ decays to a muon directly or through a tau lepton;
\item different inelasticity distribution from the SM, and shower-like events.
\end{itemize}
As mentioned at the beginning of this section, we focus on the $\nu_\mu$CC-like events, including both starting and throughgoing muon-like events, and follow the procedure therein to calculate the sensitivities of large-volume neutrino observatories to the leptogluon model.

In Fig.~\ref{fig:dist}, we show the energy fraction $x_i=E_i/E_\nu$ distributions of the final-state particles in the neutrinogluon production and its subsequent decay. We see that due to the spin correlation, we have an approximately linear behavior for $e_8$ and $W$, as direct decay products of the $\nu_8$. As a consequence, the average energy fractions behave as
\begin{equation}\label{eq:xe8}
\langle x_{e_8}\rangle=\frac{2}{3}, ~\langle x_{W}\rangle=\frac{1}{3}.
\end{equation}
Sequentially, the charged leptogluon $e_8$ will decay into a charged lepton and a gluon jet. Again, the energy fractions of charged lepton $\ell$ and jet with respect to the mother particle $e_8$ behave linearly as shown in Fig.~\ref{fig:dist} (middle), resulting in similar average fractions of $2/3$ and $1/3$, respectively. When folded in the full system of $\nu N\to\nu_8$, the corresponding distribution is shown as the green and orange lines in Fig.~\ref{fig:dist} (left). The corresponding average energy fractions are
\begin{equation}\label{eq:xl}
\langle x_{\ell}\rangle=\frac{4}{9},\quad \langle x_{j}\rangle=\frac{2}{9}.
\end{equation}
Finally, we also examine the dependence of the average energy fractions on the mass ratio $M_{e_8}/M_{\nu_8}$ in Fig.~\ref{fig:dist} (right), which show a good constant behavior as Eq.~(\ref{eq:xe8}) and (\ref{eq:xl}).

\begin{figure}[th!]
\includegraphics[width=0.96\columnwidth]{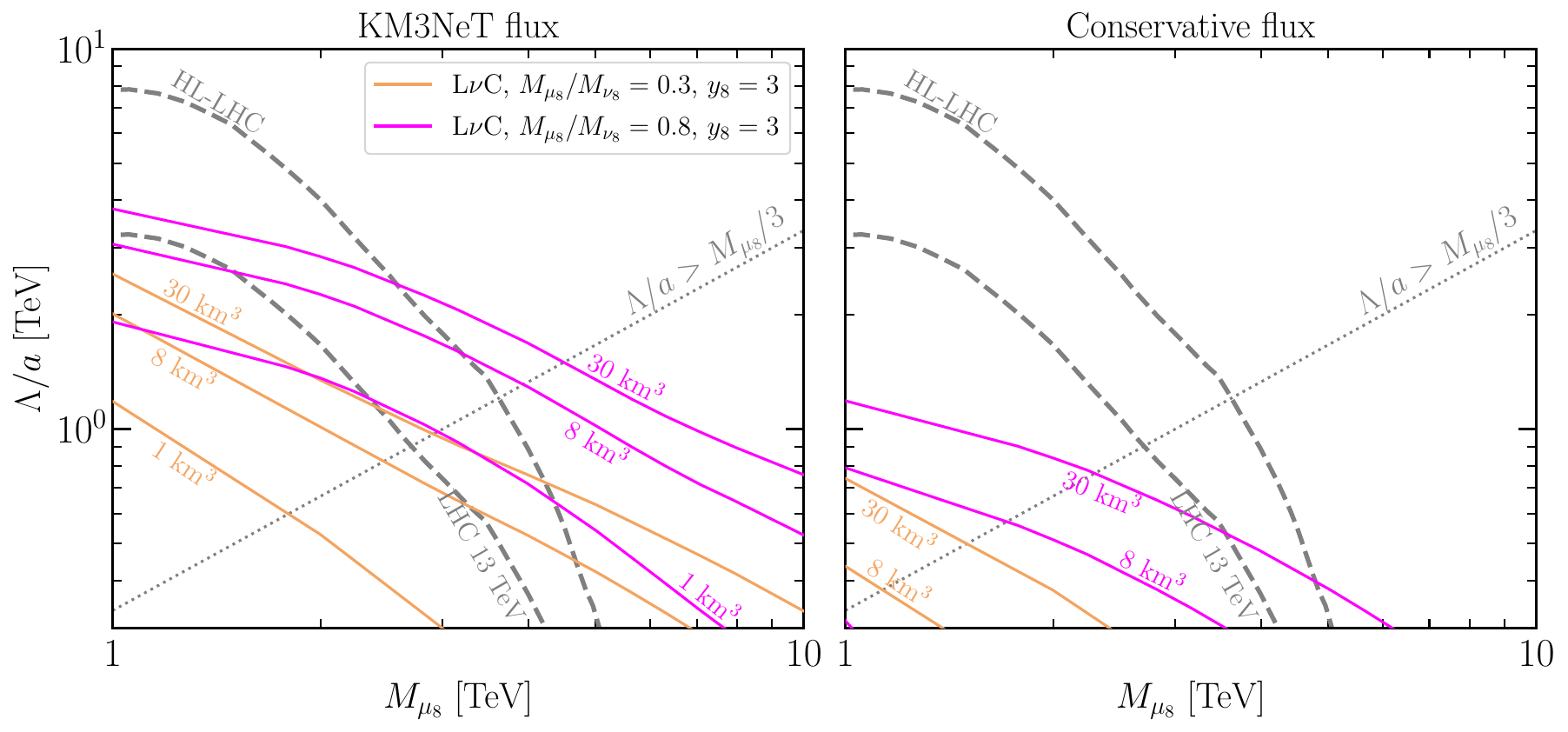}

\caption{
Sensitivity of L$\nu$Cs to the leptogluon in the $(M_{\mu_8},\Lambda/a)$ plane for $M_{e_8}=M_{\mu_8}$, $M_{\mu_8}/M_{\nu_8}=0.8$ and 0.3, and $y_8=3$. The left and right panels use the KM3NeT and conservative fluxes, respectively. The $M_{\mu_8}/M_{\nu_8}=0.8$ result is also shown in Fig.~3 of the main text. The LHC and HL-LHC bounds are recast from Ref.~\cite{Almeida:2022udp}.}
\label{Fig_LG_sensitivity_appdx}
\end{figure}

{
Similar to App.~\ref{app:4f}, we take the energy fraction $\langle E_\ell/E_\nu\rangle=4/9$ (where $\ell=\mu$ here) in Eq.~(\ref{eq:Eavg}) and estimate the starting and throughgoing muon event rates. With the likelihood measure $\sqrt{q_0}=1.96$, we present the sensitivity of L$\nu$Cs to the LGs in the parameter space $(M_{\mu_8},\Lambda/a)$ in Fig.~\ref{Fig_LG_sensitivity_appdx}, by fixing $M_{\mu_8}/M_{\nu_8}=0.8(0.3)$ and $y_8=3$. Here we choose $M_{e_8}=M_{\mu_8}$ instead of $M_{\nu_8}$ for the convenience of directly comparing with the LHC bounds on the charged leptogluon~\cite{Almeida:2022udp}. For the KM3NeT flux, the L$\nu$Cs can directly probe the neutrinogluon or leptogluon in the multi-TeV regime, depending on the chromo-dipole operator coefficient $a/\Lambda$. In comparison with the LHC and HL-LHC exclusion limits, the L$\nu$Cs have an advantage at large $M_{\mu_8}=M_{e_8}$, thanks to their large COM energy. The conservative flux gives a weaker but still complementary reach. In contrast, the LHC and HL-LHC are more sensitive at low $M_{\mu_8}$, mainly owing to their higher luminosities. The mass ratio $M_{\mu_8}/M_{\nu_8}=0.8$ gives a better probe than 0.3 as a result of the corresponding larger cross section, as shown in Fig.~\ref{fig:app:xsLG} (left). For this reason, we demonstrate the mass ratio of 0.8 in Fig.~3 of the main text. In addition, we add a dotted line for $\Lambda/a>M_{\mu_8}/3$ to denote a weak boundary of the valid region of the chromo-dipole operator, by requiring $\Lambda>M_{e_8}=M_{\mu_8}$ with the implicit assumption $a\sim3$.
With the benchmark $\Lambda/a=M_{\mu_8}/2$, we compare the probed LG mass $M_{\mu_8}$ for different L$\nu$Cs and both UHE-flux scenarios with the corresponding LHC and HL-LHC reaches in Fig.~1 of the main text.
}

\subsection{Leptoquark}
\label{app:LQ}

Similar to the leptogluon model, we can have leptoquarks (LQs) predicted by many BSM unification theories, such as the Pati-Salam model~\cite{Pati:1974yy} or superstring models~\cite{Dobado:1987pj}.
As the name suggests, leptoquarks directly couple to leptons and quarks, which can be either scalar or vector bosons.
As a consequence, LQs can be produced in the neutrino-nucleus scattering and therefore detected at L$\nu$Cs.
Some theoretical studies~\cite{Anchordoqui:2006wc,Romero:2009vu,Becirevic:2018uab,Babu:2022fje,Kirk:2023fin} about specific types of scalar LQ models have been performed. Phenomenology at the LHC have been extensively studied ~\cite{Baker:2019sli,DiLuzio:2018zxy,Cornella:2021sby,Dorsner:2018ynv,Schmaltz:2018nls,Diaz:2017lit,Borschensky:2021hbo,Borschensky:2022xsa,Bhaskar:2023ftn,Das:2025osr}.

\begin{table}[hb!]
	\centering    \renewcommand{\arraystretch}{1.5}
	\begin{tabular}{c|c|c|c|c}
		\hline
		Spin & $SU(2)_W$ & $Y$ & Notation & Interactions \\
		\hline
		0 & 1 & 1/3 & $S_1^{1/3}$ & $S_1^{1/3}(y_1^{LL}\bar{q}_L^c\ell_L+y_1^{RR}\bar{u}_R^ce_R)$\\
		0 & 2 & 7/6 & $R_2=(R_2^{5/3},R_2^{2/3})$ & $R_2(y_2^{LR}\bar{q}_L^ce_R+y_2^{RL}\bar{u}_R\ell_L)$ \\
		0 & 3 & 1/3 & $S_3=\begin{pmatrix} S_3^{1/3} & S_3^{4/3} \\ S_3^{-2/3} & -S_3^{1/3} \end{pmatrix}$ &
		$y_3^{LL}\bar{q}_L^cS_3\ell_L$ \\
		\hline
		1 & 1 & 2/3 & $V_1^{2/3}$ & $V_{1\mu}^{2/3}(g_1^{LL}\bar{q}_L\gamma^{\mu}\ell_L+g_1^{RR}\bar{d}_R\gamma^\mu e_R)$\\
		1 & 2 & 5/6 & $V_2=(V_2^{4/3},V_2^{1/3})$ & $V_{2\mu}(g_2^{LR}\bar{q}_L^c\gamma^\mu e_R+g_2^{RL}\bar{d}^c_R\gamma^\mu\ell_L)$ \\
		1 & 3 & 2/3 & $V_3=\begin{pmatrix} V_3^{2/3} & V_3^{5/3} \\ V_3^{-1/3} & -V_3^{2/3} \end{pmatrix}$ &
		$g_3^{LL}\bar{q}_LV_{3\mu}\gamma^\mu\ell_L$ \\
		\hline
	\end{tabular}
	\caption{The possible leptoquarks that can be produced in the neutrino-nucleus scattering and decay into charged leptons.}
	\label{tab:LQ}
\end{table}

Following the Particle Data Group~\cite{ParticleDataGroup:2024cfk}, we first list all the possible LQs that can produce $\nu_\mu$CC-like signals in Tab.~\ref{tab:LQ} (see the right panel of Fig.~\ref{feyn:NG}  for the Feynman diagram).
In comparison, the singlet scalar $S_1^{4/3}$ and vector $V_1^{5/3}$ can only couple to quarks and charged leptons, but not to neutrinos.
On the other hand, the doublet scalar $\tilde{R}_{2}=(\tilde{R}_2^{2/3},\tilde{R}_2^{-1/3})$ and vector $\tilde{V}_{2}=(\tilde{V}_2^{1/3},\tilde{V}_2^{-2/3})$ with hyper charges $Y=1/6$ and $-1/6$, respectively, can only give the interactions
\begin{equation}
\tilde{R}_2\bar{d}_R\ell_L=\bar{d}_R(\tilde{R}_2^{-1/3}\nu_L+\tilde{R}_2^{2/3}e_L),\quad
\tilde{V}_{2\mu}\bar{u}_R^c\gamma^\mu\ell_L=\bar{u}_R^c\gamma^\mu(\tilde{V}_{2\mu}^{1/3}e_L+\tilde{V}_{2\mu}^{-2/3}\nu_L).
\end{equation}
As a consequence, these two types of LQs can only induce the NC-like scattering $\nu q\to\textrm{LQ}\to\nu j$. The signal for this type of process will be very different from CC-like events in this study; thus, we leave for future exploration.

\begin{table}[ht!]
 \renewcommand{\arraystretch}{1.5}
	\begin{tabular}{c|c|c|c}
		\hline
		sLQ & Channel & $\Gamma~[M_{S}/16\pi]$ & $\mathcal{B}$  \\
		\hline
		\multirow{2}{*}{$S_1^{-1/3}$}  &  $d\nu$ & $(y_1^{LL})^2$ & 1/3  \\
		& $u\ell$ & $(y_1^{LL})^2+(y_1^{RR})^2$ & 2/3\\
		\hline
		\multirow{2}{*}{$R_2^{2/3}$}  &  $u\nu$ & $(y_2^{RL})^2$ & 1/2  \\
		& $d\ell$ & $(y_2^{LR})^2$ & 1/2\\
		\hline
		\multirow{2}{*}{$S_3^{-1/3}$}  &  $d\nu$ & $(y_3^{LL})^2$ & 1/2  \\
		& $u\ell$ & $(y_3^{LL})^2$ & 1/2 \\
		\hline
	\end{tabular} \hspace{2cm}
	\begin{tabular}{c|c|c|c}
		\hline
		vLQ & Channel & $\Gamma~[M_{V}/24\pi]$ & $\mathcal{B}$ \\
		\hline
		\multirow{2}{*}{$V_1^{-2/3}$}  &  $u\nu$ & $(g_1^{LL})^2$ & 1/3  \\
		& $d\ell$ & $(g_1^{LL})^2+(g_1^{RR})^2$ & 2/3 \\
		\hline
		\multirow{2}{*}{$V_2^{-1/3}$}  &  $d\nu$ & $(g_2^{RL})^2$ & 1/2  \\
		& $u\ell$ & $(g_2^{LR})^2$ & 1/2 \\
		\hline
		\multirow{2}{*}{$V_3^{-2/3}$}  &  $u\nu$ & $(g_3^{LL})^2$ & 1/2 \\
		& $d\ell$ & $(g_3^{LL})^2$ & 1/2 \\
		\hline
	\end{tabular}
	\caption{The partial decay widths $\Gamma$ and benchmark branch fractions $\mathcal{B}$ with the parameter choice in Eq.~(\ref{eq:coupling}) for the relavent scalar and vector LQs.}
	\label{tab:Width}
\end{table}

For the possible LQs listed in Tab.~\ref{tab:LQ}, we first identify the possible CC-like partonic processes of Fig.~\ref{feyn:NG} (right) at L$\nu$C as
\begin{equation}
	\begin{aligned}
		&\nu d\to S_1^{-1/3}(S_3^{-1/3})\to \ell^-u, \quad&
		&\nu\bar{u}\to R_2^{-2/3}\to \ell^-\bar{d},\\
		&\nu\bar{u}\to V_1^{-2/3}(V_3^{-2/3})\to \ell^-\bar{d}, \quad&
		&\nu d\to V_2^{-1/3}\to \ell^-u.\\
	\end{aligned}
\end{equation}
As pointed out in Refs.~\cite{Becirevic:2018uab,Kirk:2023fin}, the $t$-channel quark and gluon initiated processes also emerge starting from one QCD higher order. However, we realize that those $t$-channel processes are part of the next-to-leading order (NLO) correction to the inclusive LQ production, while the virtual correction from the loop contribution is not included. The complete NLO calculation to the inclusive LQ production only gives corrections less than 20\%, which is negligible to our analysis. Therefore, we mainly perform our simulation based on the leading-order prediction in the rest of this work.

Before presenting the corresponding cross sections, we summarize the relevant partial decay widths in Tab.~\ref{tab:Width}.
In this work, we fix the Yukawa and gauge-like couplings as
\begin{equation}\label{eq:coupling}
	\begin{aligned}
		&y^{LL}_{1,ij}=y^{RR}_{1,ij}=y^{LR}_{2,ij}=y^{RL}_{2,ij}=y^{LL}_{3,ij}=y\beta_{ij},\\
		&g^{LL}_{1,ij}=g^{RR}_{1,ij}=g^{LR}_{2,ij}=g^{RL}_{2,ij}=g^{LL}_{3,ij}=g\beta_{ij},\\
	\end{aligned}
\end{equation}
where the quark and lepton flavor index $\beta_{12}=1$ for the $\nu_\mu$CC-like signal, while other elements are zeros for simplicity.
As a consequence, the corresponding decay widths and branch fractions in the massless limit of final states are summarized in Tab.~\ref{tab:Width}.

\begin{figure}[th!]
	\centering
	\includegraphics[width=0.49\linewidth]{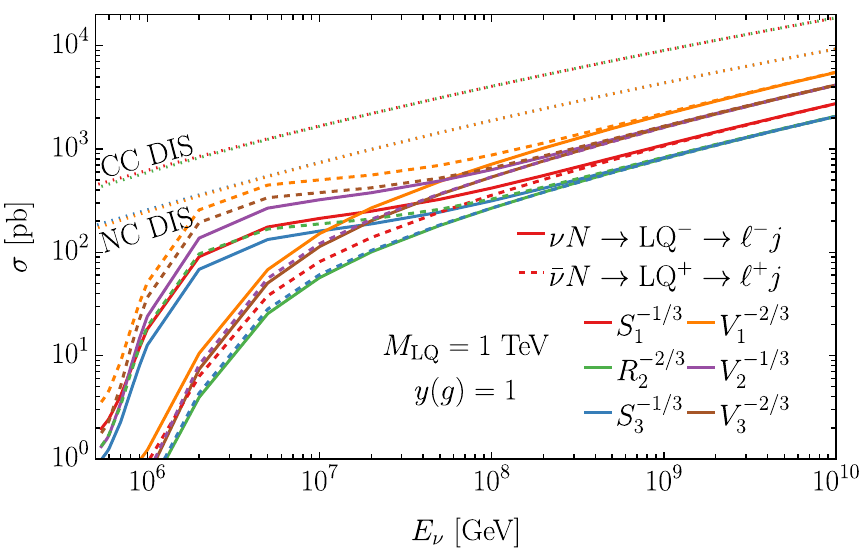}
	\includegraphics[width=0.49\textwidth]{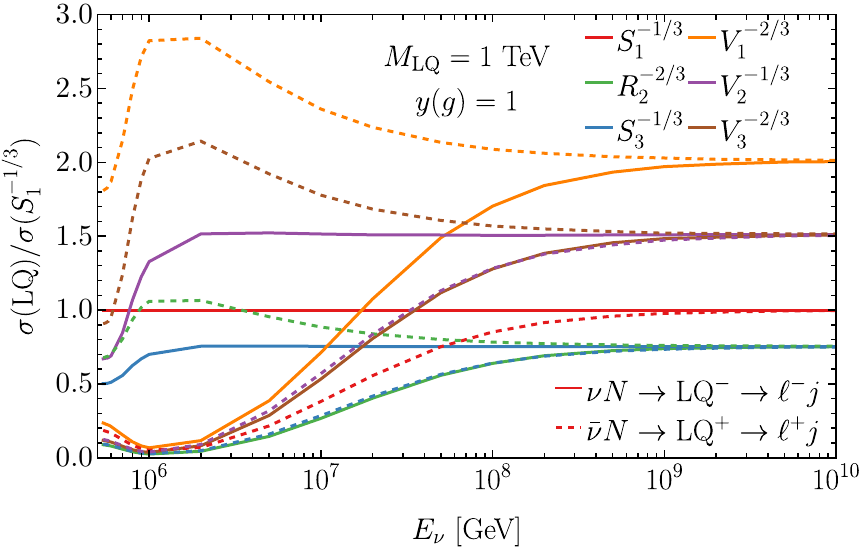}
	\caption{Benchmark production cross sections for the scalar and vector LQs with $M_{\rm LQ}=1~\TeV$ in the neutrino-nucleus scattering.}
	\label{fig:xsLQ}
\end{figure}

The LQ production cross section at the nucleon level shares a similar form as Eq.~(\ref{eq:XS}), with the corresponding branching fraction $\mathcal{B}$, spin factor $J=0$ or 1 for scalar and vector LQs, and quark PDFs. With the Yukawa parameter choice, in Eq.~(\ref{eq:coupling}), the benchmark production cross sections are presented in Fig.~\ref{fig:xsLQ}, with the corresponding ratio normalized to the $S_1^{1/3}$ shown on the right.
We see that, for a specific LQ and its antiparticle $\overline{\textrm{LQ}}$, the cross section difference mainly arises from the valence $d,u$ and sea $\bar{d},\bar{u}$ quarks. However, with $E_\nu$ increasing, LQ and $\overline{\textrm{LQ}}$ cross sections gradually become the same. It can be understood in terms of the $q\simeq\bar{q}$ at a small $x\sim M_{\rm LQ}^2/s$ in Eq.~(\ref{eq:XS}). Meanwhile, the iso-spin symmetry for the iso-scalar target gives $u\simeq d$ and ends up with a universal scaling for LQ production cross section in Fig.~\ref{fig:xsLQ} (right). The corresponding ratios can be well estimated with the $J$ factor as well as the branch fractions in terms of Tab.~\ref{tab:Width}, which only receive sizeable corrections around the threshold region.

\begin{figure}[th!]
    \centering
    \includegraphics[width=0.32\linewidth]{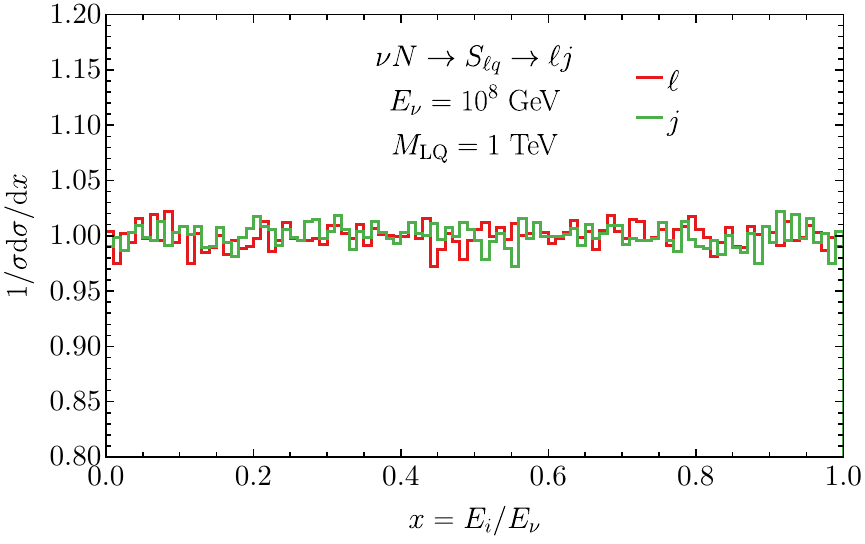}
    \includegraphics[width=0.32\linewidth]{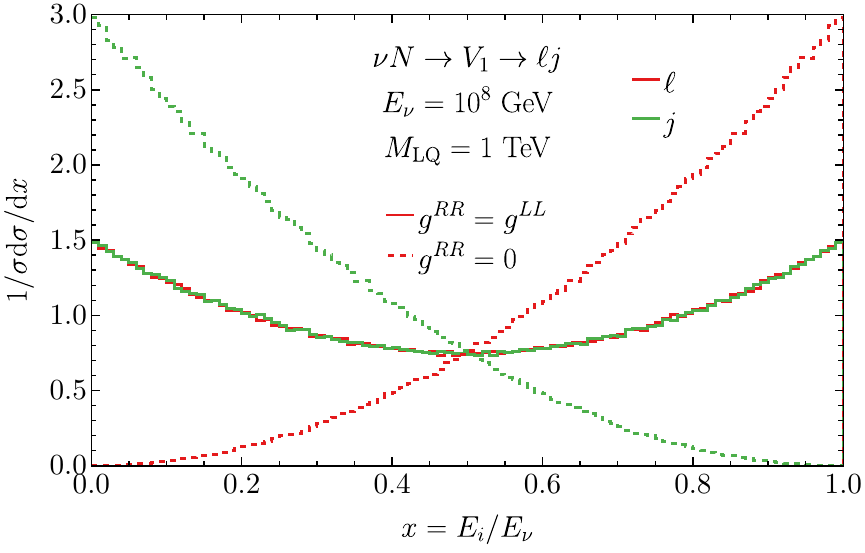}
    \includegraphics[width=0.32\linewidth]{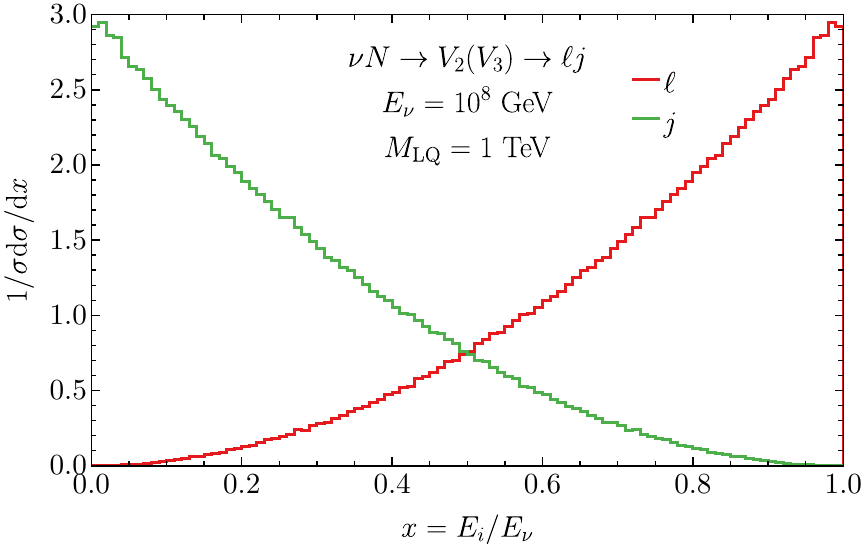}
    \caption{The energy distribution of final-state particles in the scalar and vector leptoquark production. For the singlet vector LQ, we demonstrate two parameter choices of $g^{RR}=g^{RR}$ and $g^{RR}=0$ as solid and dashed lines, respectively. The rest of them correspond to the parameter choice in Eq.~(\ref{eq:coupling}).}
    \label{fig:distLQ}
\end{figure}

In Fig.~\ref{fig:distLQ}, we present the fractional energy $x_i=E_i/E_\nu$ distribution of the final-state particles in the LQ production and its subsequent decay. The corresponding average energy fractions for charged leptons are
\begin{equation}\label{eq:xlLQ}
\langle x_\ell(S)\rangle=\langle x_\ell(V_1)\rangle=\frac{1}{2},\quad
\langle x_{\ell}(V_{2,3})\rangle=\frac{3}{4}.
\end{equation}
For the scalar LQ, the energy distribution is flat in all three types, leading to an average energy fraction as $\langle x_\ell\rangle=1/2$.
For the doublet and triplet vector LQs, $V_2$ and $V_3$, the specific chiral interaction in Tab.~\ref{tab:LQ} leads to a more energetic charged lepton, and the corresponding energy fraction is $3/4$.
In comparison, we obtain a parabolic behavior for both the lepton and the jet for the singlet vector LQ $V_1$.
It can be understood in terms of the two chiral interactions of $g^{LL}$ and $g^{RR}$ in Tab.~\ref{tab:LQ}. As a consequence, both the energy distributions of final-state particles are a superposition of the $g^{RR}$ and $g^{LL}$ interactions. In turn, the charged lepton's average energy fraction is also 1/2. If we turn off the $g^{RR}$ interaction, the charged lepton and jet energy distributions for $V_1$ is shown as the corresponding dashed lines in Fig.~\ref{fig:distLQ} (middle), which become identical to the $V_2$ and $V_3$ cases (right).

\begin{figure}[th!]
\centering
\includegraphics[width=0.48\columnwidth]{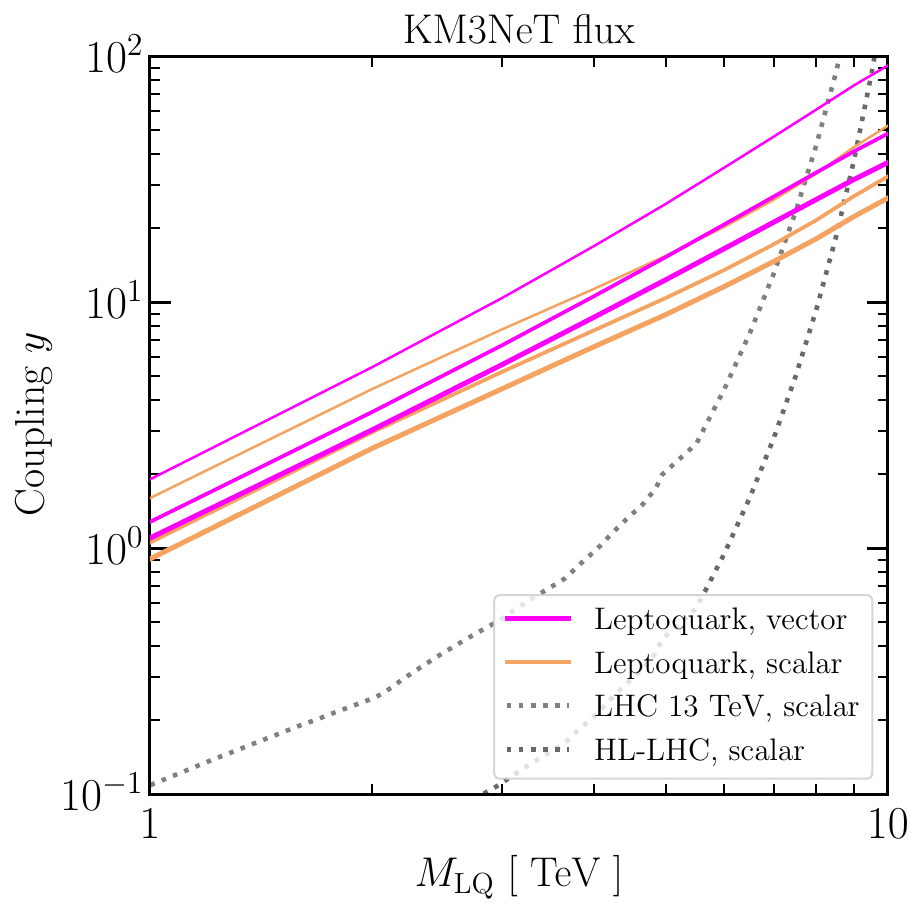}
\includegraphics[width=0.48\columnwidth]{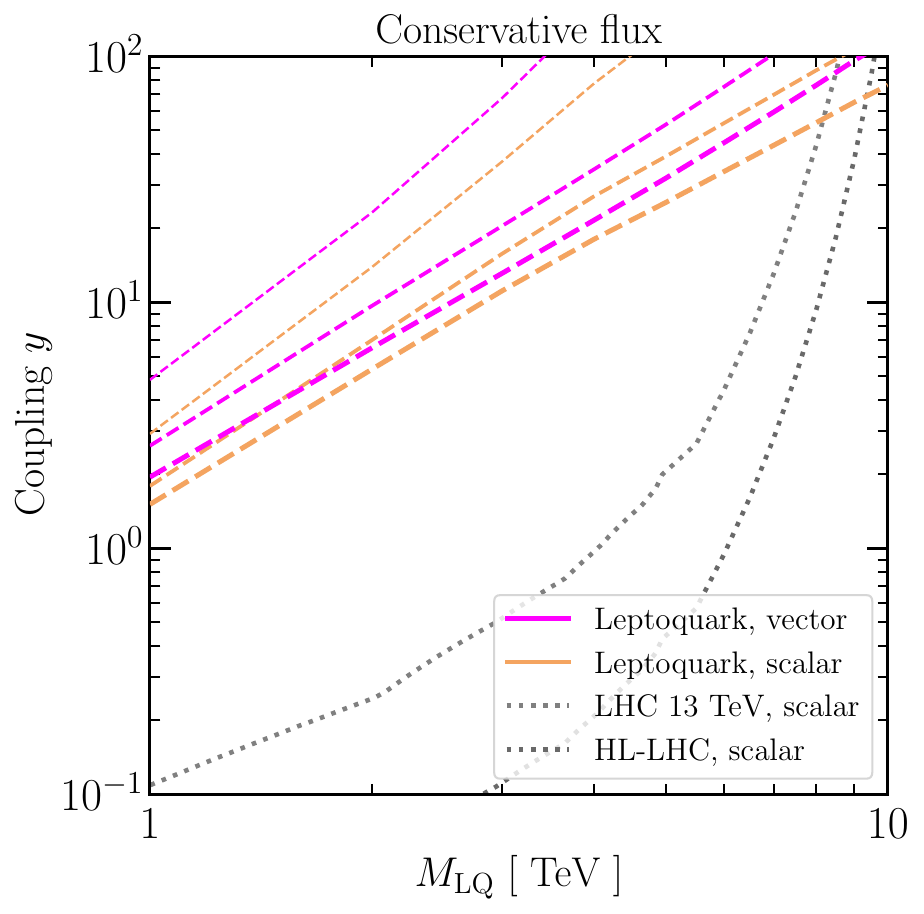}
\caption{Our projected leptoquark sensitivities using starting and throughgoing tracks in large-volume neutrino telescopes. The left and right panels use the KM3NeT and conservative fluxes, respectively. In each panel, magenta and orange curves represent the vector $V_1^{-2/3}$ and scalar $S_1^{-1/3}$ models; the three curves correspond to 1, 8, and 30~km$^3$ volumes and 20 years of observation. Gray dashed curves show the $S_1^{-1/3}$ LHC limit at 138~fb$^{-1}$~\cite{Campana:2024} and the HL-LHC projection at 3~ab$^{-1}$.}
\label{Fig_LQ_sensitivity}
\end{figure}

{
As before, we take the average energy distribution $\langle E_\ell/E_\nu\rangle$ in Eq.~(\ref{eq:xlLQ}) for Eq.~(\ref{eq:Eavg}) to estimate the event rates. Based on the $\sqrt{q_0}=1.96$ measure for 95\% CL, the L$\nu$Cs' sensitivities to the LQs under the KM3NeT and conservative flux scenarios are shown in Fig.~\ref{Fig_LQ_sensitivity}. Here we only present $S_1^{-1/3}$ and $V_1^{-2/3}$ for demonstration, while other types can be projected similarly. For the KM3NeT flux, the L$\nu$C can probe the LQ up to 10~TeV, depending on the corresponding couplings, while the conservative flux gives a weaker reach.
Interestingly, the sensitivity to the vector LQ is slightly weaker than that to the scalar one, even with a slightly larger production cross section. This can be understood from Earth absorption, which reduces the vector-LQ event statistics more significantly than in the scalar case. For the KM3NeT flux, the L$\nu$C can gain new parameter space above 8~TeV with a coupling around $4\pi$, which approaches the unitarity bound. The conservative flux shifts this accessible region toward lower masses. Similar to the leptogluon scenario, we take the scalar LQ $S_1^{-1/3}$ with the Yukawa coupling $y=3$ as a benchmark and compare the exclusion limits of L$\nu$Cs under both flux scenarios with those of the LHC~\cite{Campana:2024} and HL-LHC. The comparison is presented in Fig.~1 of the main text as an example of lepton--quark collisions.
}
```

\subsection{Neutrino-electron scattering: $W'/H^-$ resonance production}
\label{app:Wprime}

\begin{figure}[th!]
\centering
\includegraphics[width=0.49\textwidth]{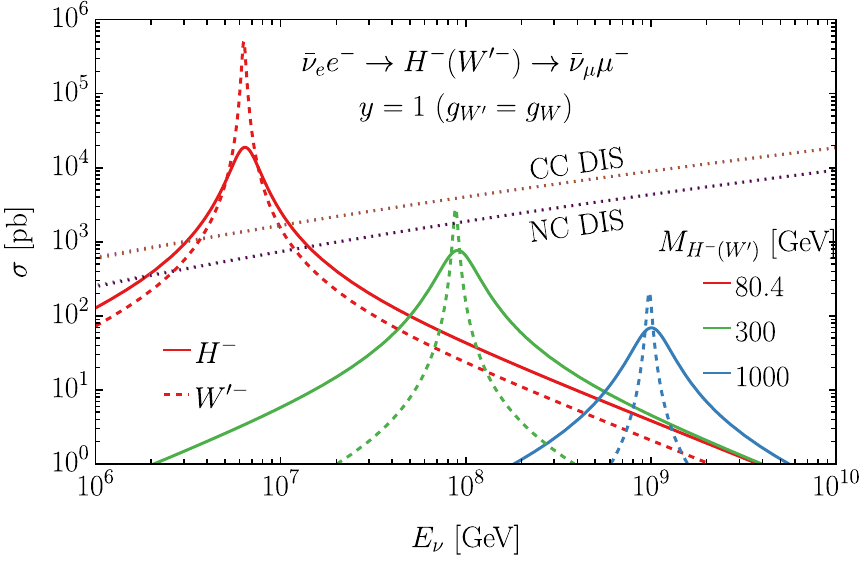}
\includegraphics[width=0.48\columnwidth]{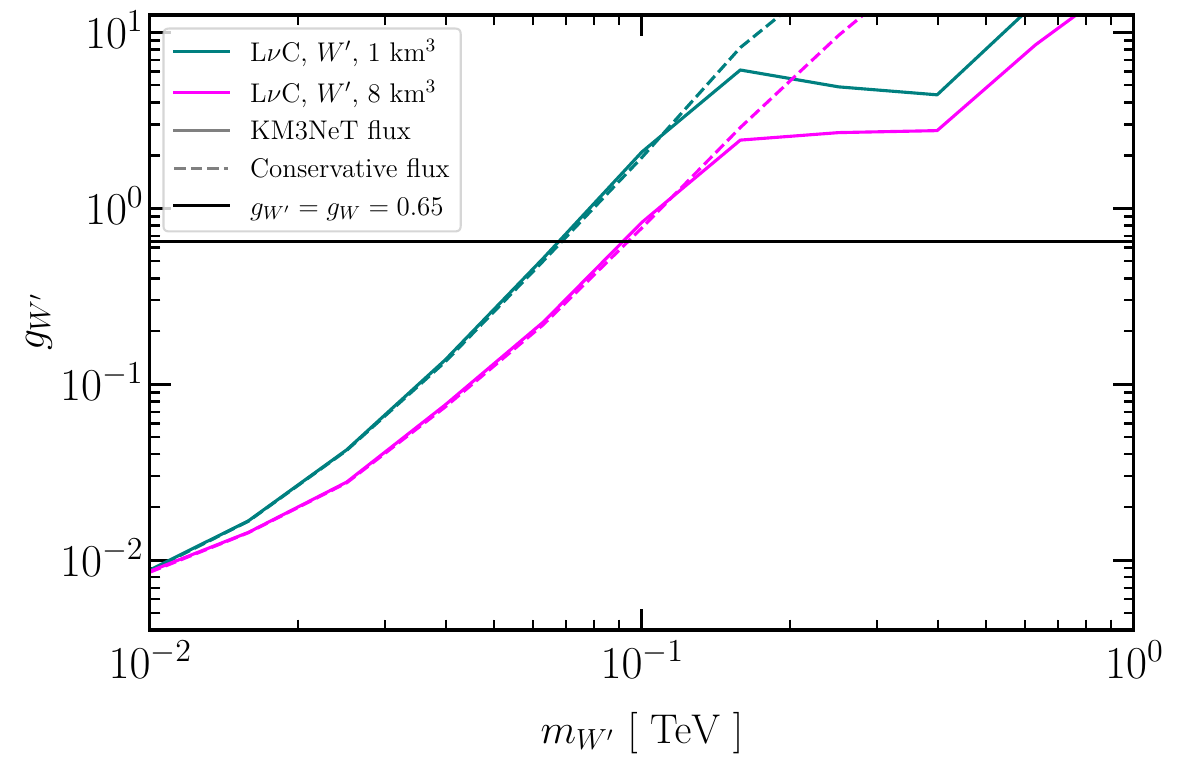}
\caption{\textbf{Left}: Production cross sections for vector $W'$ and scalar $H^-$ resonances compared with SM CC and NCDIS. \textbf{Right}: Exclusion limits in the $(M_{W'},g_{W'})$ plane for different L$\nu$C volumes. Solid and dashed curves use the KM3NeT and conservative fluxes, respectively. We omit the 30~km$^3$ configuration because its 10--100~TeV threshold is still being finalized~\cite{Huang:2023mzt}.
}
\label{fig:Wprime}
\end{figure}

For the $\bar{\nu} + e$ colliders, one could probe a similar class of models. Furthermore and analogous to the Glashow resonance for the $W$ gauge boson, one could have a vector boson $W'$, with the production cross section
\begin{eqnarray}\label{eq:XSWprime}
	\sigma(\bar{\nu}_e e^-\to W'\to\bar{\nu}\mu^-)=
	8\pi(2J+1)\frac{\Gamma(W'\to\bar{\nu}_ee^-)\Gamma(W'\to\bar{\nu}_\mu\mu^-)}{M_{W'}^2}\frac{s}{(s-M_{W'}^2)^2+M_{W'}^2\Gamma_{W'}^2}.
\end{eqnarray}
Here, the decay widths are
\begin{equation}\label{eq:WidthWprime}
	\Gamma(W'\to\bar{\nu}_ee^-/\bar{\nu}_\mu\mu^-)=\frac{g_{W'}^2M_{W'}}{(2J+1)16\pi}, \quad
	\Gamma_{W'}=(N_\ell+N_c N_q)\Gamma(W'\to\nu_ee^-),
\end{equation}
in the massless limit of the final-state fermions. $N_\ell=3$ and $N_q=2(3)$ if $M_{W'}$ is smaller (larger) than top quark mass $M_t$. The spin factor is $J=1$ for the vector boson $W'$. Among many $W^\prime$ models, we choose the ``Sequential Standard Model (SSM)"~\cite{Altarelli:1989ff}, in which the $W^\prime$ couplings to the fermions with the same strength as the $W$ boson (\emph{i.e.}, $g_{W'}=g_W$) but with suppressed couplings to SM bosons. Similarly, we can also have a charged scalar production through $\bar{\nu}_ee^-\to H^-\to\bar{\nu}_\mu\mu^-$. The cross section shares a similar form as Eq.~(\ref{eq:XSWprime}), with a spin factor $J=0$, mass $M_{H^-}$, and coupling switched to $y$ in Eq.~(\ref{eq:WidthWprime}). Here we take a universal Yukawa coupling for all SM fermions and ignore the bosonic decay channels of $\Gamma_{H^-}$ for simplicity. With a few representative $M_{W'}$ and $M_{H^-}$ values, the $W'/H^-$ resonance production cross sections are presented in the left panel of Fig.~\ref{fig:Wprime}. We see that if $M_{W'}>1~\TeV$, the resonance production is completely negligible with respect to charged- and neutral-current DIS.

The energy fraction distributions of final-state charged lepton and neutrino behave similarly to Fig.~\ref{fig:distLQ} (right) for $W'$ resonance, while the charged lepton distribution switches to the quark one.
For the scalar $H^-$, the energy fraction is flat as Fig.~\ref{fig:distLQ} (left). In turn, the average energy fractions are
\begin{equation}
\langle x_{\ell}(W')\rangle=\frac{1}{4},\quad
\langle x_\ell(H^-)\rangle=\frac{1}{2}.
\end{equation}

{
Varying $(M_{W'},g_{W'})$, we present the sensitivity of L$\nu$Cs in Fig.~\ref{fig:Wprime} (right), where the horizontal line corresponds to the SSM coupling $g_{W'}=g_W=0.65$. For the KM3NeT flux, the L$\nu$Cs can exclude the $W'$ boson up to $M_{W'}\sim70$ and 100~GeV with instrumented volumes of 1 and 8~km$^3$, respectively. The conservative flux gives a weaker reach, as shown by the dashed curves. Recently, the IceCube Collaboration observed the Glashow resonance (i.e., the SM $W$ boson) with a significance of up to $2.7\sigma$ using the shower together with secondary muons from the hadronic decay~\cite{IceCube:2021rpz}. In comparison, we only consider muon tracks in this work, consistently with the other BSM searches, while including additional event signatures could improve the sensitivity in future studies.
The current LHC has already excluded the SSM $W'$ up to $M_{W'}>6~\TeV$~\cite{ParticleDataGroup:2024cfk}, and the future HL-LHC is expected to improve the bound to $M_{W'}>8.8~\TeV$, mainly owing to its higher luminosity and larger effective center-of-mass energy compared with the $\bar{\nu}_e e^-$ system at L$\nu$Cs. Therefore, neither UHE-flux scenario provides access to new parameter space for the SSM $W'$ or conventional charged-Higgs $H^-$ models. Nevertheless, the $\bar{\nu}_e e^-$ system functions as a lepton--lepton collider with a center-of-mass energy
$\sqrt{s}\approx\sqrt{2m_eE_\nu}\approx500~\text{GeV}\,(E_\nu/250~\text{PeV})^{1/2}$,
which can surpass the LEP energy of 209~GeV and even the projected COM energies of the proposed FCC-ee and CEPC colliders~\cite{FCC:2018evy,CEPCStudyGroup:2023quu,CEPCStudyGroup:2025kmw}, and may provide an efficient probe of other BSM models.
}

\textbf{A final remark}. Before concluding this appendix, we want to emphasize that we mainly focus on the $\nu_\mu$CC-like starting and thoroughgoing events in this work, as the leading signal at the L$\nu$Cs. Other subleading signatures, in particular related to the neutral-current (NC) scattering, can be searched as well, in spite of higher backgrounds. It can expand the sensitivity to other related models, such as extra-dimension theories~\cite{Alvarez-Muniz:2001efi,Alvarez-Muniz:2002snq,Anchordoqui:2005ey} or heavy neutral leptons~\cite{BookMotzkin:2024qgd,IceCube:2025kve}, etc. For the model discussed above, the NC signal can also probe some complementary components, including more contact operators, the neutral components of LQs, and a new parameter space of the LG model (e.g., $M_{\nu_8}<M_{e_8}$). We leave all these possibilities for future exploration.

\end{document}